\documentclass{aa}
\usepackage{natbib,longtable}
\bibpunct{(}{)}{;}{a}{}{,} 
\usepackage{txfonts}
\usepackage{graphicx}
\def\nodata{\multicolumn{1}{c}{$\cdots$}}
\def\kms{\nobreak\mbox{$\;$km\,s$^{-1}$}}
\font\teneufm=eufm10
\font\seveneufm=eufm7
\font\fiveeufm=eufm5
\newfam\eufmfam
\textfont\eufmfam=\teneufm
\scriptfont\eufmfam=\seveneufm
\scriptscriptfont\eufmfam=\fiveeufm
\def\frak#1{{\fam\eufmfam\relax#1}}

\begin{document}
\title{New Period-Luminosity and Period-Color Relations of Classical
       Cepheids: II. Cepheids in LMC}
\author{A. Sandage\inst{1} \and G.A. Tammann\inst{2} \and B. Reindl\inst{2}}

\institute{%
The Observatories of the Carnegie Institution of Washington,
813 Santa Barbara Street, Pasadena, CA 91101
\and
Astronomisches Institut der Universit\"at Basel,
Venusstrasse 7, CH-4102 Binningen, Switzerland \\
\email{G-A.Tammann@unibas.ch}}

\date{Received {10 February 2004} / Accepted {11 May 2004}}
\abstract{
Photometric data for 593 Cepheids in the LMC, measured 
by Udalski et al. in the OGLE survey, augmented by 97 longer 
period Cepheids from other sources, are analyzed for the period-
color (P-C) and period-luminosity (P-L) relations, and for the 
variations of amplitude, light curve shape, and period across 
the instability strip at constant absolute magnitude. Both the    
P-C and P-L relations have different slopes for periods smaller 
and larger than 10 days. The break at 10 days is also seen in the 
period-amplitude relations, and the compound Fourier combinations 
of $R_{21}$ and $\Phi_{21}$ introduced by Simon and Lee. 

     The LMC Cepheids are bluer than Galactic Cepheids in the 
$B$,$V$, and $I$ color bands, part of which is due to differential 
Fraunhofer line blanketing and part to real differences in the 
temperature boundaries of the instability strip. The LMC strip is 
hotter by between $80\;$K and $350\;$K depending on the period. Hence,  
both the slopes and (necessarily) the zero points of the P-L 
relations in $B$, $V$, and $I$ must differ between LMC and the revised 
relations (also given here) for the Galaxy, and in fact they do. 
The LMC Cepheids are brighter by up to $0.5\;$mag at $\log P = 0.4$ (2 
days) and fainter by $0.2\;$mag at $\log P = 1.5$ (32 days). These 
facts complicate the use of Cepheid as precision distance 
indicators until the reason is found (metallicity differences or 
other unknown differences) for the non-universality of the P-L 
and P-C relations. 

     The very large data base permits mapping of various Cepheid 
properties at different positions within the instability strip, 
both at constant period and at constant absolute magnitude over 
the range of $2 < P < 40$ days and $-2 > M_{V} > -5$. Amplitude of the 
light curves are largest near the blue edge of the strip for 
periods between 2 and 7 days and longer than 15 days. The sense 
is reversed for periods between 7 and 15 days. The shape of the 
light curves varies systematically across the strip. Highly peaked 
curves (of large amplitude) that necessarily have large values of 
$R_{21}$ of about $0.5$, occur near the blue edge of the strip. More 
symmetrical (small amplitude) light curves that have, thereby, 
small values of $R_{21}$, generally occur near the red edge of the 
strip. Consequently, there is a strong correlation of $R_{21}$ with 
color within the strip at a given absolute magnitude. Strong 
correlations also exist between color and period, and color and 
amplitude at given absolute magnitudes, for the same reason that 
has long been known for RR Lyrae stars, based on the sloping 
lines of constant period in the CMD combined with the variation of 
amplitude, and now $R_{21}$, with color. The highly peaked light-curve 
shapes and large amplitudes (indicating a non-linear regime that 
is overdriven out of the linear regime) near the blue edge of 
the strip, show that the energy~driver for the pulsation (i.e.\ 
the negative dissipation) is strongest at the blue edge. 
}
\titlerunning{New P-L and P-C Relations of Classical
       Cepheids in LMC}
\maketitle

\section{Introduction}
     We showed in Paper~I \citep{Tammann:etal:03} that the 
period-luminosity (P-L) relation for Cepheids in the Galaxy, LMC, and 
SMC have significantly different slopes. If true, by necessity 
the slopes of the ridge-line (mean) period-color (P-C) relations 
in the three galaxies must also differ. We saw that this 
requirement was met, based on the extensive new photometric data 
by \citet{Udalski:etal:99b,Udalski:etal:99c} for LMC and SMC
(\citeauthor*{Tammann:etal:03}, Fig. 16; and
Sect.~\ref{sec:PL:galaxy:comparison} here) from the OGLE project and
that of \citet{Berdnikov:etal:00} for the Galaxy.      

     The problem posed by this result is severe. With the 
calibrated Galactic P-L relation (as revised in 
Sect.~\ref{sec:PL:galaxy:revised} here) fixed by two independent
methods (main sequence fittings and the Baade-Becker-Wesselink
(BBW) kinematic expansion method), and in the LMC by its distance
determined by a variety of non-Cepheid methods
(\citeauthor*{Tammann:etal:03}, Table~6), our LMC relation is 
{\em brighter\/} than that for the Galaxy by $-0\fm37$ in $V$ at 
$\log P = 0.4$ ($P = 2.5^{d}$), and {\em fainter\/} by $+0\fm11$
at $\log P = 1.6$ ($P = 40^{d}$). The differences in $I$ at the same
periods are $-0\fm32$ and $+0\fm19$. The LMC and Galactic P-L
relations cross at $\log P \approx 1.3$ ($P = 20^{d}$) {\em if\/} our
adopted distance modulus of LMC at $(m-M)^{0} = 18.54$ is correct.  

     A principal purpose for this series, of which this is the 
second paper of a projected four paper brief, is to analyze in 
tedious detail the new LMC and SMC data obtained by the OGLE 
consortium and to compare them with the Galaxy. Our purpose is to
ferret out clues for the cause of the differences in the P-L and P-C
relations and of the different position in the luminosity-temperature
diagram of the Cepheids in the three galaxies. In the present
discussion here we follow the methods of
\citeauthor*{Tammann:etal:03}. 
             
     The prime suspects for at least part of the variations from 
galaxy to galaxy are the established metallicity differences, for 
the reasons suggested in \citeauthor*{Tammann:etal:03}. At the most
elementary level, the purely technical effects on the colors of the
differential blanketing effects for different metallicities affect the
color-color relations, and color-period relations. And, as mentioned 
above, once there is a difference in the slopes of the 
color-period relations for whatever series of reasons, the P-L 
relations must also differ. 

     We saw in \citeauthor*{Tammann:etal:03} that real temperature
differences exist in the ridge-line $L_{\rm bol}$, $\log T_{\rm e}$
plane (the HR diagram) between the Galaxy and LMC Cepheids. Is this
also due to the effect of metallicity differences in the much more
complicated physics of the pulsation and the position of the
instability strip? If so, this, of course, would be a deeper reason
than the simple technical effect of Fraunhofer blanketing. One of the
purposes of this paper is to extend the discussion of these points from 
\citeauthor*{Tammann:etal:03} here in
Sect.~\ref{sec:InstabilityStrip:MTeff}.  

     A second purpose is to explore the properties of the 
instability strip by the same technique made manifest by the 
horizontal branch of globular clusters as it threads the 
instability strip nearly at constant luminosity. The large data 
base from the new precision photometric data of 
\citet{Udalski:etal:99b,Udalski:etal:99c} for the LMC Cepheids
permit a variety of studies of the Cepheid properties as a function of
position (color) in the strip, similar to such studies for the RR
Lyrae stars.          

     Section~\ref{sec:Data} discusses the OGLE data of 
\citet{Udalski:etal:99b,Udalski:etal:99c} for LMC Cepheids, defining
the sample, the Cepheid reddenings, and the errors. 
The period-color relations in $(B\!-\!V)^{0}$ and $(V\!-\!I)^{0}$, 
with comparisons with the Galaxy, are in Sect.~\ref{sec:PC}. 
The P-L relations for LMC in $B$, $V$, and $I$, with comparisons with
an updated calibration of the Galactic Cepheids are in
Sect.~\ref{sec:PL}.  
Discussion of the break in the P-L and P-C relations at $\log P = 1.0$,
and the abnormal behavior of the Fourier components and amplitudes of
the light-curves, also at 10 days, is in Sect.~\ref{sec:break}. 
Section~\ref{sec:InstabilityStrip} is an extended discussion of the
position of the instability strip in the HR diagram ($M_{V}$-color
plane). Comparison with the Galaxy (Fig.~15 of
\citeauthor*{Tammann:etal:03}) is made there. 
The break in the slope of the $M_{V}$-color relations in
$(B\!-\!V)^{0}$ and $(V\!-\!I)^{0}$ at 10 days period is manifest. 
Also in that section, the slopes of the lines of constant period in
the color-magnitude diagram (CMD) are derived in different period
ranges, as are the correlations of amplitude with position (color) in
the strip.  

     Section~\ref{sec:PLC} shows the period-luminosity-color (PLC)
relations in $B$, $V$, and $I$, based on these aforementioned
properties of the LMC Cepheids. 
The instability strip in the luminosity-temperature  
($\log L\!-\!\log T_{\rm e}$) plane is derived in
Sect.~\ref{sec:InstabilityStrip:MTeff} where the effects of  
different metallicities and helium abundances, based on models in 
the literature, are discussed. 
Section~\ref{sec:conclusion} is a summary, and states the dilemma
caused by these results in the using of Cepheids as precision distance
indicators in our program to calibrate the Hubble constant by means of
Cepheid distances to galaxies that have produced type Ia supernovae
\citep{Parodi:etal:00,Saha:etal:01}.

\section{The Data for LMC Cepheids}
\label{sec:Data}
%
\subsection{The OGLE sample}
\label{sec:Data:u99}
\citet{Udalski:etal:99b}\footnote{The revised and in $B$ extensively
  augmented catalog (updated on Apr. 24, 2003) is available via {\tt
  ftp://bulge. princeton.edu/ogle/ogle2/var\_stars/lmc/cep/catalog/}}
have presented multi-epoch CCD photometry in
the standard $BVI$ system, obtained within the OGLE program, of
hundreds of variable stars distributed along the bar of LMC and have
identified those which they consider as classical, fundamental-mode
Cepheids. After exclusion of double-entries we have also excluded,
following a scrutiny kindly provided by S.~Kanbur \& C.-C.~Ngeow,  
80 stars of the sample which have not equally well defined or atypical
light curves. We have finally excluded 14 Cepheids with $\log P<1$ and
being redder than $(B\!-\!V)^0>0.8$ or $(V\!-\!I)^0>0.9$.\footnote{All
  magnitudes quoted in this paper are magnitudes $<\!\!x\!\!>$ of
  intensity means. Colors $(x-y)$ mean $<\!\!x\!\!>-<\!\!y\!\!>$.}
At least some of these stars may be type~II Cepheids. 
This leaves 593 Cepheids with $VI$ magnitudes, 584 of which have also
$B$ magnitudes. Their period interval is restricted to
$0.4\le \log P \le 1.5$. The lower limit is set, following
\citet{Udalski:etal:99b}, to avoid confusion with stars which may not
be classical, fundamental-mode Cepheids. The upper limit is forced
by saturation effects of the OGLE observations. The clipped sample of
\citet{Udalski:etal:99b} is called U99 sample in the following.

\subsection{The additional sample}
\label{sec:Data:LMCother}
Mainly because of the upper period limitation due to image saturation,
the U99 sample was augmented by 97 LMC Cepheids for which
photoelectric $V$ photometry could be compiled in the literature.
From the numerous CCD magnitudes in $B$, $V$, and $I$ by
\citet{Sebo:etal:02} only those 20 Cepheids above $P=10$ days were
retained which have at least ten or more single observations in each
color. The very red and faint Cepheid HV\,2749 is probably a type~II
Cepheid and is excluded. -- The additional 97 Cepheids are listed in
Table~\ref{tab:LMCother}; 96 have $B$, 97 $V$, and 55 $I$
magnitudes. These Cepheids are referred to as the
``Table~\ref{tab:LMCother} sample'' in the following.

\subsection{The color excesses $E(B\!-\!V)$}
\label{sec:Data:EBV}
Individual color excesses for the U99 sample were adopted from
\citet{Udalski:etal:99b}. These authors have determined absorption
values $A_{I}$ from the mean $I$-magnitudes of adjacent red-clump
stars which makes them exceptionally independent of the Cepheids
themselves, particularly as to a possible spread in metallicity. They
then assume $E(B\!-\!V)=A_{I}/1.96$ which we adopt, although our
absorption law requires $E(B\!-\!V)=A_{I}/1.95$ (see below). The small
difference is unconsequential
because the zero point of $E(B\!-\!V)$ is tied by
\citet{Udalski:etal:99b} to several external fields in LMC.

     Most of the U99 Cepheids have $0.10<E(B\!-\!V)\le0.16$ with a
median value of $0.135$, but there are also
122 Cepheids with $0.16<E(B\!-\!V)\le0.20$. These values are higher
than the standard value of $E(B\!-\!V)=0.10$ for LMC, but it is known
that the bar is relatively dust-rich \citep{Hodge:72}.
We will return to this question in \ref{sec:Data:error}.

     The reddenings of the Table~\ref{tab:LMCother} sample were
compiled from the literature where available; in the remaining 37
cases $E(B\!-\!V)=0.10$ was assumed. The average value of $E(B\!-\!V)$
for this sample is $0.09$.

     To convert the adopted $E(B\!-\!V)$s into $E(V\!-\!I)$s and into
the corresponding absorption values $A_{BVI}$ we assume that LMC
Cepheids follow the same absorption law as Galactic Cepheids
(see \ref{sec:PL:galaxy:revised}).

\subsection{Errors}
\label{sec:Data:error}
The merging of the U99 and Table~\ref{tab:LMCother} samples raises
the question whether the independent photometries, although nominally
in the standard system, are compatible indeed. To test this question
we have derived P-C relations in $(B\!-\!V)^0$ and
$(V\!-\!I)^0$ and P-L$_{BVI}$ relations below (Eqs.
\ref{eq:PC:BV:lt1},
\ref{eq:PC:BV:ge1},
\ref{eq:PC:VI:lt1},
\ref{eq:PC:VI:ge1},
 and
\ref{eq:PL:B:lt1}$-$\ref{eq:PL:I:ge1})
from the joint sample and have determined random and systematic
deviations of either one sample from the mean relations. The
differences were combined to give the color and magnitude differences
between the two samples as shown in Table~\ref{tab:LMCsamplediff}.

\setcounter{table}{1}
\begin{table}
\begin{center}
\caption{The color and magnitude differences between the U99 and
  Table~1 samples in the sense $\Delta=\Delta_{\rm U99}-\Delta_{\rm
  Table\,1}$.}
\label{tab:LMCsamplediff}
\small
\begin{tabular}{cccccc}
\hline
\hline
\noalign{\smallskip}
 & \multicolumn{1}{c}{$(B\!-\!V)^0$} &
   \multicolumn{1}{c}{$(V\!-\!I)^0$} &
   \multicolumn{1}{c}{$B^0$} &
   \multicolumn{1}{c}{$V^0$} &
   \multicolumn{1}{c}{$I^0$} \\
\noalign{\smallskip}
\hline
\noalign{\smallskip}
$\Delta$   &   $-0.008$ &   $-0.015$ &   $+0.017$ &   $+0.013$ &   $+0.009$ \\
$\epsilon$ & $\pm0.012$ & $\pm0.013$ & $\pm0.041$ & $\pm0.032$ & $\pm0.035$ \\
\noalign{\smallskip}
\hline
\end{tabular}
\end{center}
\end{table}
The photometric agreement between the two samples is very good; the
difference of $\sim\!0\fm01$ is in all cases statistically
insignificant. The two samples can therefore be combined without
introducing any noticeable systematic errors into the photometry.

     The agreement of the photometry of the two samples is even more
astounding as the comparison is performed {\em after\/} the colors and
magnitudes have been corrected for the adopted reddenings. This
provides a strong test for the adopted values of $E(B\!-\!V)$ and
confirms that the reddenings of the U99 Cepheids in the bar of LMC
must be larger on average than of the widely distributed Cepheids in
Table~\ref{tab:LMCother}.

\section{The Period-Color Relations of LMC}
\label{sec:PC}
%
\subsection{The Period-Color (P-C) Relation in $(B\!-\!V)^0$}
\label{sec:PC:BV}
\begin{figure}[t]
\resizebox{1.0\hsize}{!}{\includegraphics{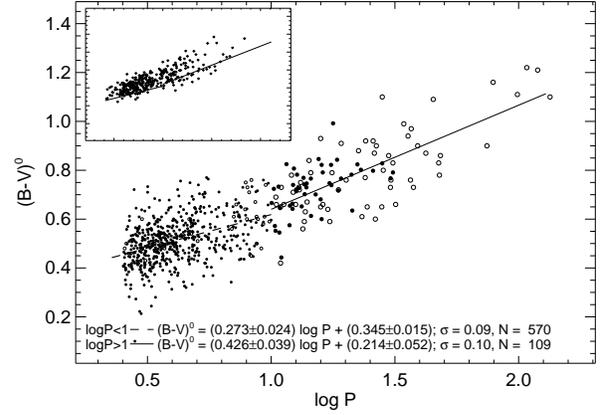}}
\caption{The P-C relation in $(B\!-\!V)^0$ of 679 LMC Cepheids. Black
  dots are from U99; small dots for $\log P<1.0$, large dots for $\log
  P>1.0$. Open symbols are from Table~\ref{tab:LMCother}. Two separate
  fits for Cepheids with $\log P {\protect\grole} 1.0$ are shown. The
  insert repeats the mean LMC relations and shows the individual
  Galactic Cepheids from Paper~I.}
\label{fig:PC:BV}
\end{figure}

The reddening-corrected colors $(B\!-\!V)^0$ of the accepted
679 Cepheids with known $(B\!-\!V)^0$ colors  from U99 and
Table~\ref{tab:LMCother} are plotted against
$\log P$ in Fig.~\ref{fig:PC:BV}a.
\makeatletter
\def\fnum@figure{\figurename\,\thefigure a}
\makeatother

     A linear fit over all periods of the data in Fig.~\ref{fig:PC:BV}a
gives $(N=679)$
\begin{equation}\label{eq:PC:BV}
   (B\!-\!V)^0 = (0.358\!\pm\!0.011)\log P + (0.294\!\pm\!0.009),\;\sigma=0.09
\end{equation}
in good statistical agreement with earlier LMC P-C relations
\citep{Caldwell:Coulson:86,Laney:Stobie:94}.
However, the linear fit
is not particularly good. A better fit is achieved if the Cepheids with
$\log P\grole 1.0$ are separately fitted.
One obtains then for $\log P<1 \;(N=570)$:
\begin{equation}\label{eq:PC:BV:lt1}
  (B\!-\!V)^0 = (0.273\!\pm\!0.024)\log P + (0.345\!\pm\!0.015),\; \sigma=0.09
\end{equation}
and for $\log P>1 \;(N=109)$:
\begin{equation}\label{eq:PC:BV:ge1}
  (B\!-\!V)^0 = (0.426\!\pm\!0.039)\log P + (0.214\!\pm\!0.052),\; \sigma=0.10.
\end{equation}
The slope of the long-period Cepheids is clearly steeper ($3.4\sigma$)
than for those with $\log P<1.0$. The breaking point at $\log
P=1$ is not well determined; it will be justified below. No attempt
was made here -- nor in the following cases -- to smooth out the jump
in color of $0\fm022$ at $\log P=1$ between Eqs.~(\ref{eq:PC:BV:lt1})
and (\ref{eq:PC:BV:ge1}); although it is within the statistical error,
it could also hint at a physical effect.

     The scatter of $\sigma=0\fm09$ in Eq.~(\ref{eq:PC:BV:lt1}) is
mainly due to the intrinsic width of the instability strip. The
scatter in Eq.~(\ref{eq:PC:BV:ge1}) is only marginally larger.

\subsection{The P-C Relation in $(V\!-\!I)^0$}
\label{sec:PC:VI}
647 accepted and dereddened LMC Cepheids with known ~~~~$(V\!-\!I)^0$
colors from U99 and Table~\ref{tab:LMCother} define the P-C relation
in $(V\!-\!I)^0$ (Fig.~\ref{fig:PC:VI}b).
\setcounter{figure}{0} 
\makeatletter
\def\fnum@figure{\figurename\,\thefigure b}
\makeatother
\begin{figure}[t]
\resizebox{1.0\hsize}{!}{\includegraphics{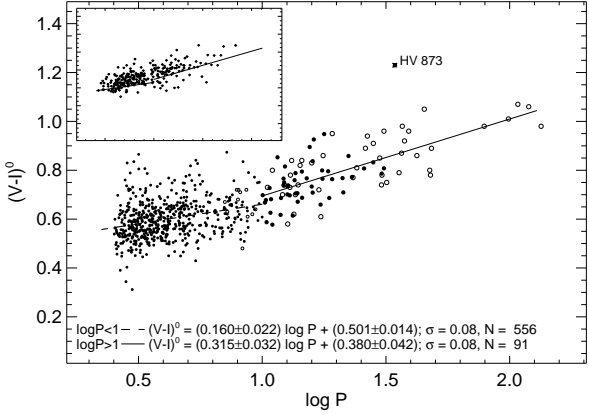}}
\caption{The P-C relation in $(V\!-\!I)^0$ of 634 LMC
  Cepheids. Symbols and insert as in Fig.~\ref{fig:PC:BV}a. Separate
  linear regressions are shown for Cepheids with $\log
  P{\protect\grole} 1.0$.} 
\label{fig:PC:VI}
\end{figure}
\makeatletter
\def\fnum@figure{\figurename\,\thefigure }
\makeatother

     A linear fit over all periods yields $(N=647)$
\begin{equation}\label{eq:PC:VI}
  (V\!-\!I)^0 = (0.256\!\pm\!0.010)\log P + (0.444\!\pm\!0.008),\;\sigma=0.08.
\end{equation}
Yet the data in Fig.~\ref{fig:PC:VI}b are not well fit through the
linear Eq.~(\ref{eq:PC:VI}).
It predicts Cepheids with $\log P<0.5$ to be redder by $0\fm05$ and
those at $\log P\sim1.6$ to be bluer by $0\fm05$ as they are actually
observed on average. Indeed eye inspection already suggests that the
P-C relation has a rather abrupt change of slope near $\log
P\approx1.0$. If, as for the $(B\!-\!V)^0$ colors above, one fits two
separate linear regressions for short- and long-period Cepheids, one
obtains for $\log P<1 \; (N=556)$:
\begin{equation}\label{eq:PC:VI:lt1}
  (V\!-\!I)^0 = (0.160\!\pm\!0.022)\log P + (0.501\!\pm\!0.014), \;\sigma=0.08
\end{equation}
and for $\log P>1 \; (N=91)$:
\begin{equation}\label{eq:PC:VI:ge1}
  (V\!-\!I)^0 = (0.315\!\pm\!0.032)\log P + (0.380\!\pm\!0.042), \;\sigma=0.08.
\end{equation}
The short-period Cepheids define a much flatter P-C relation than
long-period ones. The slope difference has a significance of
$4.1\sigma$. Another way of expressing this result is provided by an
t-test which excludes the hypothesis that the single-line solution of 
Eq.~(\ref{eq:PC:VI}) is an equally good fit as the two separate fits
of Eqs. (\ref{eq:PC:VI:lt1}) and (\ref{eq:PC:VI:ge1}) with a probability
of more than 99\% \citep{Kanbur:Ngeow:04}.
The steep slope of the P-C relation in $(V\!-\!I)^0$ of long-period
LMC Cepheids has been anticipated by \citet{Caldwell:Coulson:86}.

\subsection{Comparison of the LMC and Galactic P-C Relations}
\label{sec:PC:comparison}
The mean LMC P-C relations in $(B\!-\!V)^0$ (Eqs.~\ref{eq:PC:BV:lt1}
\& \ref{eq:PC:BV:ge1}) and $(V\!-\!I)^0$ (Eqs.~\ref{eq:PC:VI:lt1}
\& \ref{eq:PC:VI:ge1}) are repeated in the inserts of
Fig.~\ref{fig:PC:BV}a,b and
compared with the position of 321 Galactic Cepheids from
\citeauthor*{Tammann:etal:03}. The {\em overall slope\/} of the LMC
P-C relations (Eqs.~\ref{eq:PC:BV} \& \ref{eq:PC:VI}) are similar to
the Galactic P-C relations in $(B\!-\!V)^0$ and $(V\!-\!I)^0$, but the
best-fit LMC slopes, which are different for Cepheids with  $\log
P\grole 1.0$, are significantly flatter for short-period and steeper for
long-period Cepheids. In addition the bulk of the LMC Cepheids is
considerably {\em bluer\/} in $(B\!-\!V)^0$ and $(V\!-\!I)^0$ than the
Galactic Cepheids. The mean color difference $\Delta (B\!-\!V)^0_{\rm
  Gal-LMC}$ amounts to $0\fm05$ at $\log P=0.4$, to increase to
$\sim\!0\fm10$ at $\log P=1.0$, and to decrease to $0\fm04$ at $\log
P=1.8$. The corresponding, smaller color differences $\Delta
(V\!-\!I)^0_{\rm Gal-LMC}$ are $0\fm03$, $\sim\!0\fm08$, and
$0\fm01$.

     The color differences between LMC and Galactic Cepheids must be
due to blanketing and/or temperature differences. This point will be
followed up in Sect.~\ref{sec:PC:two-color}.

\subsection{The two-color diagram of LMC and Galactic Cepheids}
\label{sec:PC:two-color}
\begin{figure}[t]
\centering
\resizebox{0.9\hsize}{!}{\includegraphics{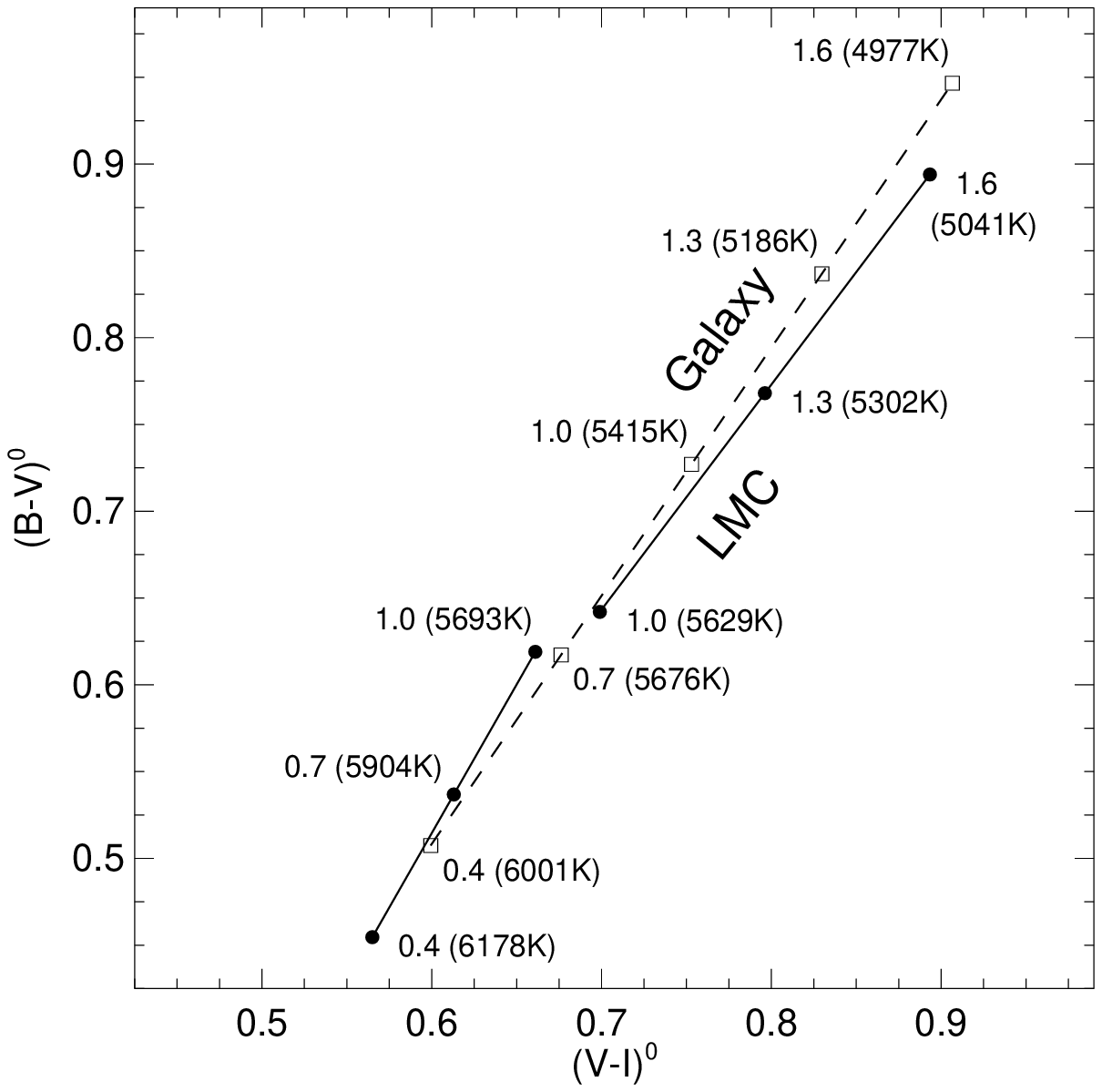}}
\caption{The two color diagram of Cepheids in LMC (full line) and the
  Galaxy (dashed). Loci of different values of $\log P$ are labelled
  as well as the corresponding temperature $T_{\rm e}$.}
\label{fig:PC:two-color}
\end{figure}

The mean position of the LMC Cepheids in a plot of $(B\!-\!V)^0$
vs. $(V\!-\!I)^0$ is shown with two full lines for $\log P\grole 1.0$
in Fig.~\ref{fig:PC:two-color}. The lines are determined from a
combination of Eqs.~(\ref{eq:PC:BV:lt1}) \& (\ref{eq:PC:VI:lt1}) and
(\ref{eq:PC:BV:ge1}) \& (\ref{eq:PC:VI:ge1}), respectively.
Also shown in the figure is the mean position of
Galactic Cepheids (dashed line), where the position of the line and of
the different $\log P$ values is taken from
\citeauthor*{Tammann:etal:03}.  

     Figure~\ref{fig:PC:two-color} is the comparison of the two color
diagrams of the Galaxy and LMC. It differs subtly from the similar
diagram 7b in \citeauthor*{Tammann:etal:03} where the LMC relation is
slightly above that for the Galaxy, whereas in
Fig.~\ref{fig:PC:two-color} here it is below the Galaxy line for $\log
P > 1$, $(B\!-\!V)^{0} > 0.65$. The difference is that the P-C
relations (Figs.~3 and 5 in \citeauthor*{Tammann:etal:03}; Figures 1a
and 1b here) were fit in \citeauthor*{Tammann:etal:03} by a straight
line, whereas here the break at $\log P = 1$ is allowed for. This now
lets LMC Cepheids with log $P > 1$ be redder in $(V\!-\!I)^{0}$ than
Galactic ones for $(B\!-\!V)^{0} > 0.65$ as the models of Bell and
Tripicco \citep[][hereafter SBT]{Sandage:etal:99} require due to the
lower metallicity of LMC (cf. Fig.~7a panel d of
\citeauthor*{Tammann:etal:03}).    

     However, a comparison of the mean relations in
Fig.~\ref{fig:PC:two-color} does not tell the full story. The points
of equal period are shifted to each other. LMC Cepheids of given
period are bluer in $(B\!-\!V)^0$ by $0\fm05-0\fm10$ and in
$(V\!-\!I)^0$ by roughly $0\fm05$ than their Galactic counterparts.
The reason is as follows. 
 
\begin{figure}[t]
\centering
\resizebox{0.95\hsize}{!}{\includegraphics{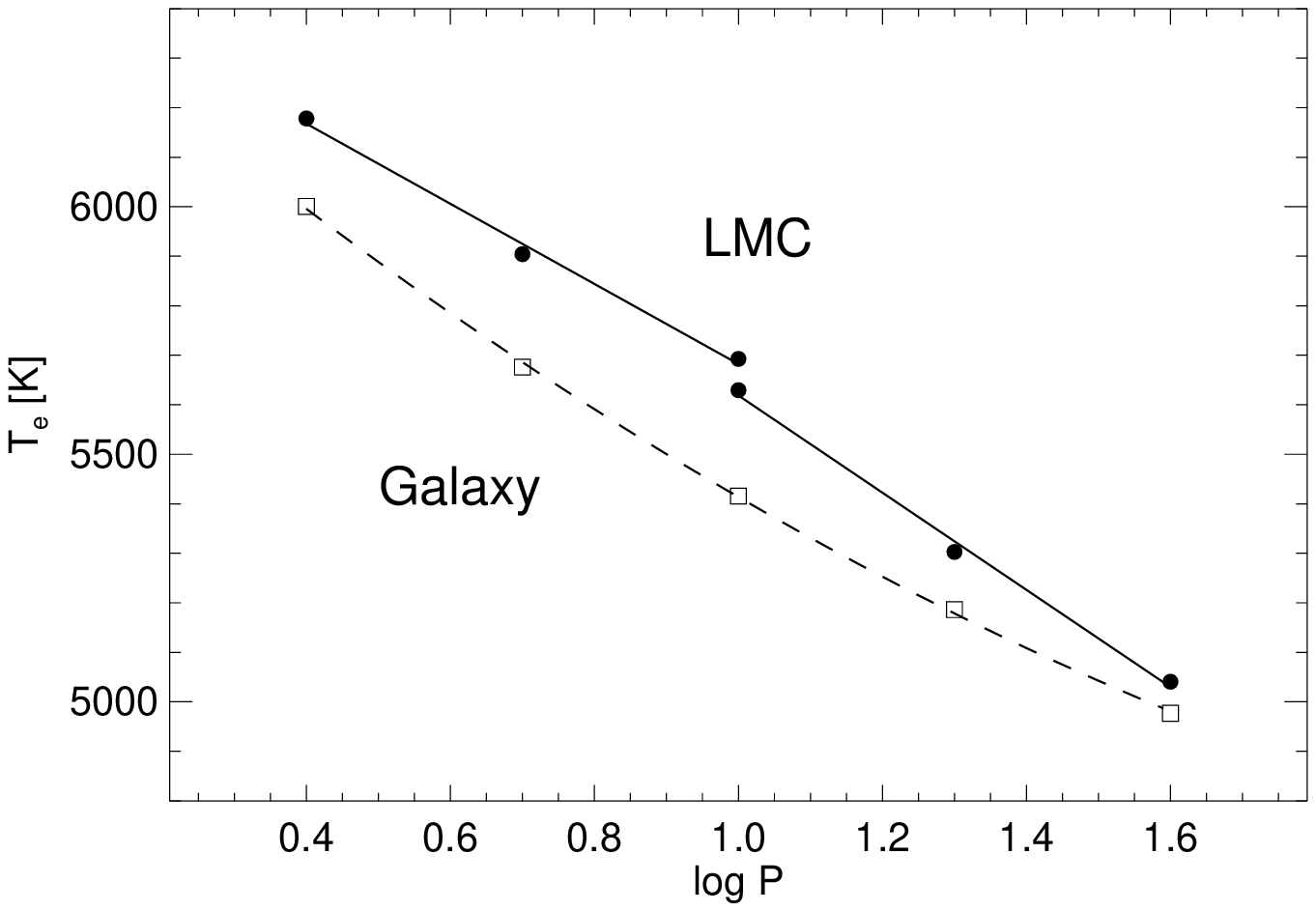}}
\caption{The $\log P$$-$$T_{\rm e}$ variation of LMC and Galactic
  Cepheids.}
\label{fig:logPTe}
\end{figure}

    There is a real temperature difference (see
Sect.~\ref{sec:InstabilityStrip:MTeff}) in the ridge-line $M_{V}$,
$\log T_{\rm e}$ diagram (the HR diagram of Fig.~\ref{fig:LMC:Teff}, 
later) for the Galaxy compared with LMC. The calculations set out
later in Sect.~\ref{sec:InstabilityStrip:MTeff} lead to
Figure~\ref{fig:logPTe}, showing an average temperature difference of
about $150^{\circ}$, ranging from $80\;$K to $350\;$K
depending on the period. Although our mean value is smaller than the
$250^{\circ}$ found by \citet{Laney:Stobie:86} using $(J\!-\!K)$
colors, it is in the same sense.  

     It is clear why this temperature shift toward the blue 
for LMC causes the period separations at given colors between 
LMC and the Galaxy in Fig.~\ref{fig:PC:two-color} (and Fig.~7b in
\citeauthor*{Tammann:etal:03}). The lines of constant period that
thread the instability strip (Fig.~\ref{fig:CPL} later) cut a hotter
central (ridge-line) position at bluer color in LMC than for the a
cooler ridge-line position in the Galaxy. Hence, for LMC, the
$(B\!-\!V)^{0}$ and $(V-I)^{0}$ colors {\em at a given period\/} are
bluer than for the Galaxy at the same period. This is the sense of the
scrambling of periods in Figure~7b of \citeauthor*{Tammann:etal:03}
and Figure~\ref{fig:PC:two-color} here. 

     Possible reasons for the shift in the temperatures of the 
ridge-line HRDs between the Galaxy and LMC are discussed in 
Sect.~\ref{sec:InstabilityStrip:MTeff}.

\section{The Period-Luminosity Relations of LMC}
\label{sec:PL}
%
\subsection{The P-L Relation of LMC in $B$, $V$, and $I$}
\label{sec:PL:LMC}
It would be possible to discuss the P-L relation of LMC using the
dereddened apparent magnitudes only. We prefer, however, to use
absolute magnitudes which allow also a comparison with the Galactic
zero points. We adopt $(m-M)^0_{\rm LMC}=18.54$ as the best
Cepheid-independent distance modulus as justified in
\citeauthor*{Tammann:etal:03}. This choice has no qualitative
consequences for the following conclusions, and the reader can adjust
the quantitative results to any preferred distance of LMC.

\begin{figure*}[t]
\centering
     \includegraphics[width=14.0cm]{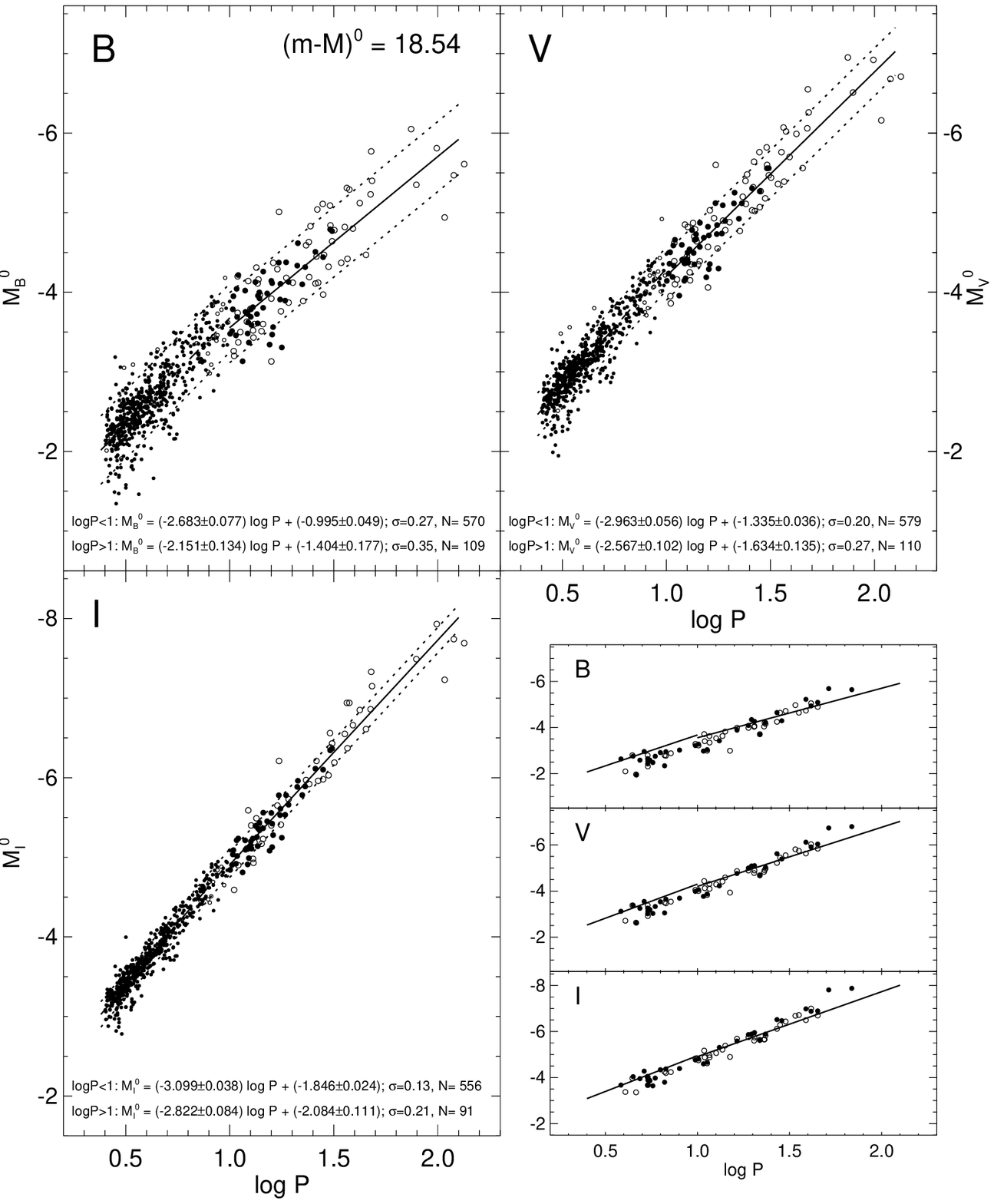}
     \caption{The P-L relations in $B$, $V$, and $I$ of LMC Cepheids.
     The data in each color are fitted with two linear regressions
     breaking at $\log P=1.0$. Symbols as in Fig.~\ref{fig:PC:BV}a.
     For the dashed intrinsic boundaries see text
     (\ref{sec:InstabilityStrip:CPL}).
     Comparison with the revised Galactic calibration in
     Sect.~\ref{sec:PL:galaxy:revised}
     (Eqs.~\ref{eq:gal:PL:B}\,$-$\,\ref{eq:gal:PL:I}) are in the
     lower right panel. The individual Galactic Cepheids with known
     absolute magnitudes (cf. \ref{sec:PL:galaxy:revised}) are the
     data points. The LMC mean relations are the solid lines.} 
\label{fig:PL:final}
\end{figure*}
     The absolute magnitudes $M^0_{BVI}$ of the sample Cepheids
follow directly from the adopted LMC modulus and the color excesses
discussed in Sect.~\ref{sec:Data:EBV}. The adopted absorption law
follows in Sect.~\ref{sec:PL:galaxy:revised}. The resulting absolute
magnitudes are plotted against $\log P$ in Fig.~\ref{fig:PL:final}.
If the data in the three panels of Fig.~\ref{fig:PL:final} are fit
over the entire period range by linear regressions, one obtains 
$(N=679, 689, 647)$
\begin{eqnarray}
 \label{eq:PL:B}
   M^{0}_{B} & = & -(2.340\!\pm\!0.037)\log P - (1.200\!\pm\!0.029), \;
   \sigma=0.29, \\
 \label{eq:PL:V}
   M^{0}_{V} & = & -(2.702\!\pm\!0.028)\log P - (1.491\!\pm\!0.022), \;
   \sigma=0.22, \\
 \label{eq:PL:I}
   M^{0}_{I} & = & -(2.949\!\pm\!0.020)\log P - (1.936\!\pm\!0.015), \;
   \sigma=0.14.
\end{eqnarray}
These fits,which are in reasonable agreement with previous P-L
relations of LMC 
\citep[e.g.][]{Kraft:61,Sandage:Tammann:68,Laney:Stobie:94,Gieren:etal:98}
and close to the solutions in $V$ and $I$ by
\citet{Udalski:etal:99a}, systematically overestimate the luminosity
of Cepheids around $\log P=1.0$, and are hence unsatisfactory.
A better fit is obtained when the P-L relations in $B$, $V$, and $I$
are each fit with two linear regressions breaking at $\log P=1.0$. The
two-slopes solutions are shown in Fig.~\ref{fig:PL:final}. The
significance of the slope differences between short- and long-period
Cepheids varies between $2.6\sigma$ ($I$) and $3.1\sigma$ ($V$). The
reality of the break is supported by the broken P-C relations in
Sects.~\ref{sec:PC:BV} and \ref{sec:PC:VI} since any non-linearity of
the P-C relations {\em must\/} be reflected in the P-L relations.

     Least-squares solutions to the data in Fig.~\ref{fig:PL:final}
with {\em two\/} linear regressions, breaking at $\log P=1.0$,  give in
$M_{B}$ and for
\begin{eqnarray}
\nonumber
   \lefteqn{\log P<1 \;(N=570):} \\ 
\label{eq:PL:B:lt1}
  \lefteqn{M^0_{B} = -(2.683\!\pm\!0.077)\log P - (0.995\!\pm\!0.049), \; \sigma=0.27,}\\
\nonumber
   \lefteqn{\log P>1 \;(N=109):} \\ 
\label{eq:PL:B:ge1}
  \lefteqn{M^0_{B} = -(2.151\!\pm\!0.134)\log P - (1.404\!\pm\!0.177), \; \sigma=0.35,}
\end{eqnarray}
in $M_{V}$ and for
\begin{eqnarray}
\nonumber
   \lefteqn{\log P<1 \;(N=579):} \\ 
 \label{eq:PL:V:lt1}
   \lefteqn{M^0_{V} = -(2.963\!\pm\!0.056)\log P - (1.335\!\pm\!0.036), \; \sigma=0.20,} \\
\nonumber
   \lefteqn{\log P>1 \;(N=110):} \\ 
\label{eq:PL:V:ge1}
   \lefteqn{M^0_{V} = -(2.567\!\pm\!0.102)\log P - (1.634\!\pm\!0.135), \; \sigma=0.27,}
\end{eqnarray}
and in $M_{I}$ and for
\begin{eqnarray}
\nonumber
   \lefteqn{\log P<1 \;(N=556):} \\ 
\label{eq:PL:I:lt1}
   \lefteqn{M^0_{I} = -(3.099\!\pm\!0.038)\log P - (1.846\!\pm\!0.024), \; \sigma=0.13,} \\
\nonumber
   \lefteqn{\log P>1 \;(N=91):} \\ 
\label{eq:PL:I:ge1}
   \lefteqn{M^0_{I} = -(2.822\!\pm\!0.084)\log P - (2.084\!\pm\!0.111), \; \sigma=0.21.}
\end{eqnarray}
The luminosity scatter of the long-period Cepheids is larger than at
periods less than 10 days. This could be due to a broadening of the
instability strip in the CMD at longer periods or to steeper
constant-period lines
(see Sect.~\ref{sec:InstabilityStrip:CPL} below). But it could also be
due to the Cepheids in Table~\ref{tab:LMCother} which carry much of
the weight at longer periods. Their extinctions $E(B\!-\!V)$ are less
uniformly determined than the U99 sample, and, scattered over the face
of LMC, they may be more affected by the depth effect of the galaxy
than the U99 Cepheids which lie along the bar of LMC. It is therefore
not possible to identify a single reason for the larger scatter of
long-period Cepheids.

\subsection{Comparison with the Galaxy}
\label{sec:PL:Galaxy}
%
\subsubsection{A revision to the Galactic P-L relations}
\label{sec:PL:galaxy:revised}
In \citeauthor*{Tammann:etal:03} we have derived the Galactic P-L
relation in $B$, $V$, and $I$ from the then available 25
fundamental-mode Cepheids in clusters and associations from
\citet{Feast:99} and 28 Cepheids with Baade-Becker-Wesselink
(BBW) expansion parallaxes from \citet{Gieren:etal:98}. 
In the meantime additional distances of cluster/association Cepheids
have become available
\citep{Turner:Burke:02,Hoyle:etal:03}. Overlapping cluster distances
of the three sources show no systematic difference after they have
been reduced to a common zero point. The zero point was based on the
Pleiades at $(m-M)^0=5.61$, a value which was determined from careful
photometry by \citet{Stello:Nissen:01}. 
This value agrees within $\pm0\fm05$ with the photometric distance of
\citet{Pinsonneault:etal:98}, the distance of the double star Atlas
\citep{Pan:etal:04}, the moving-cluster distance
\citep{Narayanan:Gould:99}, the rotation period method of late-type
members \citep{ODell:etal:94}, as well as now also with the latest
HIPPARCOS parallax \citep{Makarov:02}. Distances are now known for 33
Cepheids which are members of clusters or associations (the deviating
CG\,Cas and AQ\,Pup are excluded as too uncertain cluster members). In
cases where more than one independent distance determination is
available weighted means were adopted.

     The BBW distances of \citet{Gieren:etal:98} have been superceded
by the revised distances of \citet{Fouque:etal:03} who have also added
four new Cepheids.
We have augmented their list by the BBW distances of
\citet{Barnes:etal:03}. Also included are the distances from
interferometric diameter observations of 
$\delta$\,Cep \citep{Benedict:etal:02,Nordgren:etal:02}, 
$\eta$\,Aql \citep{Lane:etal:02,Kervella:etal:04}, 
$\zeta$\,Gem \citep{Lane:etal:02}, and 
l\,Car \citep{Kervella:etal:04}.
Unweighted means have been adopted for Cepheids with more than one BBW
distance. This brings the total of BBW distances to 36 Cepheids.

     \citet{Kovacs:03} has published independent
BBW absolute magnitudes of 25 Galactic Cepheids. His results for $\log
P>1$ are in excellent agreement on average with
the above data, yet his magnitudes for $\log P<1$ are $\sim\!\!0\fm5$
brighter on average than the latter. Correspondingly his suggested
slope of the Galactic P-L relation is much flatter than from either
Fouqu{\'e}'s et~al.\ results or the cluster Cepheids. For this reason
his results are not used here. 

\begin{figure}[t]
\centering
\resizebox{1.0\hsize}{!}{\includegraphics{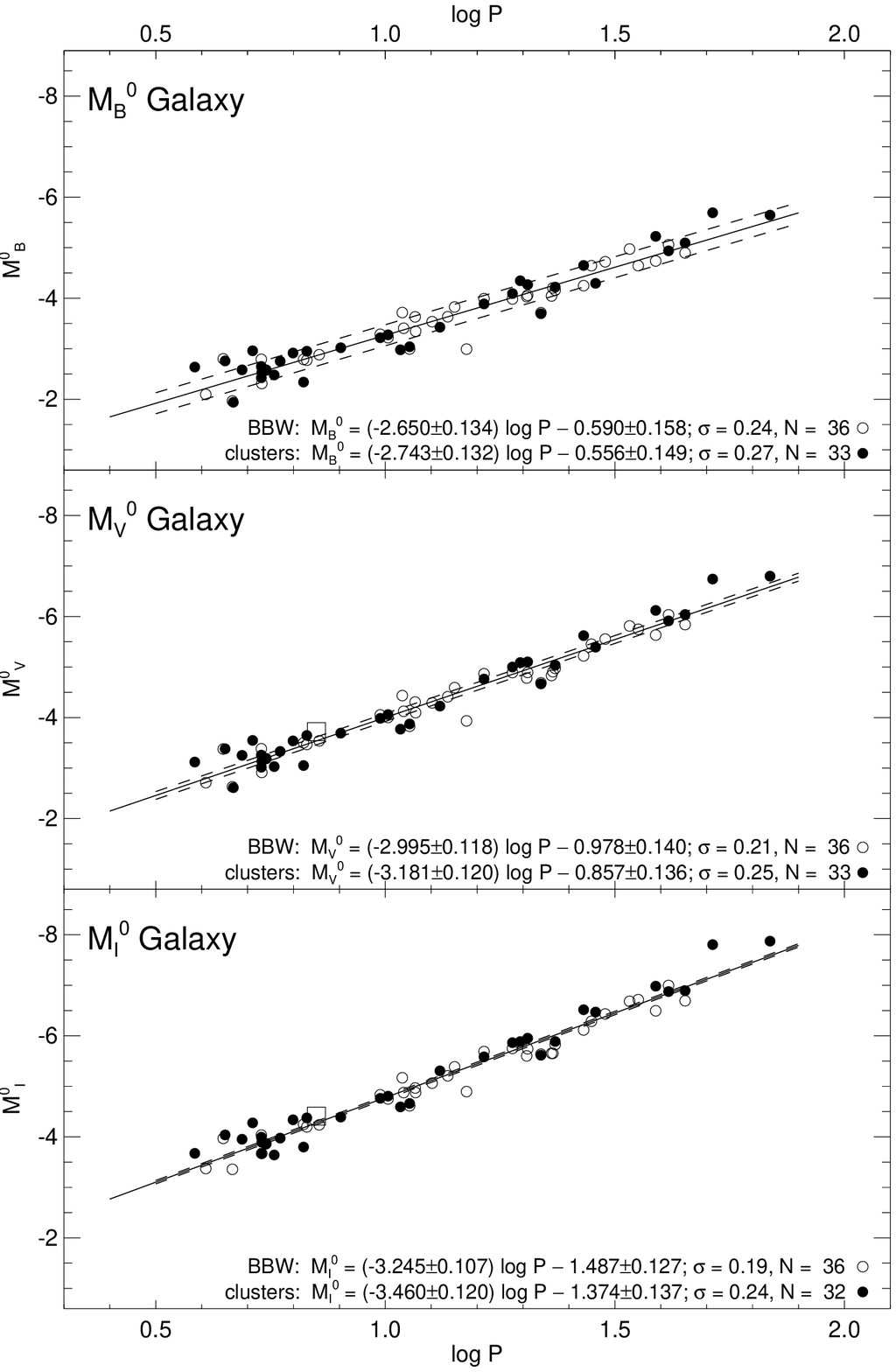}}
\caption{The Galactic P-L relations in $B,V$, and $I$ from 36 Cepheids
  with cluster distances and 33\,(32) Cepheids with BBW distances. The
  mean relations from Eqs.~(\ref{eq:gal:PL:B})$-$(\ref{eq:gal:PL:I})
  are shown as full lines. The open square is from the HIPPARCOS
  calibration by \citet{Groenewegen:Oudmaijer:00}. The dashed
  boundaries are explained in the text.}
\label{fig:PL:galaxy}
\end{figure}

     The Galactic P-L relations in $B,V$, and $I$ as defined by the 33
cluster Cepheids\footnote{$I$-magnitudes of new calibrators not in
  Table~1 of Paper~I are taken from \citet{Lanoix:etal:99}. No
  $I$-magnitude is available for TV\,CMa.} and the 36 Cepheids with
BBW distances are shown in Fig.~\ref{fig:PL:galaxy}. Separate linear
regressions of either set are given at the bottom of each panel in
Fig.~\ref{fig:PL:galaxy}. The agreement of the P-L relations from
entirely independent distance determinations is impressive. The
cluster data give marginally steeper slopes. If the BBW distances of
the longest-period Cepheids GY\,Sgr and V\,Vul had been included from
\citet{Gieren:etal:98} the agreement in slope would be nearly perfect;
however they have been excluded by \citet{Fouque:etal:03} because of
their variable period which impairs the match between photometric and
radial-velocity data. -- The independent pairs of P-L relations in
$B,V$, and $I$ agree to within $0\fm01$ at $\log P=0.5$. They diverge
by up to $0\fm11$, $0\fm14$, and $0\fm21$ in the three colors at long
periods ($\log P=1.5$), the cluster data being brighter.

     One may suspect that the (small) slope difference is rather due
to errors of the BBW distances than of the cluster distances, because
it is hard to explain why the latter should depend on the period of a
Cepheid. However, since it is impossible to objectively judge the {\em
  systematic\/} errors which affect the two independent methods we
have joined the cluster and BBW Cepheids with equal weights, which
leads to the following combined $\mbox{P-L}$ relations for our Galaxy
$(N=69, 69, 68)$:
\begin{eqnarray}
 \label{eq:gal:PL:B}
   M^{0}_{B} & = & -(2.692\!\pm\!0.093)\log P - (0.575\!\pm\!0.107),\;
   \sigma=0.25 \\
 \label{eq:gal:PL:V}
   M^{0}_{V} & = & -(3.087\!\pm\!0.085)\log P - (0.914\!\pm\!0.098),\;
   \sigma=0.23 \\
 \label{eq:gal:PL:I}
   M^{0}_{I} & = & -(3.348\!\pm\!0.083)\log P - (1.429\!\pm\!0.097),\;
   \sigma=0.23.
\end{eqnarray}
The main difference of these revised {\em Galactic\/} P-L relations
with those derived in \citeauthor*{Tammann:etal:03} is that they are
based now on a significantly broader base with correspondingly smaller
statistical errors of the coefficients and even marginally reduced
scatter. The new relations are brighter in $B,V$, and $I$ for all
periods, but nowhere by more than $0\fm07$. They agree exceedingly
well with the Galactic P-L relations of \citet{Ngeow:Kanbur:04} and
\citet{Storm:etal:04} which are derived on somewhat different
precepts. The deviations in $B$, $V$, and $I$ are never larger than
$0\fm07$ over the entire interval of $0.4<\log P<1.5$.

     As discussed in \citeauthor*{Tammann:etal:03} the HIPPARCOS zero
point in $V$ and $I$ by \citet{Groenewegen:Oudmaijer:00} is somewhat
{\em brighter\/} than the combined cluster and BBW calibration, but
the difference is now reduced to $\Delta V=0\fm18\pm0\fm11$ and
$\Delta I=0\fm14\pm0\fm12$. 

     Also the HST parallax \citep{Benedict:etal:02} of $\delta$\,Cep
($\log P=0.730$) is rather {\em bright\/} with
$M_{V}=-3.47\pm0.10$ (or $-3.41$ according to \citealt{Feast:02}),
while the P-L relation of Eq.~(\ref{eq:gal:PL:V}) predicts
$M_{V}=-3.17\pm0.10$ and the Galactic P-L-C relation as given in
Table~\ref{tab:PLC:coeff} (below) $M_{V}=-3.19\pm0.10$.
The difference amounts to $\sim\!0.25\pm0.14$.
The agreement is somewhat better with the HIPPARCOS zeropoint of
\citet{Feast:Catchpole:97}, giving $M_{V}=-3.55\pm0.10$ at a median
period of $<\!\!\log P\!\!>\,=0.80$, which is to be compared with
$M_{V}=-3.38\pm0.10$ from Eq.~(\ref{eq:gal:PL:V}), and also with the
zeropoint of \citet{Lanoix:etal:99} of $M_{V}=-3.59\pm0.10$ at 
$<\!\log P\!>\,=0.82$ instead of $M_{V}=-3.45\pm0.10$ here.
(The values of \citeauthor{Feast:Catchpole:97} and
\citeauthor{Lanoix:etal:99} are here reduced to the $E(B\!-\!V)_{\rm
corr}$ as described in \citeauthor*{Tammann:etal:03}). Thus,
considering the parallax data one could conclude that the present
zeropoint was too {\em faint\/} by $\sim\!0.15$ in $V$, -- a
possibility which on statistical grounds cannot be excluded. It
should be noted, however, that the BBW and cluster distances
agree formally to within $0\fm04$ at the relevant periods of 
$<\!\log P\!>\approx0.80$. 

     The absorption-to-reddening ratios ${\cal R}_{BVI}$ were derived
in \citeauthor*{Tammann:etal:03} from the then available Cepheid
calibrators by requiring that their residuals from the mean P-L
relations be not a function of $E(B\!-\!V)$. The now available
69\,(68) calibrators give slightly revised values, i.e.
\begin{equation}\label{eq:PL:RBVI}
     {\cal R}_{B}=4.23, \qquad {\cal R}_{V}=3.23, \qquad {\cal R}_{I}=1.95. 
\end{equation}
To be exact these values apply for Cepheids with median color of
$(B\!-\!V)=0.766$ and median color excess $E(B\!-\!V)=0.382$.
The dependence of ${\cal R}_{BVI}$ on Cepheid color and $E(B\!-\!V)$
remains the same as in Eq.~(8) in \citeauthor*{Tammann:etal:03}.

     In the remainder of this paper
Eqs.~(\ref{eq:gal:PL:B})$-$(\ref{eq:gal:PL:I}) are referred to as the
(revised) ``Galactic P-L relations''. The adopted P-C relations of
\citeauthor*{Tammann:etal:03} remain unaffected by this revision.

     The new Galactic P-L relation in $V$ gives in combination with
the Galactic P-C relations in \citeauthor*{Tammann:etal:03}
(Eqs.~3\&5) the ridge line equations of the instability strip in the
CMD. One obtains (in revision of \citeauthor*{Tammann:etal:03}):  
\begin{eqnarray}
 \label{eq:gal:inst:BV}
   M_{V} & = & -(8.43\pm0.41)(B\!-\!V) + 2.13\pm0.21, \\
 \label{eq:gal:inst:VI}
   M_{V} & = & -(12.04\pm0.86)(V\!-\!I) + 5.08\pm0.48.
\end{eqnarray}

    In \citeauthor*{Tammann:etal:03} the slope $\beta$ of the
constant-period lines in the CMD was not determined. With the
increased sample of Galactic calibrators this is attempted here. The
result for the $M_{V}\!-\!(B\!-\!V)$ plane, averaging over all periods,
is $\beta_{V,B\!-\!V}=0.60\pm0.40$ (see Table~\ref{tab:PLC:coeff}
below). This slope has still a considerable error, but it is
significantly flatter than the canonical semi-theoretical value of
$\beta_{V,B\!-\!V}\approx2.5$ \citep{Sandage:Tammann:69,Sandage:72}.
Also the slope of the constant-period lines in the $M_{V}\!-\!(V\!-\!I)$
plane is quite flat, i.e. $\beta_{V,V\!-\!I}=0.66\pm0.41$
(Table~\ref{tab:PLC:coeff}). Since the slope in the $M_{I}\!-\!(V\!-\!I)$
plane is one unit smaller (see Eq.~\ref{eq:alpha:I:VI}) it becomes {\em
  negative}, i.e. $\beta_{I,V\!-\!I}=-0.34$. Cepheids near the red
edge of the instability strip are therefore (slightly) {\em brighter}
in $I$ than their bluer counterparts.

     The flat constant-period lines of Galactic Cepheids have an
interesting consequence in as much as the intrinsic width of the
Galactic P-L relation becomes quite narrow. The width is given by the
product of the width of the instability strip and of the slope of the
constant-period lines. The strip width was found to be
$\Delta(B\!-\!V)=\pm0.13$ and $\Delta(V\!-\!I)=\pm0.10$ in
\citeauthor*{Tammann:etal:03} from only 53 Galactic calibrators with
independently known absolute magnitudes, but they are supported by the
16 additional calibrators considered here. Multiplying the strip width
with the appropriate constant-period line slopes
($\beta_{B,B\!-\!V}=1.60$, $\beta_{V,B\!-\!V}=0.60$,
$\beta_{I,V\!-\!I}=-0.34$) gives predicted half-widths of the P-L
relation of $\pm0\fm21$, $\pm0\fm08$, and $\pm0\fm03$ in $B$, $V$, and
$I$. The corresponding boundaries of the P-L relation in $B$, $V$, and
$I$ are shown in Fig.~\ref{fig:PL:galaxy} as dashed lines.
The tight P-L relations found here for the Galaxy with its relatively
high metallicity are opposite to \citet{Allen:Shanks:04} who have
suggested that the intrinsic dispersion increased with the metallicity.

     The expected narrow boundaries of the P-L relations in
Fig.~\ref{fig:PL:galaxy} are seemingly in conflict with the individual
Cepheids defining the mean relations, but their scatter is mainly
caused by the random errors of the cluster and BBW distances and is
therefore about equal in $B$, $V$, and $I$ (see
Eqs.~\ref{eq:gal:PL:B}$-$\ref{eq:gal:PL:I}).

\subsubsection{Comparisons of the Galaxy and LMC P-L relations}
\label{sec:PL:galaxy:comparison}
A comparison between the P-L$_{BVI}$ relations in LMC and in the
Galaxy is shown in the lower right panel of
Fig.~\ref{fig:PL:final}, where the {\em mean\/} P-L relations of LMC
are repeated together with the individual Galactic Cepheids with known
absolute magnitudes from Sect.~\ref{sec:PL:galaxy:revised}. 
It is obvious that the LMC relations above 10 days are much flatter
than in the Galaxy. This causes the luminosity difference between LMC
and Galactic Cepheids to be a function of period. 
At $\log P=0.4$ LMC Cepheids are {\em brighter\/} by $0\fm42$ to
$0\fm32$ in $B$, $V$, and $I$ than their Galactic counterparts. 
The luminosity difference decreases towards longer periods and
vanishes at $\log P=1.5$ in $B$, $\log P=1.4$ in $V$, 
and $\log P=1.25$ in $I$. At still larger periods Galactic
Cepheids are more luminous. The hypothesis that the P-L slopes of the
Galaxy and LMC are the same is rejected by a t-test with a probability
of more than 95\% \citep{Ngeow:Kanbur:04}.

     The practical consequence of this astonishing result is that {\em
it is in principle impossible to derive a reliable Cepheid
distance of LMC by means of a Galactic P-L relation}.
Any derived distance depends necessarily on the period distribution of
the Cepheids under consideration. Even the above statement, that the
LMC and Galactic Cepheids have equal luminosities somewhere near $\log
P\approx1.4$, offers no consolation because it depends entirely on the
{\em adopted\/} distance of $(m-M)_{\rm LMC}=18.54$.

     It was shown in \citeauthor*{Tammann:etal:03} that at least part
of the observed differences of slope of the P-L relations of the
Galaxy and LMC is a necessary consequence of metallicity
differences. We will return to this point in
Sect.~\ref{sec:PLC}.

\subsection{Comparison with model calculations}
\label{sec:PL:model}
The model P-L relations of LMC in $M_{V}$ of
\citet{Baraffe:Alibert:01} and -- in their {\em linear\/} solution --
of \citet{Caputo:etal:00}, based on
independent stellar evolution and pulsation codes, agree well
with the observed relations in Eqs.~(\ref{eq:PL:V:lt1}),
(\ref{eq:PL:V:ge1}) and Fig.~\ref{fig:PL:final}b, yet they do not
predict the break at $\log P=1$. In spite of this the difference
between theoretical and observed mean magnitudes is nowhere larger
than $0\fm2$ over the entire interval $0.4<\log P<1.5$. The same
statement holds for the additional $M_{I}$ magnitudes by
\citet{Caputo:etal:00}. Thus their $M_{V}$ and $M_{I}$ magnitudes fit
the LMC observations much better than in the case of the
Galaxy (cf. \citeauthor*{Tammann:etal:03}).

     However, Baraffe \& Alibert's \citeyearpar{Baraffe:Alibert:01}
$M_{B}$ magnitudes are too faint by $\sim\!0\fm3$ for shorter-period
Cepheids, while those of Caputo's et~al. \citeyearpar{Caputo:etal:00}
{\em quadratic\/} solution are too faint by the same amount at short
and long periods and by less at intermediate periods. The faintness of
the predicted theoretical $M_{B}$ magnitudes causes the LMC Cepheids to
appear too red at the corresponding periods. 

     The steep P-L$_{BVI}$ relations derived from various
evolution and pulsation codes by \citeauthor*{Sandage:etal:99} refer
to the {\em blue edge\/} of the instability strip. 
No attempt was made to reduce them to the {\em  mean\/} relations
because additional assumptions would be needed. 

     The only theoretical P-L relations which predict a flattening
above $\log P=1$ are the quadratic solutions by
\citet{Caputo:etal:00}, although they predict a gradual flattening
{\em instead of a break}, which leads to too faint magnitudes at very
long periods.

\begin{figure*}[t]
\centering
\includegraphics[width=14cm]{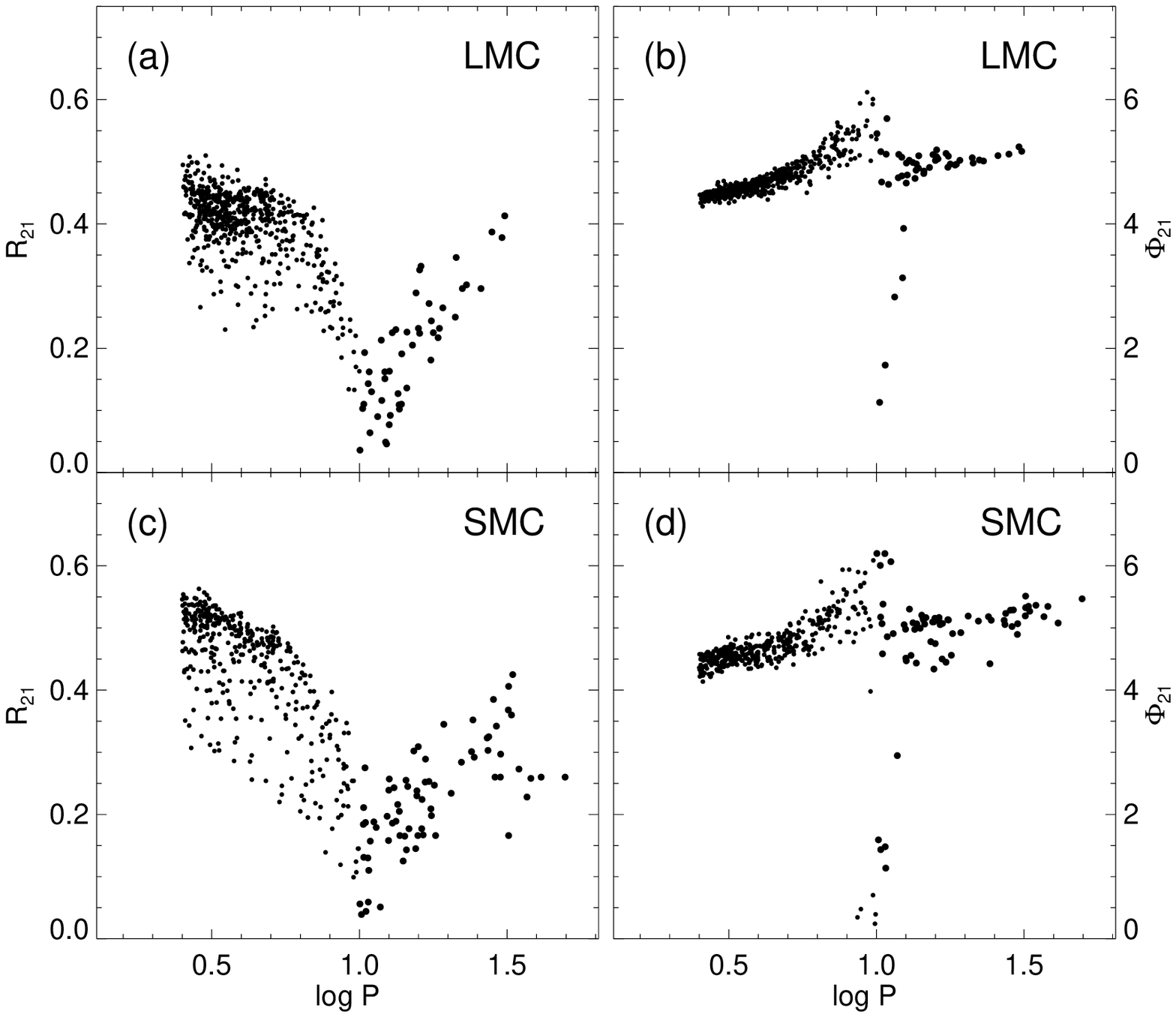}
\caption{The Fourier coefficients $R_{21}$ and $\Phi_{21}$ in function
  of $\log P$ for LMC (upper panel) and SMC (lower panel) Cepheids. 
  In spite of some systematic differences in $R_{21}$, the minimum of
  $R_{21}$ and $\Phi_{21}$ occurs in either galaxy at $\log
  P=1$. Small and large symbols are for $\log P{\protect\grole}1$.}
\label{fig:break:Rphi}
\end{figure*}
\section{The Significance of the Break at \boldmath{$\log P=1$}}
\label{sec:break}
A gradual change of slope of the Cepheid P-L relation with increasing
period was suggested several times 
(e.g. \citealt{Sandage:Tammann:68}; \citealt{Sandage:72}; 
\citeauthor*{Sandage:etal:99}; \citealt{Caputo:etal:00}), 
but a sudden change of slope of the P-C and P-L relations of LMC, as
already suggested by \citet{Tammann:etal:02} and
\citet{Tammann:Reindl:02}, came as a surprise. 
In Sects.~\ref{sec:PC} and \ref{sec:PL} we have re-analyzed
the situation and found that the break is statistically
significant. The significance of the break  becomes overwhelming from
the shape of the LMC instability strip in the CMD (see
Sect.~\ref{sec:InstabilityStrip:CMD} later).
Since the break has not been predicted by any models, and since it has
not been found in the Galaxy -- perhaps only because of the fewer
number of Galactic Cepheids with known absolute magnitudes -- it may
still appear as an ad hoc result.

     Although we cannot offer a physical explanation for the break, we
compile in the following evidence that several Cepheid parameters
undergo a rather abrupt change at $\log\!P\!=\!1$.

\noindent
(1)     In Fig.~\ref{fig:break:Rphi} the Fourier coefficients
$R_{21}=A_{2}/A_{1}$ and $\Phi_{21}=\Phi_{2}-2\Phi_{1}$ of the
$I$-band light curves of LMC and SMC, as given by
\citet{Udalski:etal:99b,Udalski:etal:99c}, are 
plotted versus $\log P$ \citep[cf. also Fig.~3
of][]{Udalski:etal:99b}. Here $A_{i}$ and $\Phi_{i}$ are the amplitude
and phase of the $(i-1)$th harmonic \citep[cf.][]{Simon:Lee:81}.
The diagrams, both for $R_{21}$ and $\Phi_{21}$, show a striking
discontinuity close to $\log P=1$. 
This is the period where the hump of the well known
Ludendorff-Hertzsprung progression 
\citep{Ludendorff:19,Hertzsprung:24,Hertzsprung:26}
reaches  the phase of maximum light, and from that period and longer
the hump appears on the ascending branch of the light curve. 
[Note: In calculating the $\Phi_{21}$ values from the listed Fourier 
coefficients by \citeauthor{Simon:Lee:81} one must add progressive
increments of $2\pi$ until one gets a positive number, in order to
reproduce the discussion by  \citeauthor{Simon:Lee:81}. Even so, they
have added one extra $2\pi$ increment to Cepheids near $\log P = 1$ to
produce an uprising $\Phi_{21}$ curve near $\log P = 1$ instead of the
downward progression we show here, using one factor of $2\pi$
smaller]. 

     It is now useful to take a side excursion to discuss a few 
properties of the Fourier components. What do they mean? 

     \citet{Schaltenbrand:Tammann:71} introduced Fourier series to 
describe the light curves of Galactic Cepheids simply as an 
objective way to derive many properties of the curves such as the 
amplitude, phase of maximum light, and fraction of the period between 
minimum and maximum light. They did not list the Fourier 
coefficients because these authors used them only as an 
intermediate step to derive accurate values of these observed 
parameters that are manifest from the observations. 

     On the other hand, \citet{Simon:Lee:81} needed an objective 
way characterize light curve shapes to make easier comparisons 
with Simon's (and a multitude of others; see \citeauthor{Simon:Lee:81}
for a summary to 1981) hydrodynamic calculations of theoretical light 
curves, circumventing the need to describe in words the ``bumps'', 
``shoulders'' and ``standstills'' in fitting the calculations with 
the observations. The comparisons between theory and the 
observations, argued Simon and Lee, would best be made by 
deriving selected Fourier coefficients from the observed light curves,
and comparing them with the Fourier decompositions of theoretical
light curves. This was the reason for the initial work by Simon and
Lee in decomposing the observations of Galactic Cepheids in Fourier
series.   
       
     Yet the $R_{21}$ and $\Phi_{21}$ values shown in
Figure~\ref{fig:break:Rphi}, beg for some description useful for
observers of what is actually seen by visual inspection of the light
curves. However, because higher order terms are not used for $R_{21}$
or $\Phi_{21}$, we cannot expect an exact description of light curve
shapes using only the first two terms. Nevertheless, a few conclusions
concerning the correspondence with the observations are evident,
especially for $R_{21}$. 

     Clearly, if a light curve is a perfect sine or cosine 
curve, all $R_{i}$ in a Fourier series will be zero for all $i > 2$. 
Therefore, $R_{21} = 0$ in that case. It is also easily shown (any 
text on Fourier analysis) that all {\em symmetrical curves\/} with
finite first derivatives (no cusps), where the rise from minimum to 
maximum is half the period, also have $R_{21}$ near 0. Hence, small
$R_{21}$ means more symmetrical light curves than large $R_{21}$
values. {\em This is precisely the first fact of the
Ludendorff-Hertzsprung progression of light curve shapes}.  

     \citet{Hertzsprung:24} writes: ``It has been pointed out by 
Ludendorff -- that for [Cepheid] variable stars [those] of 
periods between one and two weeks have the most symmetrical 
lightcurves''. This is shown decisively by Ludendorff's Figure~1 
where the rise times of 91 Cepheids with periods greater than 1 
day show nearly symmetrical (rise times$ = 0.5$ phase units) light 
curves from 7 to 12 days, whereas the light curve shapes for 
longer and shorter period are much more asymmetrical (high $R_{21}$ 
values). It is this fact that is evident in the dip where $R_{21}$ 
approaches zero near 10 days in Fig.~\ref{fig:break:Rphi}.
 
     We can also understand why the maximum value of $R_{21}$ is 
approximately 0.5 for shorter periods. At these periods, the 
light curves are most highly peaked (i.e. very asymmetrical). In 
the limit of large ``peakedness'', the curves can be approximated 
by a saw tooth with very short rise times (actually zero for a 
pure sawtooth with a vertical rise at $0$ and $2\pi$ phases). It is 
well known that for this extreme saw tooth, the first term in a 
Fourier series is $\sin x$ and the second term is $0.5\sin 2x$. 
Hence $R_{1} = 1$ and $R_{2} = 0.5$. 
Hence, $R_{21} = R_{2}/R_{1} = 0.5$ by the definition of $R_{21}$ by
\citeauthor{Simon:Lee:81}.
  
     It is also known that small amplitude Cepheids generally 
have more symmetrical light curves than those of large amplitude. 
Therefore, a relation is expected between $R_{21}$ and amplitude. We 
show in Sect.~\ref{sec:InstabilityStrip} that this indeed is the case,
and as a consequence, that the amplitudes of Cepheids in the upper left 
panel of Fig.~\ref{fig:break:Rphi} are a continuous function of the
place within the scatter of $R_{21}$ at a given period for 
$\log P < 1$ (Fig.~\ref{fig:R21BampBV} later in
Sect.~\ref{sec:InstabilityStrip:Rphi21}). $R_{21}$ is largest 
(highest asymmetry) at large amplitude for all period intervals from 3
to 7 days and again from 13 to greater than 20 days, explaining the
scatter at a given period in the upper left of
Fig.~\ref{fig:break:Rphi}. 

     We also show (Fig.~\ref{fig:BVRphi21} in
Sect.~\ref{sec:InstabilityStrip:Rphi21}) 
that $R_{21}$ varies systematically with the $(B\!-\!V)^{0}$ color when
the data are binned in intervals of absolute magnitude. 
These correlations prove that $R_{21}$ varies systematically with
position in the instability strip, just as is do amplitude and color
at fixed absolute magnitude.  
These correlations are nearly identical in character with the 
same systematic correlations for RR Lyrae variables in globular 
clusters where the instability strip is cut at nearly constant $V$ 
absolute magnitude by the horizontal branch.  

     The interpretation of the $\Phi_{21}$ phase parameter is not as 
clear, but the variations shown in Fig.~\ref{fig:break:Rphi} (right
panel), and from conversations with Norman Simon, suggest that $\Phi_{21}$ 
measures in some approximate way the progression of the 
"Hertzsprung" \citeyearpar{Hertzsprung:26} bump, occurring before
maximum light near 7 day periods to after maximum for periods near 13
days. The discontinuity in $\Phi_{21}$ near 10 days is again
pronounced.        

     A behavior very similar to LMC is known for Galactic Cepheids
\citep{Simon:Lee:81,Simon:Moffett:85,Fernie:Ehlers:99,Zakrzewski:etal:00},
where the variation in $\Phi_{21}$ is even somewhat larger and where
again a few Cepheids with $\Phi_{21}\la0.3$ are
known. \citet{Simon:Moffett:85} note also that in some phase-phase
diagrams (particularly $\Phi_{31}$ vs. $\Phi_{21}$ and $\Phi_{41}$
vs. $\Phi_{31}$) a clear separation of short- and long-period Cepheids
in the Galaxy exists, and they favor the idea ``that the Cepheids with
periods above ten days are somehow `different' from those with periods
below ten days''.

\begin{figure}[t]
\centering
\resizebox{0.95\hsize}{!}{\includegraphics{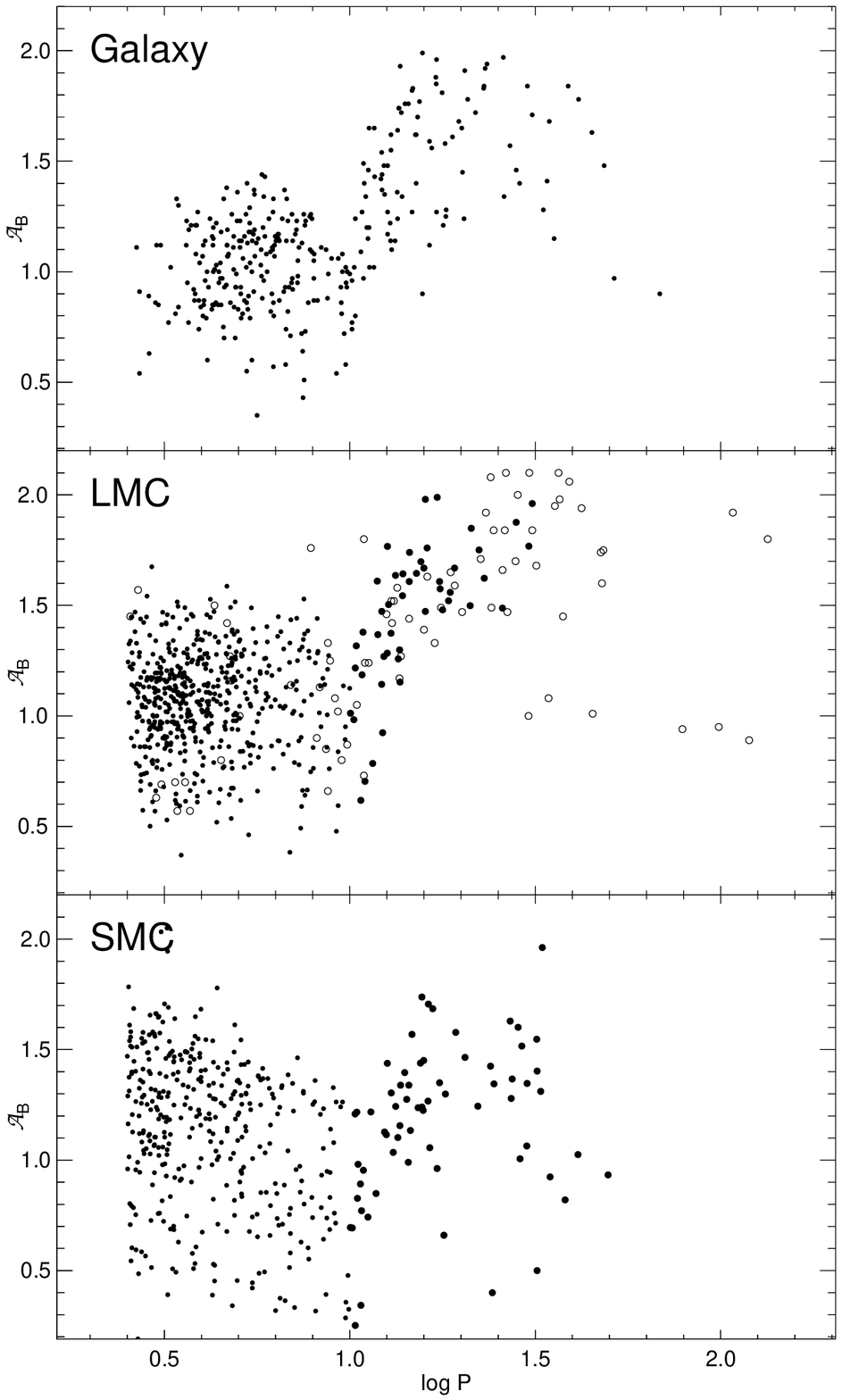}}
\caption{The distribution of the $B$ amplitude ${\cal A}_{B}$ over
  period for the Galaxy (upper panel), LMC (middle panel), and SMC
  (lower panel). Symbols for LMC as in Fig.~\ref{fig:PC:BV}a.} 
\label{fig:break:Vamp}
\end{figure}

\noindent
(2)     An amplitude\,$-$\,$\log P$ plot of LMC Cepheids is shown in
Fig.~\ref{fig:break:Vamp}. The $B$ amplitudes ${\cal A}_{B}$ are derived
from the U99 and Table~\ref{tab:LMCother} samples. For
comparison, corresponding plots are also shown for the Galaxy
\citep{Berdnikov:etal:00} and SMC with ${\cal A}_{B}$ derived from
data in \citet{Udalski:etal:99c}. In all three galaxies
 the maximum values of the amplitudes for $\log P<0.9$ scatter widely 
to reach a rather pronounced minimum just shortwards of $\log P=1$ and
then to increase sharply beyond this point.
The characteristic dip of the maximum amplitudes
around $\log P \ga1$ is common to all three galaxies
\citep[see also][]{Schaltenbrand:Tammann:70,Berdnikov:Ivanov:86},
except that the dip shortwards of $\log P=1$ is less
pronounced in LMC. LMC Cepheids in general have somewhat larger
amplitudes than in the Galaxy. -- The amplitudes ${\cal A}_{1}$ to
${\cal A}_{4}$ of the individual Fourier components in $V$ and $I$,
plotted against $\log P$, show patterns quite similar to
Fig.~\ref{fig:break:Vamp} \citep{Ngeow:etal:03}.

\noindent
(3)     We show in the next section that the slope of the ridge line
of the instability strip of the color-magnitude diagram
(Fig.~\ref{fig:InstabilityStrip}) has a break at $\log P=1$.

\noindent
(4)     Additional, unquestionable changes of the properties of short-
and long-period Cepheids in LMC are shown in the next Section, where
the slope of the constant-period lines suddenly steepens at  $\log P=1$,
where the trend of the amplitudes to decrease with increasing
color $(B\!-\!V)^0$ becomes chaotic, and where the behavior
of the Fourier coefficients $R_{21}$ and $\Phi_{21}$ changes across
the instability strip. \citet{Kanbur:Ngeow:04} have also made the
interesting observation that the {\em Galactic\/} P-C relation, which
is linear at {\em mean\/} light, flattens significantly beyond
$\log P=1$ at {\em maximum\/} light and marginally so at {\em
  minimum\/} light. Thus lineary at {\em mean\/} light would be
coincidental. 

\noindent
(5)     Also the color-amplitude relations, i.e. $(V\!-\!I)^0$
vs. ${\cal A}_{V}$ at maximum, mean, and minimum light are steeper for
long-period than for short-period LMC Cepheids; the same effect is
also observed in SMC and -- less pronounced -- in the Galaxy
\citep{Kanbur:Ngeow:04}. 

     The conclusion is that several parameters and various
correlations between these parameters to be set out in the next sections
change their behavior near a break point at $\log P=1$. 
This holds for the Cepheids in LMC and SMC as well as in the
Galaxy. The existence of the break therefore does not seem to be a
question of metallicity, but rather to indicate that short- and
long-period Cepheids of given metallicity form a non-homologous
sequence. 

     Another phenomenon seemingly supporting the change of character
of Cepheids near $P\approx10\;$days does not withstand the analysis of
modern data. It had been suggested that the number of Cepheids in the
Galaxy (and in M\,31) is exceptionally low at the break period of
$\log P=1$ \citep{Becker:etal:77}, which was supported by
\citet{Buchler:etal:97}. However, the distribution of Berdnikov's
et~al. \citeyearpar{Berdnikov:etal:00} large sample of Galactic Cepheids
shows only a mild minimum at somewhat lower periods ($\log P=0.925$). 
In the case of LMC and SMC \citet{Becker:etal:77}, using very large
Cepheid samples, did not see any significant dip, and this is
confirmed by the well searched OGLE fields in LMC and SMC
\citep{Antonello:etal:02}. The latter authors make also plausible that
the overtone pulsators should be included in the frequency histograms
plotting their presumed original fundamental periods.

     An additional distinction of the break period is that the
fundamental mode and the second-overtone mode are in resonance,
i.e. $P_{0}=2P_{2}$, for Cepheids with a period of ten days, as
pulsation calculations show
\citep{Simon:Lee:81,Moskalik:etal:92}. Ten-day Cepheids 
with the metallicity of LMC have during the second crossing of the
strip masses between roughly $6.5 \frak{M}_{\sun}$ (red edge) and $7.8
\frak{M}_{\sun}$ (blue edge) \citep{Alibert:etal:99}. This mass range
embraces -- coincidentally(?) -- the mass above which stars develop no
degenerate core avoiding thus a helium flash, but leading eventually
to a supernova explosion \citep{Woosley:Weaver:86}.

\section{Properties of the Instability Strip}
\label{sec:InstabilityStrip}
%
\subsection{The instability strip in the $M_{V}$ vs. $(B\!-\!V)^0$ and
  $M_{V}$ vs. $(V\!-\!I)^0$ plane (the CMD)}
\label{sec:InstabilityStrip:CMD}
\begin{figure*}[t]
\centering
\includegraphics[width=13cm]{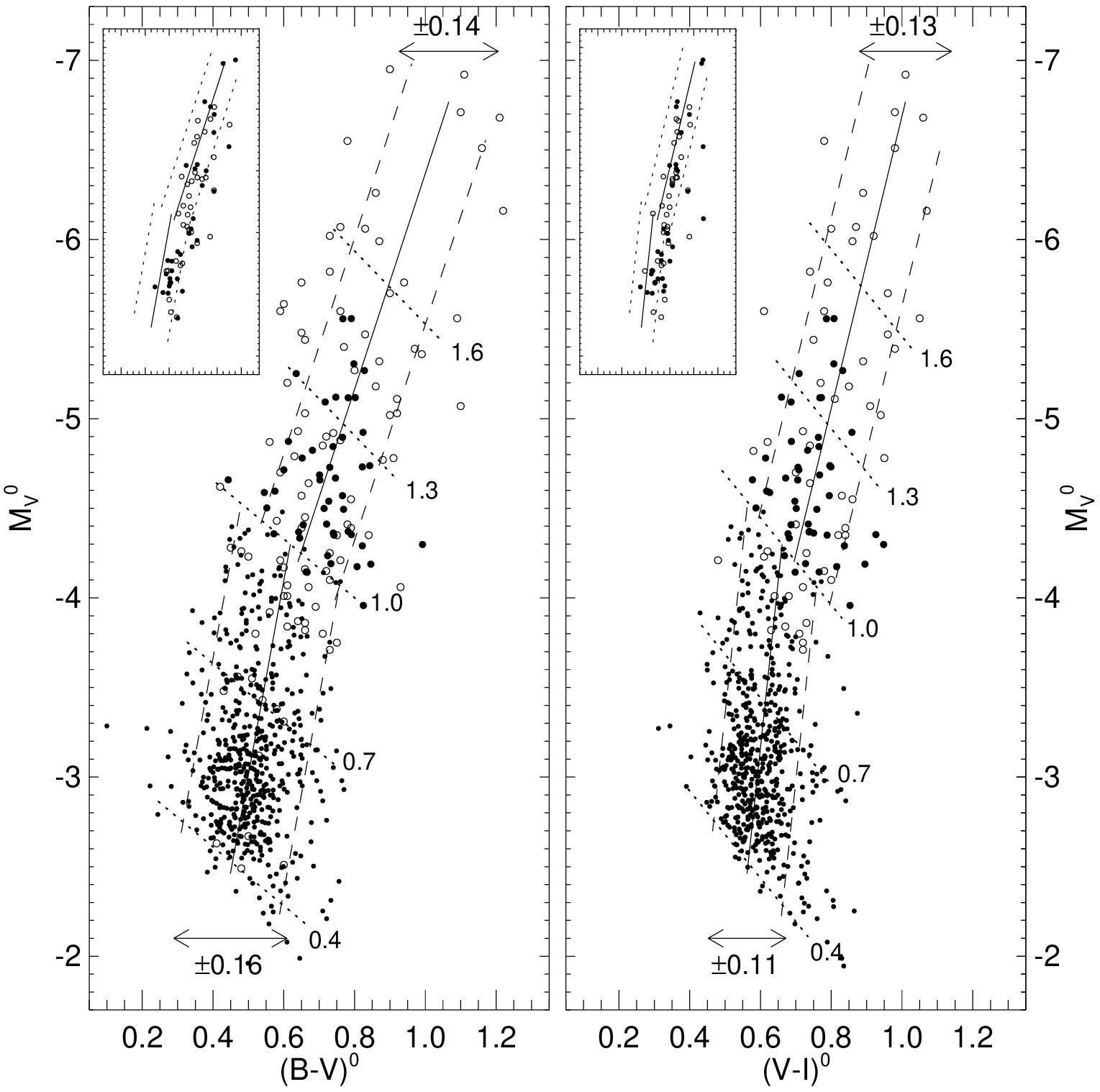}
\caption{The CMDs with $M^0_{V}$ versus $(B\!-\!V)^0$ and $M^0_{V}$
  versus $(V\!-\!I)^0$. The ridge lines of the instability
  strip, resulting from Eqs. (\ref{eq:PC:BV:lt1}, \ref{eq:PC:BV:ge1},
  \ref{eq:PC:VI:lt1}, \ref{eq:PC:VI:ge1}, \ref{eq:PL:V:lt1} and
  \ref{eq:PL:V:ge1}) are shown for $\log P{\protect\grole}1.0$. 
  The 679 (647 for $(V\!-\!I)^0$) dereddened sample Cepheids are
  plotted. Symbols as in Fig.~\ref{fig:PC:BV}a. Approximate blue and
  red edges of the instability strip are fitted to the data (see
  text). Also shown are five constant-period lines from
  Sect.~\ref{sec:InstabilityStrip:CPL}. The inserts repeat the adopted
  instability strip of LMC and show in addition the 53 Galactic
  Cepheids with known absolute magnitude $M_{V}$ taken from Paper~I
  (Fig.~15 there).} 
\label{fig:InstabilityStrip}
\end{figure*}

The absolute magnitudes $M_{V}^0$, based on an adopted value of
$(m-M)^0_{\rm LMC}=18.54$, of 679 Cepheids are plotted against their
dereddened colors $(B\!-\!V)^0$ and $(V\!-\!I)^0$ in
Figure~\ref{fig:InstabilityStrip}. The data outline the instability strip 
with considerable clarity. In particular it is obvious that the ridge
line of the strip is non-linear. However, the points are well
represented by two linear fits whose slopes hold for the cases
$\log P\grole 1$.

   The ridge lines shown in Fig.~\ref{fig:InstabilityStrip} are
determined by combining the P-C relations
(Eqs. \ref{eq:PC:BV:lt1}, \ref{eq:PC:BV:ge1} and
\ref{eq:PC:VI:lt1}, \ref{eq:PC:VI:ge1}) with the P-L relations in $V$
(Eqs. \ref{eq:PL:V:lt1} and \ref{eq:PL:V:ge1}). The resulting ridge
line equations are in $(B\!-\!V)^0$ for
\begin{eqnarray}
\label{eq:CMD:BV:lt1}
\log P<1\!: && \!\!M^0_{V} = -(10.83\!\pm\!0.98)(B\!-\!V)^0 + (2.41\!\pm\!0.38) \\
\label{eq:CMD:BV:ge1}
\log P>1\!: && \!\!M^0_{V} = -(6.02\!\pm\!0.60)(B\!-\!V)^0 - (0.35\!\pm\!0.36)
\end{eqnarray}
and in $(V\!-\!I)^0$ for 
\begin{eqnarray}
\label{eq:CMD:VI:lt1}
\log P<1\!: && \!\!M^0_{V} = -(18.56\!\pm\!2.58)(V\!-\!I)^0 + (7.97\!\pm\!1.32) \\
\label{eq:CMD:VI:ge1}
\log P>1\!: && \!\!M^0_{V} = -(8.15\!\pm\!0.88)(V\!-\!I)^0 - (1.46\!\pm\!0.50).
\end{eqnarray}
The difference of slope between short- and long-period Cepheids has a
$4\sigma$ significance in $(B\!-\!V)^0$ and $(V\!-\!I)^0$.
-- It may be noted that a determination of the ridge line through a
direct regression of $M_{V}$ on color $(B\!-\!V)^0$ or $(V\!-\!I)^0$
would be inappropriate because the slopes of the constant-period lines
define a highly biased sample in the CMD. The resulting slope would be
much too steep particularly for the short-period Cepheids.

     The blue and red boundaries of the instability strip in
$(B\!-\!V)^0$ are drawn by eye such that the strip contains
$\sim\!90\%$ of the points. The corresponding strip half-widths are
$0\fm14$ below and $0\fm16$ above $\log P=1$.

     Once the strip width in $(B\!-\!V)^0$ is fixed, the width in
$(V\!-\!I)^0$ follows by necessity. This is because the absolute
magnitude $M_{V}$ changes by $\Delta M_{V}$ as one goes from the blue
edge to the red edge along a constant-period line. The magnitude
difference $\Delta M_{V}$ must be the same in the $(B\!-\!V)^0$ and
$(V\!-\!I)^0$ strips. Therefore the strip width is inversely
proportional to the slopes of the constant period lines, i.e.\
$1.69:2.43$ for short- and $2.13:2.23$ for long-period Cepheids (for
the slope $\beta$ of the constant-period lines see
Table~\ref{tab:PLC:coeff}). The resulting strip 
half-widths in $(V\!-\!I)^0$ are  $\pm0\fm11$ and $\pm0\fm13$,
respectively. They are shown in Fig.~\ref{fig:InstabilityStrip}. 
The deduced blue and red boundaries in $(V\!-\!I)^0$ comprise the
observed points quite well, except an unexplained number of faint
Cepheids ($M_{V}> -3.0$) which are quite red ($V\!-\!I>0.7$). -- 
The somewhat narrower strip width in $(V\!-\!I)^0$ than in
$(B\!-\!V)^0$ is consistent with the marginally smaller scatter of the
$\log P - (V\!-\!I)^0$ relation (Fig.~\ref{fig:PC:VI}b) than in the
$\log P - (B\!-\!V)^0$ relation (Fig.~\ref{fig:PC:BV}a).

     An important question is whether the LMC Cepheids in
Fig.~\ref{fig:InstabilityStrip} outline the instability strip itself,
or whether they only show the distribution of the Cepheids within a
possibly much wider complete instability strip. The latter possibility
has often been discussed because the loops of present evolutionary
tracks of low-mass Cepheids \citep[e.g.][]{Baraffe:Alibert:01} do not
cross the entire instability strip, but only feed its red part. This
would have far-reaching consequences; e.g.\ one could postulate that
the true instability strip was wider than shown in
Fig.~\ref{fig:InstabilityStrip} and perfectly linear over 
the entire period range, and that its apparent break at $\log P=1$
were merely a reflection of how the evolutionary tracks populate the
strip \citep{Simon:Young:97}.

     However, there are three arguments which suggest that the
short-period Cepheids in Fig.~\ref{fig:InstabilityStrip} outline the
true LMC instability strip {\em at constant period}.
(1) If any blue Cepheids at low luminosities were missing one would
expect a blue cutoff in the $\log P - (B\!-\!V)^0$ relation
(Fig.~\ref{fig:PC:BV}a) for say $\log P<0.6$. This is not seen.
(2) It will be shown in Sect.~\ref{sec:InstabilityStrip:amp} that the 
amplitudes ${\cal A}_{B}$ at any given period are largest close to the
blue boundary of the strip, and as seen in Fig.~\ref{fig:break:Vamp}
the values of ${\cal A}_{B}$ of the Cepheids with the shortest periods
are larger, if anything, than the amplitudes of Cepheids with
intermediate periods. With other words, the amplitudes of the LMC
Cepheids with $0.4<\log P<0.5$ are fully consistent with their being
close to the true blue boundary of the strip 
(cf.\ Fig.~\ref{fig:Bamp}).
(3) The observed width of the strip in temperature
(Fig.~\ref{fig:LMC:Teff}) is $\Delta \log T_{\rm e}=0.08$, which is the
adopted width in most models (cf.\ \citeauthor*{Sandage:etal:99}).
-- If it is indeed correct that LMC Cepheids with $\log P=0.4$ reveal
their true distribution in the instability strip, one can draw
two conclusions. 
(1) A Cepheid with $\log P=0.4$ at the {\em blue\/} strip boundary has
a mass of about $4.8 \frak{M}_{\sun}$ (\citeauthor*{Sandage:etal:99},
Tables 1-5; \citealt{Alibert:etal:99}, Table~7). It must cross the
{\em entire\/} instability strip during its second excursion into the
strip. (2) A Cepheid with $\log P=0.4$ at the {\em red\/} boundary has
a mass of only about $2.9 \frak{M}_{\sun}$. The evolutionary loop of
such a star must at least touch the red strip boundary. These
conclusions are consistent with model calculations of Cepheids of
appropriate metallicity \citep{Saio:Gautschy:98,Alibert:etal:99}.

     The strip width for very luminous Cepheids is poorly determined;
only 22 Cepheids with $M_{V}< -5.5$ are available with known
$(B\!-\!V)^0$, even less with $(V\!-\!I)^0$. Ten of these Cepheids
fall outside the drawn boundaries of the $(B\!-\!V)^0$ strip. Moreover
the ratio of Cepheids lying bluewards and redwards of the ridge line
of the strip is $2:1$ in $(B\!-\!V)^0$ and $3:1$ in
$(V\!-\!I)^0$. This can be taken as a confirmation of the model
prediction \citep[][Fig.~4]{Alibert:etal:99} that the evolution
of luminous Cepheids is much faster on the red side of the strip than
on its blue side.

     A comparison of the LMC and Galactic (see
\ref{sec:PL:galaxy:revised}) instability strips is 
given by the inserts in Fig.~\ref{fig:InstabilityStrip}. The
width of the strips is the same to within the accuracy it can be
determined. 
For Cepheids with $\log P<1$ the slope
of the strip ridge line is about the same for LMC and the Galaxy, but
the Galactic Cepheids are {\em redder\/} 
by $\sim\!0.12$ in $(B\!-\!V)^0$ and $\sim\!0.09$ in $(V\!-\!I)^0$.
This color shift at $M_{V}={\rm const.}$ is not enough to separate the
two strips. It lets them overlap in part. -- LMC Cepheids
with $\log P>1$ define a so much flatter strip ridge line that it
crosses the Galactic ridge line at $M_{V}\sim -6.5$. At still higher
luminosities the LMC strip is redder than in the Galaxy.

\subsection{The slope of the constant-period lines}
\label{sec:InstabilityStrip:CPL}
From the time that the fundamental pulsation theory was combined with
the {\em finite\/} width of the instability strip \citep{Sandage:58}
it was clear, that a blue Cepheid must be brighter than a red Cepheid
of the same period. With other words the lines of constant period are
sloped downward from blue to red in the CMD. The slope
$\beta_{V,B\!-\!V}$ was determined from 
semi-theoretical considerations to be \citep{Sandage:Gratton:63}
\begin{equation}\label{eq:alphaBV:theo}
 \beta_{V,B\!-\!V}\equiv \frac{\Delta M_{V}^{0}}{\Delta (B\!-\!V)^{0}}=2.52,
\end{equation}
where $\Delta M_{V}^{0}$ and $\Delta (B\!-\!V)^{0}$ are the deviations
of a Cepheid from the ridge lines of the P-L and P-C relations,
respectively.\footnote{The deviations are defined throughout this
             paper as $\Delta M=M_{\rm obs}-M_{\rm PL}$ and $\Delta
             (B\!-\!V)=(B\!-\!V)_{\rm obs} - (B\!-\!V)_{\rm PC}$.} 
Values of $\beta_{V,B\!-\!V}$ close to 2.52 were empirically 
confirmed by \citet{Sandage:Tammann:68}, \citet{Sandage:72},
\citet{Martin:etal:79} and
\citet{Caldwell:Coulson:86}, and similar values were widely used
\citep[e.g.][]{Madore:Freedman:91}. However, the assumption that
$\beta_{V,B\!-\!V}$ is constant for all periods was shown to be
inaccurate by \citet{Saio:Gautschy:98} who concluded from stellar
evolution models and pulsation theory that $\beta$ increases with
period for bolometric magnitudes.
\begin{figure*}[t]
\centering
\includegraphics[width=13.5cm]{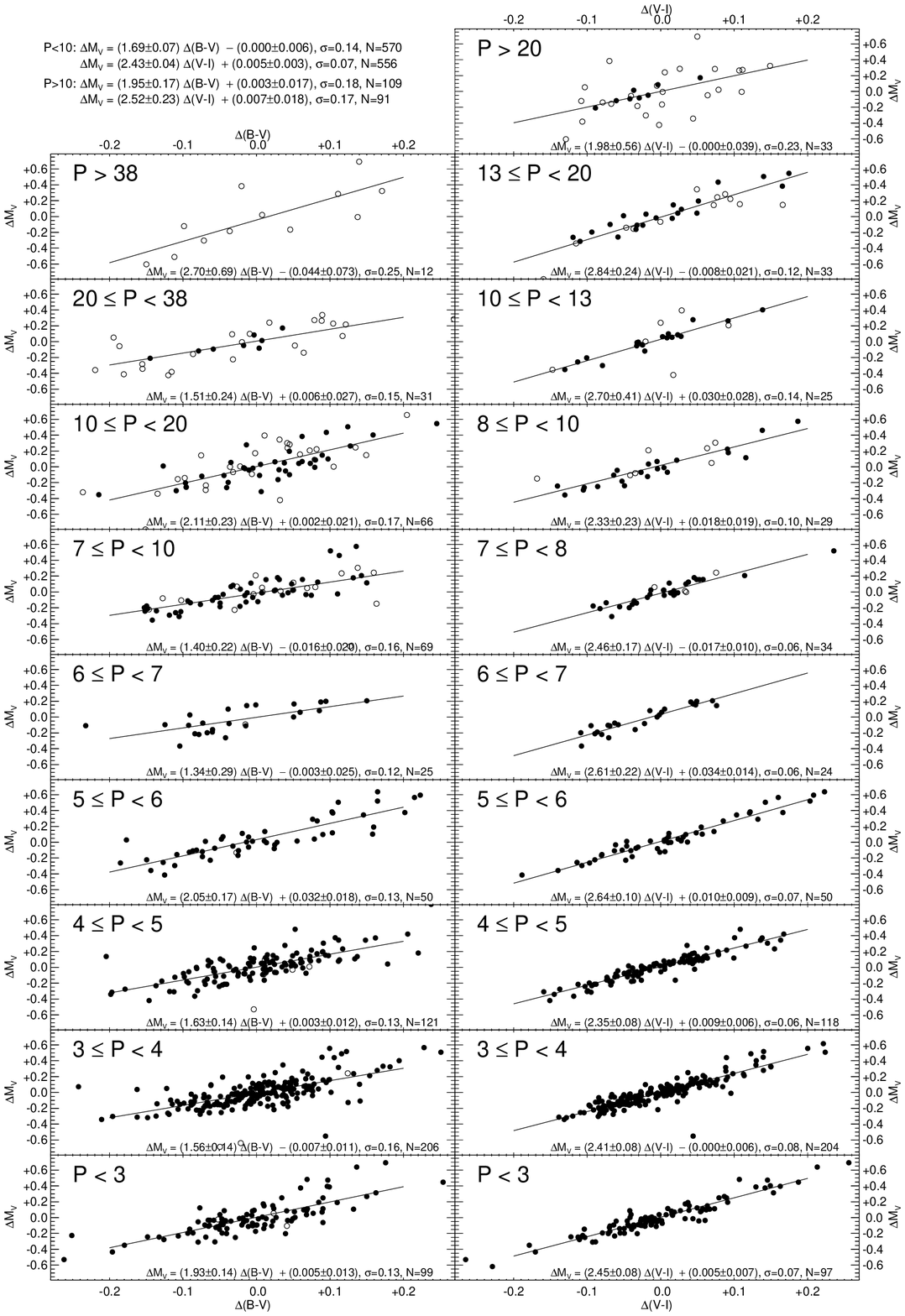}
\caption{The slope of the constant-period lines for different period
  intervals. Plotted are the magnitude residuals $\Delta M_{V}$ from
  the mean P-L relation (Eqs. \ref{eq:PL:V:lt1} and \ref{eq:PL:V:ge1})
  versus the color residuals $\Delta(B\!-\!V)$ and $\Delta(V\!-\!I)$
  from the respective mean P-C relation (Eqs. \ref{eq:PC:BV:lt1},
  \ref{eq:PC:BV:ge1}, \ref{eq:PC:VI:lt1} and
  \ref{eq:PC:VI:ge1}). Symbols as in Fig.~\ref{fig:PC:BV}a. Positive
  $\Delta M_{V}$ residuals are fainter than the P-L ridge
  line. Positive $\Delta(B\!-\!V)$ residuals are redder than the P-C
  ridge line. Hence, brighter magnitudes are bluer along a constant
  period line.} 
\label{fig:CPL}
\end{figure*}

\begin{figure}[t]
\centering
\resizebox{0.90\hsize}{!}{\includegraphics{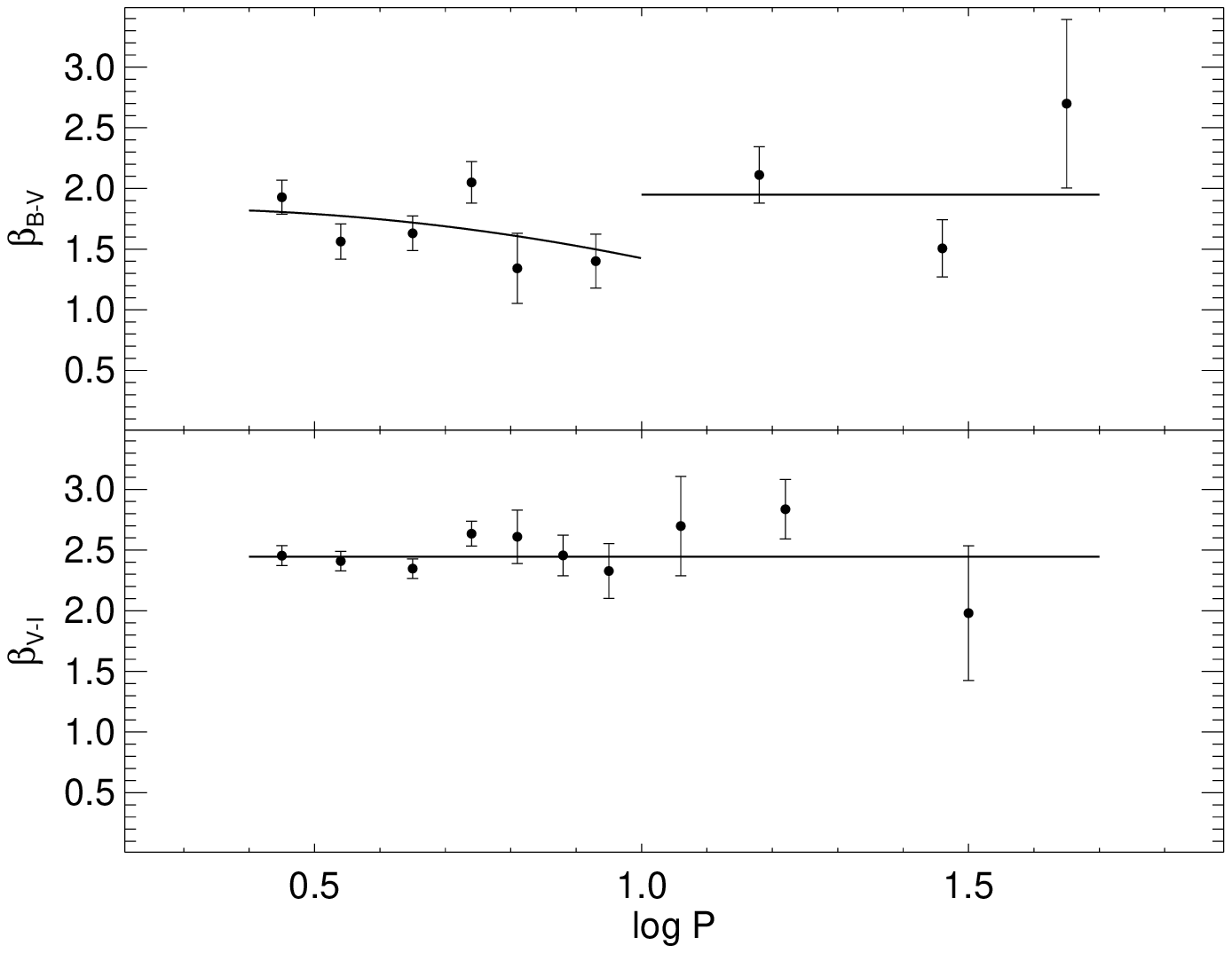}}
\caption{The LMC slopes $\beta_{V,B\!-\!V}$ and $\beta_{V,V\!-\!I}$ of
  the constant-period lines in the CMD plotted against $\log P$. The
  mean lines are drawn by eye.}
\label{fig:alpha}
\end{figure}

   The large present data set of LMC Cepheids allows to determine
$\beta$ empirically for different period intervals. The Cepheids
were binned by period as indicated in Fig.~\ref{fig:CPL} and $\beta$
was determined in each bin. This was done for the CMDs with $M_{V}^{0}$
versus $(B\!-\!V)^{0}$ and $M_{V}^{0}$ versus $(V\!-\!I)^{0}$. The
resulting slopes $\beta_{V,B\!-\!V}$ and $\beta_{V,V\!-\!I}$ are plotted
with their error bars against $\log P$ in Fig.~\ref{fig:alpha}. As can
be seen in the Figure the values of $\beta_{V,B\!-\!V}$ decline
marginally over the interval $0.4<\log P<1$ to jump to
$\beta_{V,B\!-\!V}>1.9$ for $\log P>1$. The behavior of
$\beta_{V,B\!-\!V}$ can be approximated with sufficient accuracy for
the two period intervals by
\begin{eqnarray}
 \label{eq:CPL:BV:lt1}
  \mbox{for}\; \log P<1: && \quad
   \beta_{V,B\!-\!V} = 1.69\pm0.07, \\
 \label{eq:CPL:BV:ge1}
   \mbox{and for}\; \log P>1: && \quad
   \beta_{V,B\!-\!V} = 1.95\pm0.17.
\end{eqnarray}
The values of $\beta_{V,V\!-\!I}$ vary only little over the entire
period interval and they are well represented by a mean value of
\begin{equation}\label{eq:CPL:VI}
   \beta_{V,V\!-\!I} = 2.43\pm0.04.
\end{equation}
The corresponding constant period lines are shown for five different
values of $\log P$ in the instability strip of
Fig.~\ref{fig:InstabilityStrip}.

     The slope $\beta$ of the constant-period lines together with the
strip half-widths in Fig.~\ref{fig:InstabilityStrip} yields the
predicted intrinsic half-widths of the LMC P-L relations in
Fig.~\ref{fig:PL:final}. Evaluating the appropriate values of $\beta$
from Table~\ref{tab:PLC:coeff} and relations given in
\ref{sec:PLC:PLC} yield half-widths of the P-L relation of
$\pm0\fm43$ in $B$, $\pm0\fm28$ in $V$, and $\pm0\fm16$ in $I$. They
are almost the same for $P\grole 10$ days and are significantly wider
than in the Galaxy (see below \ref{sec:PL:galaxy:revised}). The
corresponding boundaries for LMC are shown in Fig.~\ref{fig:PL:final}.

\subsection{Amplitude and the position within the instability strip}
\label{sec:InstabilityStrip:amp}
The systematic behavior -- or lack thereof -- of the amplitudes as a
Cepheid evolves from right to left, or left to right, through the
instability strip is a long-standing question
\citep[see e.g.][]{Hofmeister:67,Payne-Gaposchkin:74}.
\citet{Sandage:Tammann:71} have concluded from the data available at
the time for the Galaxy, LMC, SMC, and M\,31 that the amplitudes of
Cepheids with $\log P<0.86$ are largest near the blue boundary of the
strip and decrease towards the red boundary in a similar pattern as
for RR\,Lyrae variables. This yet unconfirmed result was less clear
for longer periods in these earlier results. 

     The now available large sample of LMC Cepheids renders itself to
a re-analysis of the amplitudes in function of the strip position.

\begin{figure}[t]
\centering
\resizebox{1.0\hsize}{!}{\includegraphics{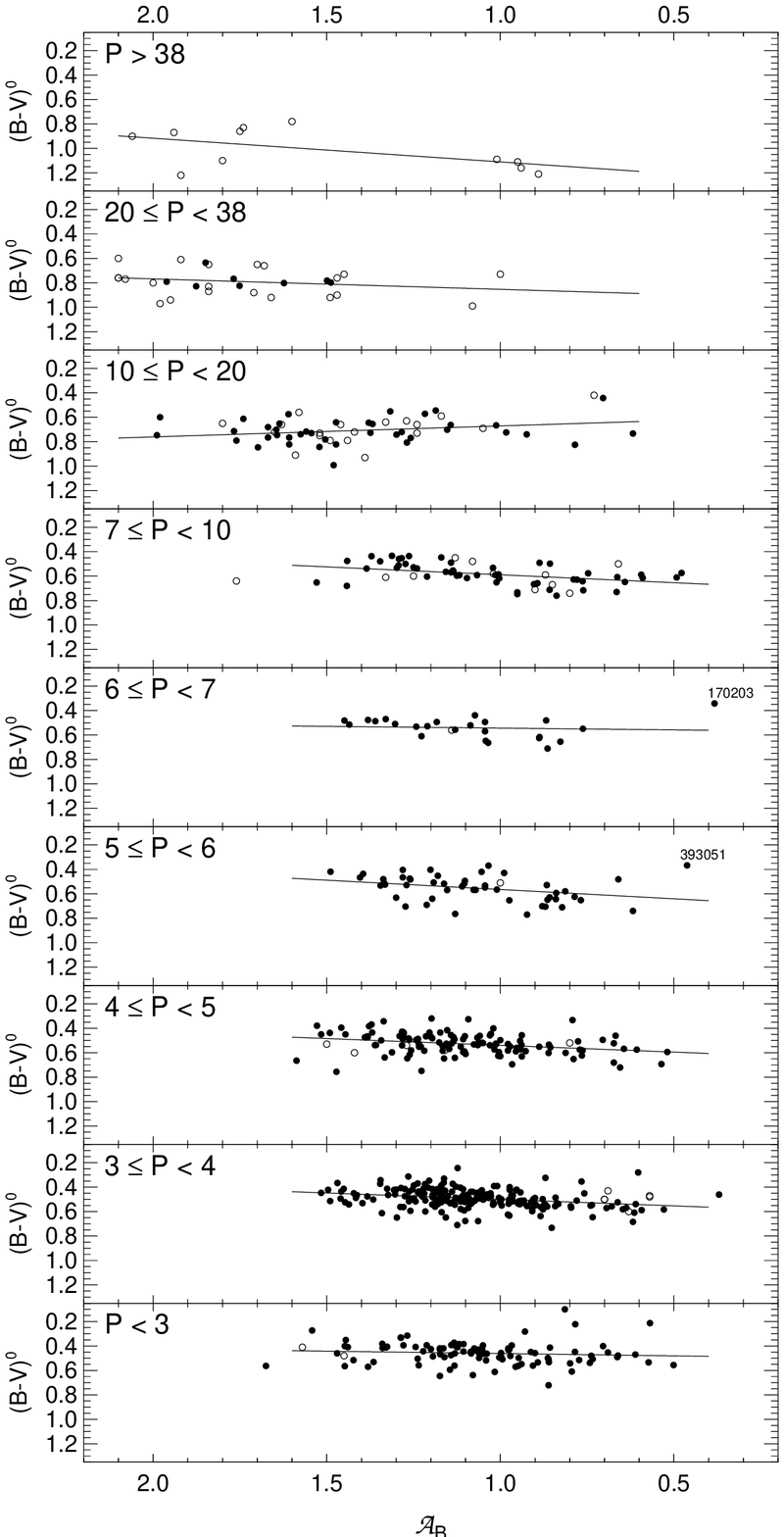}}
\caption{The correlation of the color $(B\!-\!V)$ and
  the $B$ amplitudes ${\cal A}_{B}$ for different period
  intervals. The largest amplitude occur near the blue boundary of the
  instability strip, except in the range $10<P<20^{\rm d}$. Symbols as
  in Fig.~\ref{fig:PC:BV}a.} 
\label{fig:Bamp}
\end{figure}
     The Cepheids are subdivided into nine period intervals in
Fig.~\ref{fig:Bamp}. Their colors $(B\!-\!V)$ are plotted
against their amplitude ${\cal A}_{B}$. There is a definite trend of
the blue Cepheids having larger amplitudes. The slope is steepest in
the interval $3<P<10^{\rm d}$; it may be somewhat flatter for
$P<3^{\rm d}$. Above $P=10^{\rm d}$ the color-amplitude relation
becomes marginal:
\begin{eqnarray}
\nonumber
  \lefteqn{\mbox{for}\; 3<P<10^{\rm d} \quad (N=464, 453)} \\
\label{eq:Bamp:BV:lt1}
  \lefteqn{\Delta (B\!-\!V) = (-0.109\!\pm\!0.016) {\cal A}_{B} 
                              + 0.117\!\pm\!0.018, \,\sigma =0.08}\\
\label{eq:Vamp:BV:lt1}
  \lefteqn{\Delta (B\!-\!V) = (-0.238\!\pm\!0.023) {\cal A}_{V} 
                              + 0.173\!\pm\!0.017, \,\sigma =0.08}\\ 
\nonumber
  \lefteqn{\mbox{and for}\; P>10^{\rm d} \quad (N=99, 79)} \\
\label{eq:Bamp:BV:ge1}
  \lefteqn{\Delta (B\!-\!V) = (-0.059\!\pm\!0.029) {\cal A}_{B} 
                                + 0.087\!\pm\!0.044, \,\sigma =0.10}\\
 \label{eq:Vamp:BV:ge1}
  \lefteqn{\Delta (B\!-\!V) = (-0.121\!\pm\!0.047) {\cal A}_{V} 
                                + 0.108\!\pm\!0.045, \,\sigma =0.10.}
\end{eqnarray}

The data in \citeauthor*{Tammann:etal:03} give similar
relations between the color residuals $\Delta(B\!-\!V)$ and amplitude
${\cal A}_{B}$ for the Galaxy, i.e. 
\begin{eqnarray}
\nonumber
  \lefteqn{\mbox{for}\; 3<P<10^{\rm d}}\\
\label{eq:Bamp:GAL:lt1}
  \lefteqn{\Delta (B\!-\!V) = (-0.113\!\pm\!0.017) {\cal A}_{B} + 0.115\!\pm\!0.018,}  \\
\nonumber
  \lefteqn{\mbox{and for}\; P>10^{\rm d}}\\
\label{eq:Bamp:GAL:ge1}
  \lefteqn{\Delta (B\!-\!V) = (-0.043\!\pm\!0.031) {\cal A}_{B} + 0.062\!\pm\!0.046.}
\end{eqnarray}
Two of the 563 Cepheids in Fig.~\ref{fig:Bamp} are near to the blue
boundary as drawn in Fig.~\ref{fig:InstabilityStrip} and have
exceptionally small amplitudes. Their parameters are given in
Table~\ref{tab:twoLMCceph}. They seem to belong to the rare class of
blue Cepheids with small, rapidly changing amplitudes. Of course, the
possibility remains that they are instead overtone pulsators.
\begin{table}[t]
\begin{center}
\caption{Two LMC Cepheids near the blue boundary of the instability
  strip yet with small amplitudes}
\label{tab:twoLMCceph}
\small
\begin{tabular}{lcccc}
\hline
\hline
\noalign{\smallskip}
   \multicolumn{1}{c}{HV} &
   \multicolumn{1}{c}{$\log P$} &
   \multicolumn{1}{c}{$(B\!-\!V)^0$} &
   \multicolumn{1}{c}{$(V\!-\!I)^0$} &
   \multicolumn{1}{c}{${\cal A}_{B}$} \\
\noalign{\smallskip}
\hline
\noalign{\smallskip}
 170203 (SC3) & $0.838$ & $0.34$ & $0.54$ & $0.38$ \\
 393051 (SC3) & $0.727$ & $0.37$ & $0.62$ & $0.46$ \\
\noalign{\smallskip}
\hline
\end{tabular}
\end{center}
\end{table}

     The color-amplitude relations in Fig.~\ref{fig:Bamp}
clearly show that, statistically, the amplitudes are largest near the
blue side of the instability strip, decreasing toward the red along
lines of constant period, except for the period bin of 10 to 20
days, where the relation changes sign. This was the original result of
\citet{Sandage:Tammann:71} where they concluded that for
$0.4 < \log P < 0.86$, the amplitude is largest near the blue edge
of the strip, decreasing toward the red. But in the period
interval $0.86 < \log P < 1.3$ the trend is reversed, returning to
the original sense for $\log P > 1.3$. Fig.~\ref{fig:Bamp} confirms
this early result, based now on a much larger sample than was
available in 1971.

     It is easier to visualize the trends in the strip by binning
the data by absolute magnitude rather then, as in Fig.~\ref{fig:Bamp},
by period. This maps the amplitude properties of the strip
horizontally in the HRD (i.e. nearly at constant absolute
magnitude). This is the way the properties of the RR Lyrae
instability strip has traditionally been mapped because the
horizontal branch of the clusters cuts the instability strip at
nearly constant absolute $V$ magnitude.

\begin{figure}[t]
\centering
\resizebox{0.9\hsize}{!}{\includegraphics{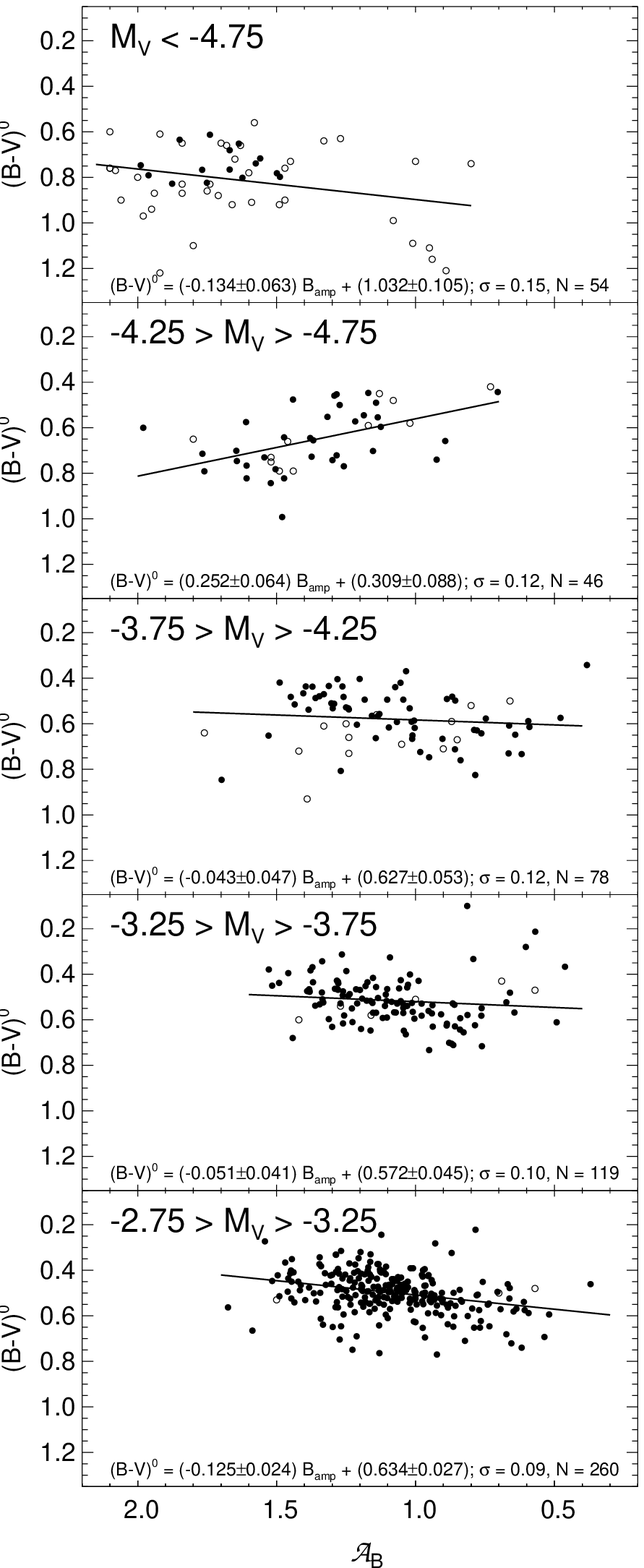}}
\caption{The color-amplitude relation in five intervals of absolute
  magnitude showing a regular progression except in absolute
  magnitude interval of $-4.25 > M_{V} > -4.75$ which is in the period
  interval near 10 days. Symbols as in Fig.~\ref{fig:PC:BV}a.}
\label{fig:BampBV}
\end{figure}

    Fig.~\ref{fig:BampBV} shows such a binning in five intervals of
absolute magnitude ranging from $M_{V} = -2.75$ to $M_{V} <
-4.75$. Clearly the largest amplitudes occur at the blue side of the
instability strip for all absolute magnitudes except, again, in the
magnitude range where the periods are near 10 days 
($-4.25 > M_{V} > -4.75$). 
Here the relation becomes chaotic or perhaps even reversed as in
\citet{Sandage:Tammann:71}.

     Because amplitude varies systematically with color across the
strip (Fig.~\ref{fig:Bamp}) and because of the correlation between
color residuals $\Delta(B\!-\!V)$ and magnitude residuals $\Delta
M_{V}=M_{V}^0 - M_{V,{\rm ridge}}$ (Fig.~\ref{fig:CPL}) there must be
also a correlation between magnitude residuals $\Delta M$ and
amplitudes. By means of a diagram analogous to Fig.~\ref{fig:Bamp} the
relation is found to be
\begin{eqnarray}
\nonumber
 \lefteqn{\mbox{for}\; 3<P<10^{\rm d} \quad (N=464, 453, 452)} \\
\label{eq:Bamp:Mv:lt1}
 \lefteqn{\Delta M_{V} = (-0.223\pm0.040) {\cal A}_{B} + 0.239\pm0.043,\; \sigma =0.20} \\
\label{eq:Vamp:Mv:lt1}
 \lefteqn{\Delta M_{V} = (-0.479\pm0.055) {\cal A}_{V} + 0.352\pm0.041,\; \sigma =0.18} \\
\label{eq:Bamp:Mi:lt1}
 \lefteqn{\Delta M_{I}\, = (-0.066\pm0.024) {\cal A}_{B} + 0.072\pm0.027,\; \sigma =0.13} \\
\nonumber
 \lefteqn{\mbox{and for}\; P>10^{\rm d} \quad (N=99, 79, 82)} \\
\label{eq:Bamp:Mv:ge1}
 \lefteqn{\Delta M_{V} = (-0.091\pm0.073) {\cal A}_{B} + 0.146\pm0.113,\; \sigma =0.25} \\
\label{eq:Vamp:Mv:ge1}
 \lefteqn{\Delta M_{V} = (-0.160\pm0.121) {\cal A}_{V} + 0.146\pm0.117,\; \sigma =0.26} \\
\label{eq:Bamp:Mi:ge1}
 \lefteqn{\Delta M_{I}\, = (-0.005\pm0.057) {\cal A}_{B} + 0.008\pm0.082,\; \sigma =0.19.} 
\end{eqnarray}
With these equations one can derive period-luminosity-amplitude
relations, but their practical use remains restricted even for the
short-period interval $3<P<10^{\rm d}$, because the scatter is hardly
reduced compared to the straight P-L relations (Eqs.\ref{eq:PL:V:lt1}
and \ref{eq:PL:V:ge1}). 

     Eqs.~(\ref{eq:Bamp:Mv:lt1})$-$(\ref{eq:Bamp:Mi:ge1}) express
a warning against systematic distance errors. The importance of
unbiased Cepheid samples has been pointed out by several authors 
\citep[e.g.][]{Sandage:88,Paturel:etal:02}.
Small-amplitude (${\cal A}_{B}\approx0\fm5$) Cepheids turn now out to
be fainter on average than their long-period counterparts (${\cal
  A}_{B}\approx2^{\rm m}$) by $\sim\!0\fm2$. To the extent that small
amplitude Cepheids are harder to detect, they may be underrepresented
in available Cepheid samples. Yet a fair amplitude distribution here
turns out to be prerequisit of an unbiased distance
determination. (Because of the correlation of amplitude with color, a
fair color distribution could also assure an unbiased Cepheid sample,
but dust reddening and typically large color errors make this kind of
test notoriously difficult).

     To estimate the amplitude bias we consider first the period
 interval $3<P<10^{\rm d}$ and make the rather extreme assumption that
 in an external galaxy like LMC only the 50 Cepheids with the largest
 amplitudes (out of a total of 464) were discovered. Their median
 amplitude is ${\cal A}_{B}=1.51$ to be compared with the true median
 value of ${\cal A}_{B}=1.09$. If the amplitude difference
 $\Delta{\cal A}_{B}=0.42$ is inserted in Eq.~(\ref{eq:Bamp:Mv:lt1})
 we find that the detected sample is overluminous by $0\fm09$,
 corresponding to a distance overestimate of 4\%.
-- For the interval $P>10^{\rm d}$ the situation is still more
favorable because the coefficient of ${\cal A}_{B}$ in
Eq.~(\ref{eq:Bamp:Mv:ge1}) is about three times smaller and in fact
insignificant. Because essentially all known Cepheids outside the
Local Group have periods larger than $10^{\rm d}$, the amplitude bias
affects the published Cepheid distances beyond $\sim\!1\;$Mpc only at
the 1\% level. In any case amplitude bias cannot be the cause of the
large bias of Cepheid distances suggested by
\citet{Teerikorpi:Paturel:02}.

\subsection{Variations in the $R_{21}$ and $\Phi_{21}$ Fourier
  components across the strip at constant absolute magnitude}
\label{sec:InstabilityStrip:Rphi21}
%
\begin{figure*}[t]
\centering
\includegraphics[width=13.0cm]{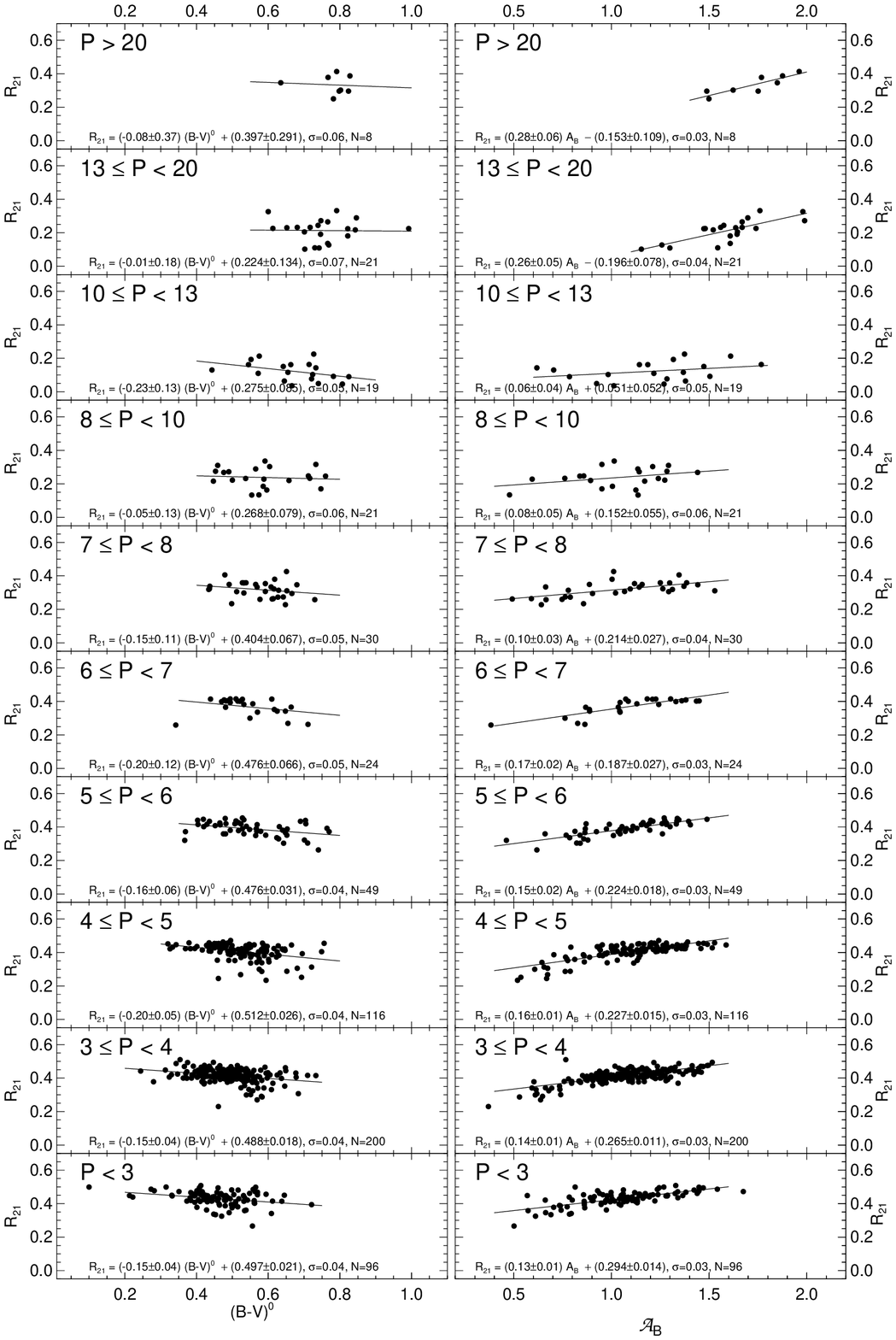}
\caption{Correlations of $R_{21}$ with the $(B\,-\,V)^{0}$ color
  (left) and with the amplitude${\cal A}_{B}$ of the $B$ light curves
  (right) in narrow bins of period. High $R_{21}$ values occur at high
  amplitude and blue color, showing that $R_{21}$ varies
  systematically across the instability strip, with highest values
  near the blue edge of the strip.}
\label{fig:R21BampBV}
\end{figure*}
     The position of a Cepheid in the strip at a given absolute
magnitude is, of course, measured by its unreddened color.
Therefore, a systematic trend of any property (amplitude, period,
asymmetry, bump position relative to the maximum phase, etc.)
with color, maps that property across the strip.
 
    $R_{21}$ is a strong function of amplitude, proved in
Fig.~\ref{fig:R21BampBV} (right panel) where the data are binned in
small intervals of period so as to understand the scatter in
Fig.~\ref{fig:break:Rphi} (left panel) in $R_{21}$ at a given
period. The correlations between $R_{21}$ and amplitude are
strong. High values of $R_{21}$ (high asymmetry in the light curve) go
with high amplitude. Hence, following the discussion of
Fig.~\ref{fig:break:Rphi} in Sect.~\ref{sec:break}, the lower envelope
to the $R_{21}$-period distribution is populated by low amplitude
variables and the upper envelope by high amplitude variables in each
period interval. 

\begin{figure*}[t]
\centering
\includegraphics[width=13cm]{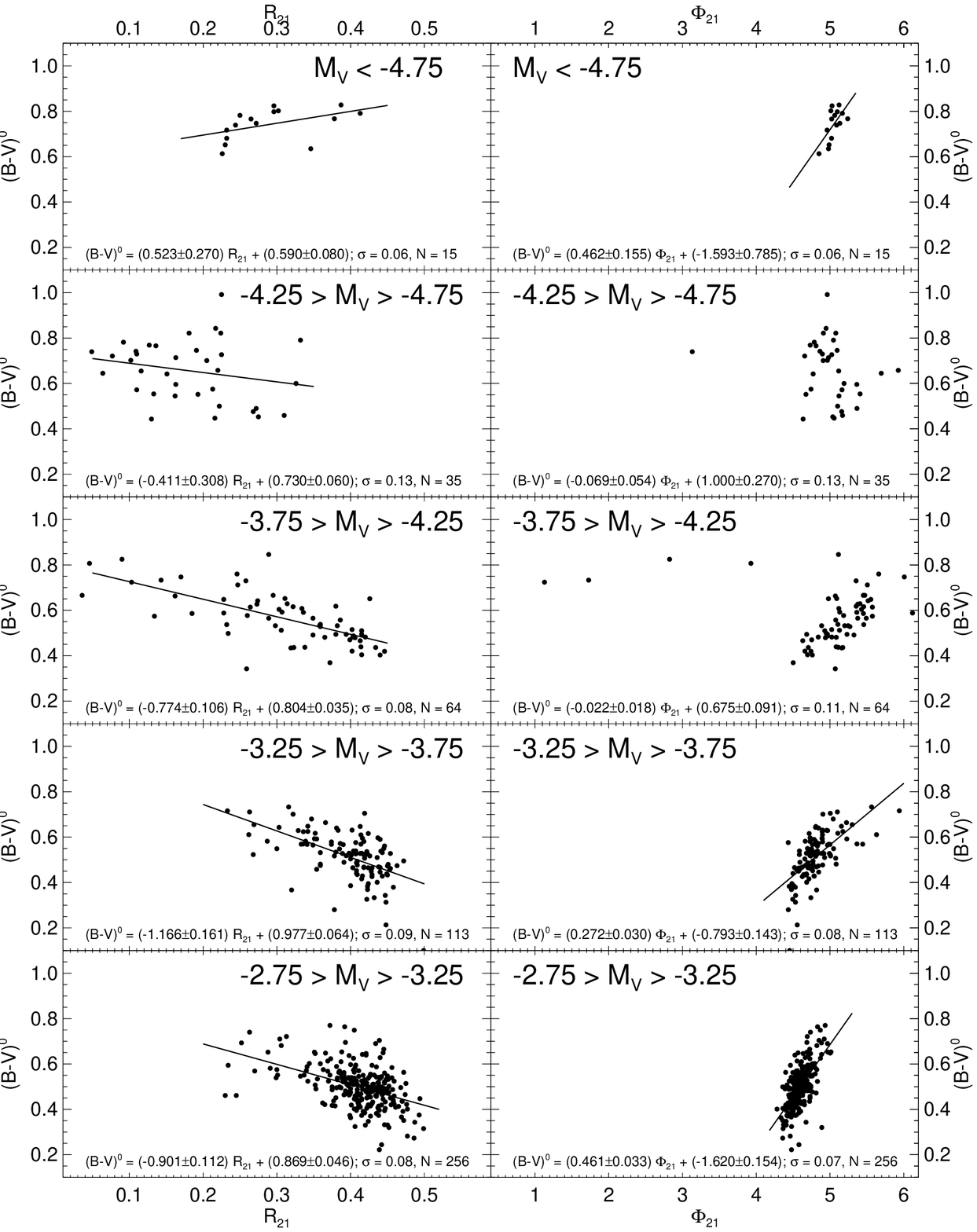}
\caption{Correlations of $R_{21}$ and $\Phi_{21}$ with $(B\,-\,V)^{0}$
  color in narrow intervals of absolute magnitude. This maps the
  $R_{21}$ and $\Phi_{21}$ Fourier combinations across the strip at
  constant luminosity.}
\label{fig:BVRphi21}
\end{figure*}
     Because $R_{21}$ is such a strong function of amplitude, and
because amplitude varies systematically with color across the
strip (NB again; high amplitude variables are bluest), there must
be a correlation of $R_{21}$ with color. If so, $R_{21}$ is expected
to be largest at bluest colors at a given period. The left panel of
Fig.~\ref{fig:R21BampBV} shows this to be the case. And, again, by
analogy with the correlations known for the RR Lyrae variables at
constant absolute magnitude, the above conclusion is again evident in
Fig.~\ref{fig:BVRphi21} (left column) where $R_{21}$ values are
correlated with $(B\!-\!V)^{0}$ color at constant absolute
magnitude. The highest $R_{21}$ values (highly peaked light curves)
occur in all absolute magnitude intervals at the bluest
colors. {\em Clearly, $R_{21}$ varies
systematically across the gap at all absolute magnitudes.}
\begin{figure*}[t]
\centering
\includegraphics[width=13.0cm]{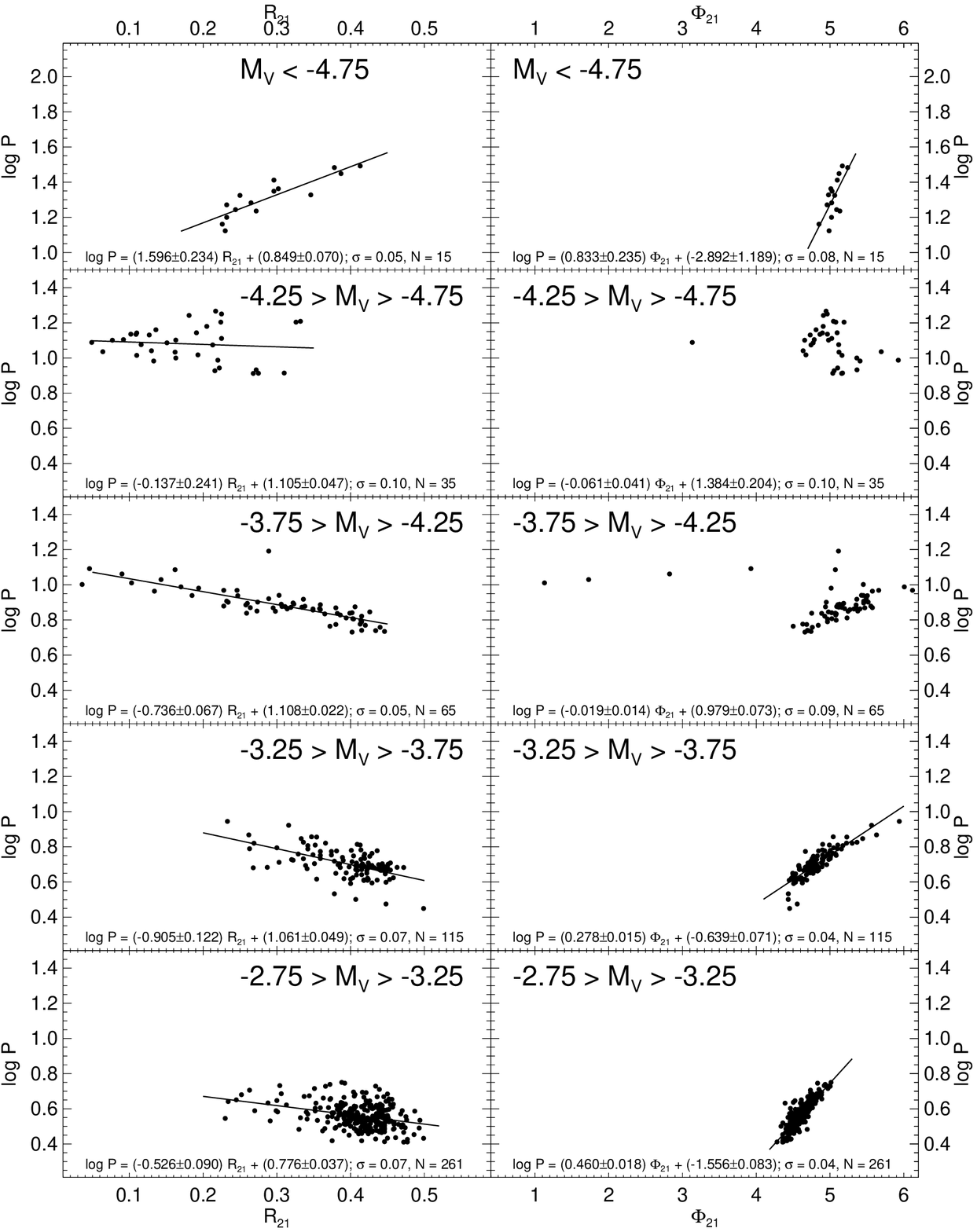}
\caption{The $R_{21}$ and $\Phi_{21}$ Fourier components correlated
  with period in intervals of absolute magnitude.}
\label{fig:logPRphi21}
\end{figure*}

     The meaning of the systematic variation of $\Phi_{21}$ is not
as clear as that of $R_{21}$, yet the right panel of
Fig.~\ref{fig:BVRphi21} shows that this phase parameter also
varies systematically across the strip.

     To continue the analogy with the RR Lyrae correlations of
parameters at constant magnitude across the RR Lyrae strip, we
show in Fig.~\ref{fig:logPRphi21} the correlations between $R_{21}$
and period, again at fixed absolute magnitude. If one draws in 
mind's eye a horizontal line across the strip in the CMD
(Fig.~\ref{fig:InstabilityStrip}), say at absolute magnitude $M_{V} =
-4$, then a family of lines of constant period (not shown) with $\log P$
between about $0.8$ and $0.85$ will cut across this constant absolute
magnitude line at different colors within the strip boundaries. 
And because $R_{21}$ is strongly correlated with color
(Fig.~\ref{fig:BVRphi21}), necessarily, the shortest period lines (at
bluer colors) will also cut the $R_{21}$ correlation with color at
higher $R_{21}$ values. There {\em must}, then, be a period-$R_{21}$
relation at each fixed absolute magnitude if the lines of constant
period slope in the $M_{V}$-color plane.
Fig.~\ref{fig:logPRphi21} shows the $R_{21}$-$\log P$ relation
for the LMC data binned in intervals of fixed absolute magnitude. It
is in the expected sense for the three faintest absolute magnitude
intervals.

     Because of the correlation of $\Phi_{21}$ with color in
Fig.~\ref{fig:BVRphi21} (right panels) and the period-color relations
at constant $M_{V}$ (Fig.~\ref{fig:logPBVI} in the next section),
$\Phi_{21}$ must necessarily be correlated with period (at constant
$M_{V}$) for the same reasons as $R_{21}$. 
The right panels of Fig.~\ref{fig:logPRphi21} show this to be true,
but again the interpretation in terms of the Hertzsprung ``bumps'' is
not obvious.

\subsection{The color-period and period-amplitude relations at constant 
            absolute magnitude}
\label{sec:InstabilityStrip:caM}
\begin{figure*}[t]
\centering
\includegraphics[width=13.0cm]{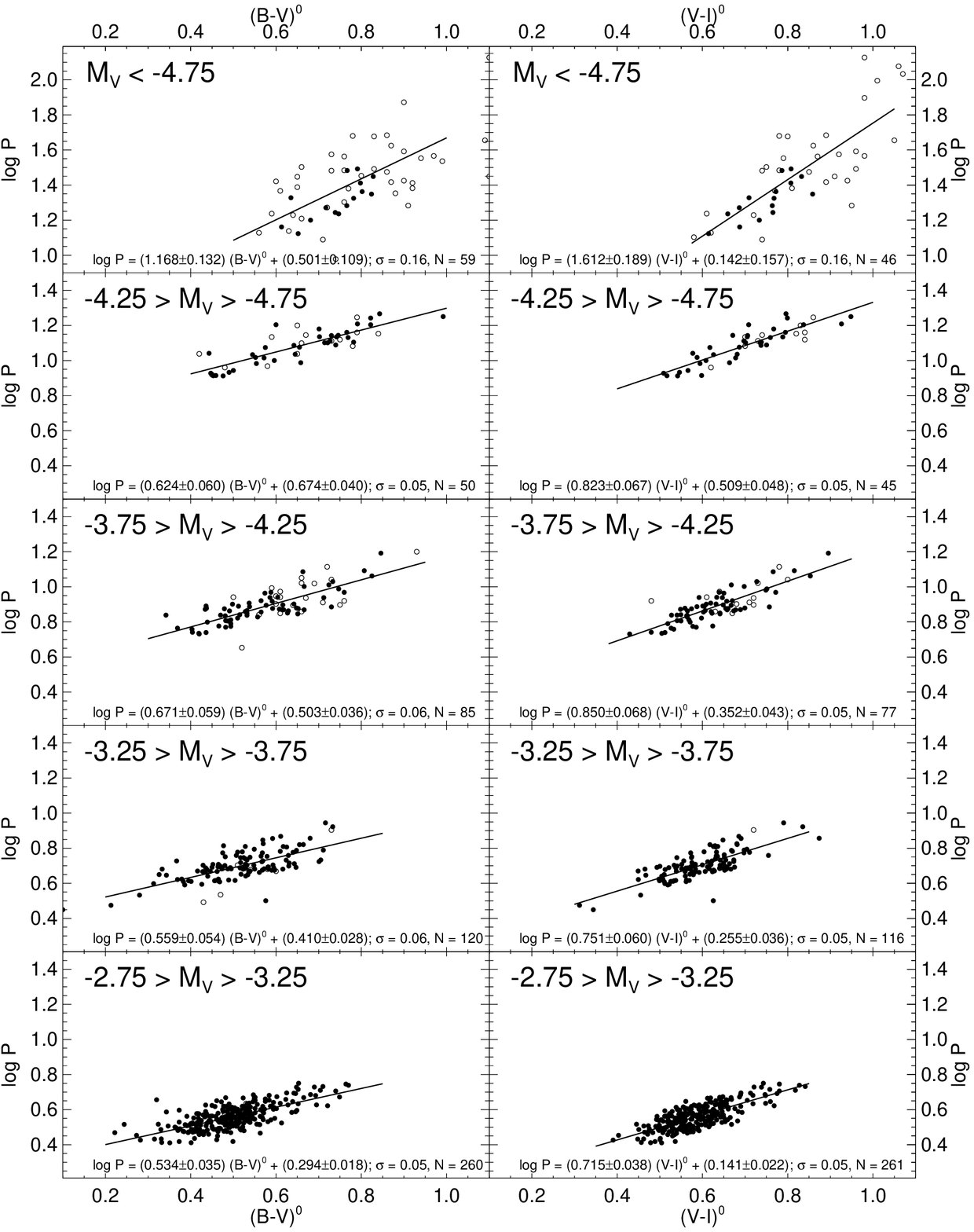}
\caption{The period-color correlations at fixed absolute luminosity
intervals. The correlations are similar and in the same sense as
for RR Lyrae stars (because the lines of constant period in the
CMD have similar slopes here and for the RR Lyrae variables). Symbols
as in Fig.~\ref{fig:PC:BV}a.}
\label{fig:logPBVI}
\end{figure*}

In analogy with the RR Lyrae correlations, and for the same
reason just given (lines of constant period cutting a constant
absolute magnitude line in the CMD at different color), there
must be a period-color relation at constant absolute magnitude. 
In an obvious representation, the period-color relations in
$(B\!-\!V)^{0}$ and $(V\!-\!I)^{0}$ are shown in
Fig.~\ref{fig:logPBVI}. As expected, and as in the RR Lyrae
correlation, shorter periods occur at bluer color at fixed absolute
magnitude for all absolute magnitude intervals from the faintest to
the brightest. There is no break at 10 days because 
Fig.~\ref{fig:CPL} for the lines of constant period show that the
slope of these lines always has the same sense for all periods,
including the intervals that embrace 10 days. If the slope had
anywhere been zero or negative, then the correlations in
Fig.~\ref{fig:logPBVI} would either be flat or would slope in the
opposite direction than shown.
\begin{figure*}[t]
\centering
\includegraphics[width=13.0cm]{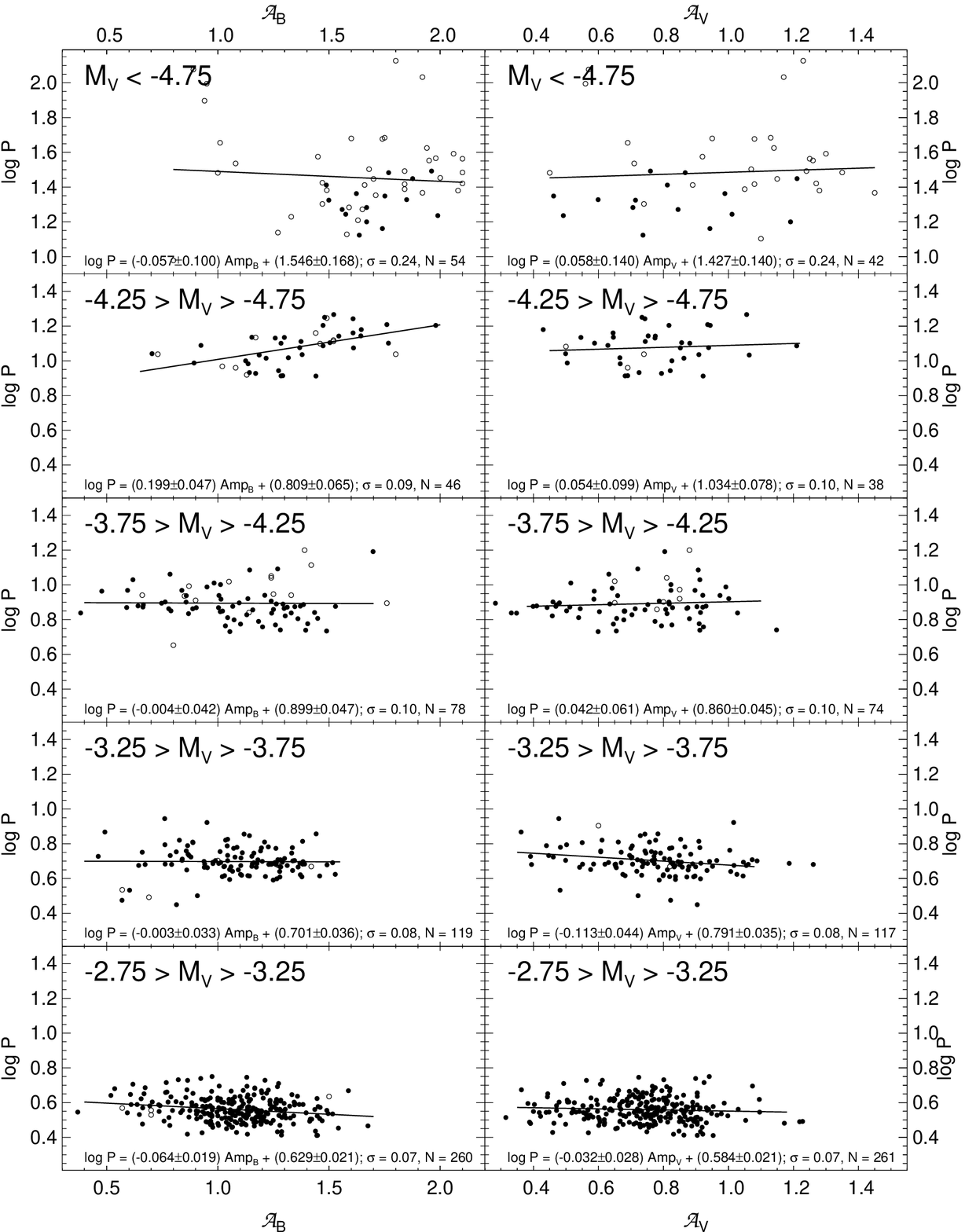}
\caption{The period-amplitude correlations at fixed absolute
magnitude intervals as similarly known for RR\,Lyrae stars in globular
clusters. Symbols as in Fig.~\ref{fig:PC:BV}a.} 
\label{fig:logPAmpBV}
\end{figure*}

     In a similar way, again as in the RR Lyrae correlations,
there are correlations between period and amplitude for fixed
intervals of absolute magnitude, shown in Fig.~\ref{fig:logPAmpBV}. 
The reverse slope for the absolute magnitude
interval of $-4.25 > M_{V} > -4.75$ and the reversal again to the
original is a consequence of the reverse color-amplitude relations in
the relevant period ranges (i.e.\ $\log P <0.86$, $0.86<\log P<1.3$,
and  $\log P > 1.3$) discussed earlier (Fig.~\ref{fig:BampBV})
and by \citet{Sandage:Tammann:71}.

\section{The Period-Luminosity-Color Relation of LMC}
\label{sec:PLC}
%
\subsection{The P-L-C relation}
\label{sec:PLC:PLC}
The intrinsic scatter of the P-L relation is due to the finite
width of the instability strip and the sloping of the constant-period
lines. The scatter can be reduced by introducing a P-L-C relation of
the form 
\begin{equation}\label{eq:PLC:theo}
M_{\lambda}^{\rm R}=a_{\lambda}\log P -
\beta_{\lambda}\,[(B\!-\!V) \;{\rm or}\; (V\!-\!I)] + \gamma_{\lambda},
\end{equation}
where $\beta$ is the slope of the constant-period lines as discussed
in Sect.~\ref{sec:InstabilityStrip:CPL}.

     Eq.~(\ref{eq:PLC:theo}) translates the magnitude to the value an
unreddened Cepheid would have if it were lying on the ridge line of
the P-C relation and hence -- except for observational errors -- also
on the ridge line of the P-L relation. 
The application of the P-L-C relation has been tried many times with
values of $\beta_{V,B\!-\!V}\sim 2.5$, but the results were generally
unsatisfactory. The reason is now clear, because the value was too
high and it is strongly variable from galaxy to galaxy as will be
shown in the following.

     The coefficients $\alpha$, $\beta$, and $\gamma$ for $M_{V}$ can be
determined by combining the P-L relation (Eqs.~\ref{eq:PL:V:lt1} and
\ref{eq:PL:V:ge1}) with the appropriate P-C relations
(Eqs.~\ref{eq:PC:BV:lt1}, \ref{eq:PC:BV:ge1}, \ref{eq:PC:VI:lt1}, and
\ref{eq:PC:VI:ge1}) or by direct regression over the available
Cepheids. Either way yields the same results with only minute
differences. The results for LMC are shown in
Table~\ref{tab:PLC:coeff}. 
\begin{table}
\begin{center}
\caption{The coefficients in the P-L-C relation
  (Eq.~\ref{eq:PLC:theo}) for different wavelength combinations
  for LMC, SMC,  and the Galaxy.}
\label{tab:PLC:coeff}
\small
\begin{tabular}{lccc}
\hline
\hline
\noalign{\smallskip}
 & 
   \multicolumn{1}{c}{$\alpha$} &
   \multicolumn{1}{c}{$ \beta$} &
   \multicolumn{1}{c}{$\gamma$} \\
\noalign{\smallskip}
\hline
\noalign{\smallskip}
 &   \multicolumn{3}{c}{Galaxy} \\
$M_{V}$, $(B\!-\!V)$ &
          $-3.324\pm0.180$ & $0.601\pm0.404$ & $-1.117\pm0.168$ \\
$M_{V}$, $(V\!-\!I)$ &
          $-3.253\pm0.143$ & $0.666\pm0.410$ & $-1.266\pm0.241$ \\
\noalign{\smallskip}
\hline
\noalign{\smallskip}
$M_{V}$, $(B\!-\!V)$ & \multicolumn{3}{c}{LMC}  \\
  $~~~\log P<1$ & $-3.417\pm0.045$ & $1.686\pm0.071$ & $-1.922\pm0.036$ \\
  $~~~\log P>1$ & $-3.407\pm0.099$ & $1.948\pm0.169$ & $-2.034\pm0.097$ \\
$M_{V}$, $(V\!-\!I)$ & & & \\
  $~~~\log P<1$ & $-3.328\pm0.022$ & $2.434\pm0.041$ & $-2.565\pm0.025$ \\
  $~~~\log P>1$ & $-3.300\pm0.101$ & $2.517\pm0.233$ & $-2.660\pm0.128$ \\
\noalign{\smallskip}
\hline
\noalign{\smallskip}
$M_{V}$, $(B\!-\!V)$ & \multicolumn{3}{c}{SMC}  \\
  $~~~\log P<1$ & $-3.086\pm0.075$ & $1.701\pm0.126$ & $-1.948\pm0.056$ \\
  $~~~\log P>1$ & $-3.457\pm0.141$ & $1.431\pm0.184$ & $-1.371\pm0.144$ \\
$M_{V}$, $(V\!-\!I)$ & & & \\
  $~~~\log P<1$ & $-3.313\pm0.053$ & $2.813\pm0.094$ & $-2.760\pm0.053$ \\
  $~~~\log P>1$ & $-3.548\pm0.094$ & $2.825\pm0.175$ & $-2.513\pm0.130$ \\
\noalign{\smallskip}
\hline
 \end{tabular}
\end{center}
\end{table}

     The values $\beta_{V,B\!-\!V}$ and $\beta_{V,V\!-\!I}$
transform $V$ magnitudes onto the ridge line of the P-L relation in
$V$. It follows trivially that the coefficients to bring the $B$ and
$I$ magnitudes onto the respective ridge lines are
\begin{equation}\label{eq:alpha:B:BV}
  \beta_{B,B\!-\!V} \equiv  \frac{\Delta
  M_{B}}{\Delta(B\!-\!V)}=\beta_{V,B\!-\!V} +1
\end{equation}
and
\begin{equation}\label{eq:alpha:I:VI}
  \beta_{I,V\!-\!I}  \equiv   \frac{\Delta
  M_{I}}{\Delta(V\!-\!I)}=\beta_{V,V\!-\!I} -1.
\end{equation}
With these equations it is easy to convert the P-L-C relation in
$M_{V}$ and $(B\!-\!V)$ into those in $M_{B}$ and $(B\!-\!V)$ or 
$M_{I}$ and $(V\!-\!I)$.

     While we have for LMC $\alpha_{V,V\!-\!I} = \alpha_{I,V\!-\!I} =
-3.328$ and $-3.300$ for short- and long-period Cepheids,
respectively, Udalski's et~al. \citeyearpar{Udalski:etal:99a} overall
value of $\alpha_{I,V\!-\!I}=-3.246\pm0.015$ is slightly flatter; this
is explained by the additional long-period Cepheids used here from
Table~\ref{tab:LMCother}. 
Their value $\beta_{I,V\!-\!I}=1.409\pm0.026$, implying 
$\beta_{V,V\!-\!I}=2.409\pm0.026$, is somewhat smaller than
the two corresponding values given in Table~\ref{tab:PLC:coeff} for
short- and long-period Cepheids.

     The coefficients in Table~\ref{tab:PLC:coeff} are directly
determined by linear regressions in two variables. An alternative way
is to combine the P-L relations as given above (\ref{sec:PL:LMC} and
\ref{sec:PL:galaxy:revised}; the values of SMC still being
unpublished) with the appropiate P-C relations (Sect.~\ref{sec:PC} and
\citeauthor*{Tammann:etal:03}, Eq.~3 and 5). This leads to the same
coefficients as in Table~\ref{tab:PLC:coeff} with only minute
differences. In addition the coefficients $\beta$ of LMC are nearly
identical -- as required -- with the slope of the constant period
lines (Fig.~\ref{fig:CPL} and \ref{fig:alpha}, 
Eq.~\ref{eq:CPL:BV:lt1}$-$\ref{eq:CPL:VI}). This strengthens the
confidence in the P-L-C relations of Table~\ref{tab:PLC:coeff} for
LMC; the same holds also for SMC. In case of the Galaxy the
coefficients have large errors because they rest on only 69
calibrators, which in addition carry the errors of their individual
BBW or cluster distances.

\begin{figure}[t]
\centering
\resizebox{1.0\hsize}{!}{\includegraphics{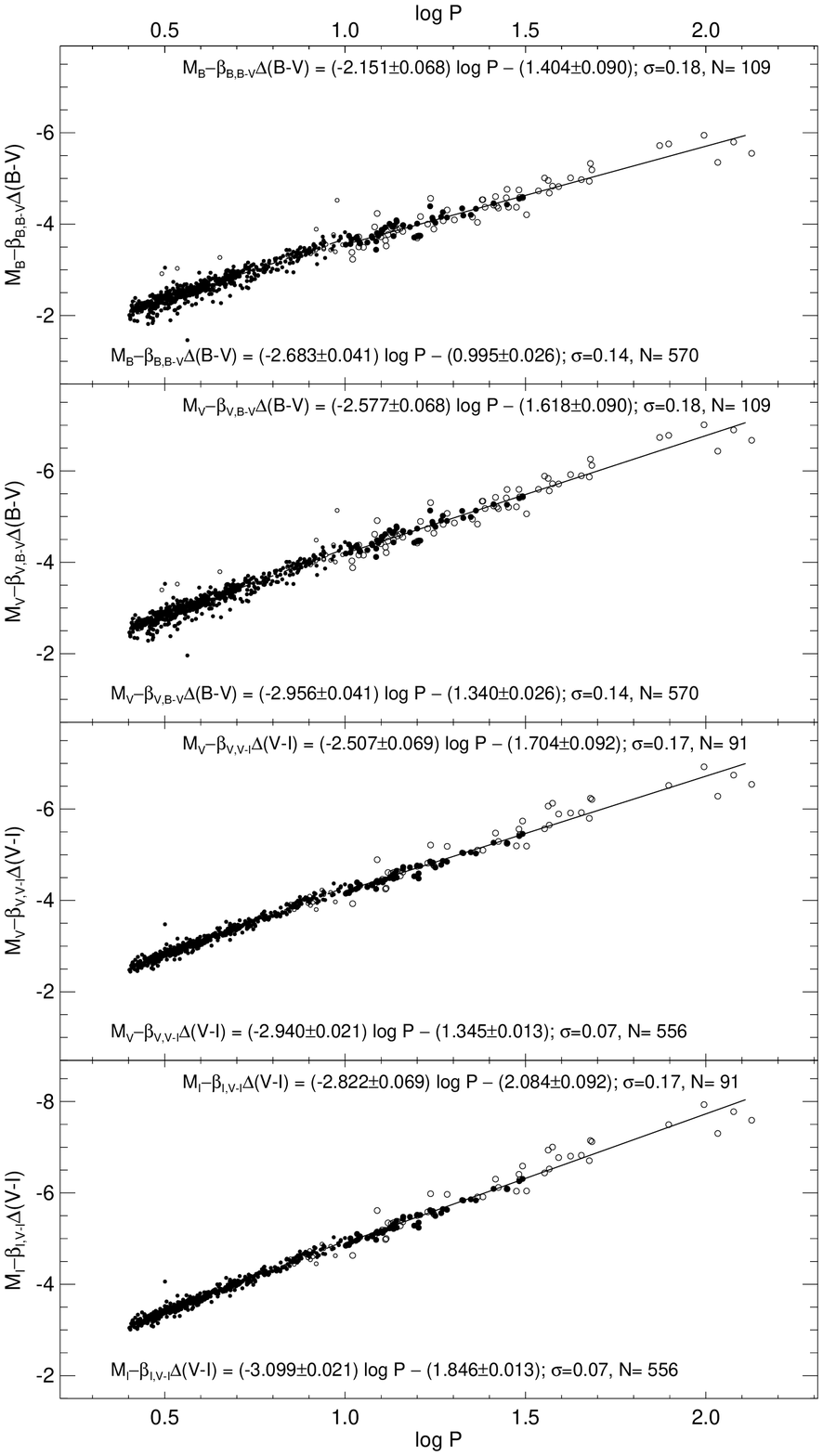}}
\caption{The P-L relations of LMC after correcting $M_{\lambda}({\rm
    obs})$ for the color term $\beta_{\lambda}\Delta(B\!-\!V)$ and
  $\beta_{\lambda}\Delta(V\!-\!I)$, respectively; for definitions see
  \ref{sec:InstabilityStrip:CPL}. The small scatter may be noted. The
  least-squares P-L relations for $\log P{\protect\grole} 1.0$ are
  given in the lower and upper part, respectively, of each panel.}
\label{fig:PLC}
\end{figure}

     The ridge line-corrected LMC magnitudes $M_{B,V,I}^{\rm R}$ from
Eq.~(\ref{eq:PLC:theo}) and Table~\ref{tab:PLC:coeff} are plotted
versus $\log P$ in Fig.~\ref{fig:PLC}. The $M_{V}^{\rm R}$ are plotted
twice, once based on $(B\!-\!V)^0$ and a second time on
$(V\!-\!I)^0$. It turns out that using the $(V\!-\!I)^0$ colors as
second parameter leads to about half the scatter compared to using the
$(B\!-\!V)^0$ colors. This can be understood because the simple P-L
relation has considerably less scatter in $M_{I}$ than in $M_{B,V}$
(see Fig.~\ref{fig:PL:final}), and also the scatter of the P-C
relation is somewhat larger in $(B\!-\!V)^0$ than in $(V\!-\!I)^0$
(cf. Figs.~\ref{fig:PC:BV}a and \ref{fig:PC:VI}b). -- Averaging the
different $M^{\rm R}$ values does not reduce the scatter indicating
that the errors are correlated.

     The equation for the ridge line of the
$\log P - M_{B,V,I}^{\rm R}$ relations (shown in Fig.~\ref{fig:PLC})
are closely the same -- as must be required -- as the P-L relations in
Eqs.~(\ref{eq:PL:B:lt1}$-$\ref{eq:PL:I:ge1}). The only significant
difference is that the scatter in $M^{\rm R}$ is reduced by a factor
of $\sim\!2$. This is of course an important practical advantage for
the determination of Cepheid distances, provided the {\em
  appropriate\/} P-L-C relation is known.

     Shown in Table~\ref{tab:PLC:coeff} are also the coefficients of
the {\em Galactic\/} P-L-C relation. The coefficients have here larger
errors because only 69 Galactic Cepheids are available with absolute
magnitudes either from cluster membership or from purely physical BBW
distances (see Sect.~\ref{sec:PL:galaxy:revised}). The P-L-C relation
of SMC, based on data by \citet{Udalski:etal:99c} has somewhat
provisional character; a more thorough discussion is deferred to
Paper~III. 

     It can be seen in Table~\ref{tab:PLC:coeff} that the coefficient
$\alpha$ of the period term of the P-L-C relation varies slightly, but
significantly between the Galaxy, LMC, and SMC. In contrast, the
variation of the coefficient $\beta$ of the color term is very
important; it is always (much) larger in LMC than in the Galaxy and is
generally larger in LMC than in SMC except
$\beta_{V,B\!-\!V}$ which is smaller in SMC than in LMC.

\subsection{An application of the P-L-C relation}
\label{sec:PLC:application}
\begin{figure}[t]
\centering
\resizebox{0.9\hsize}{!}{\includegraphics{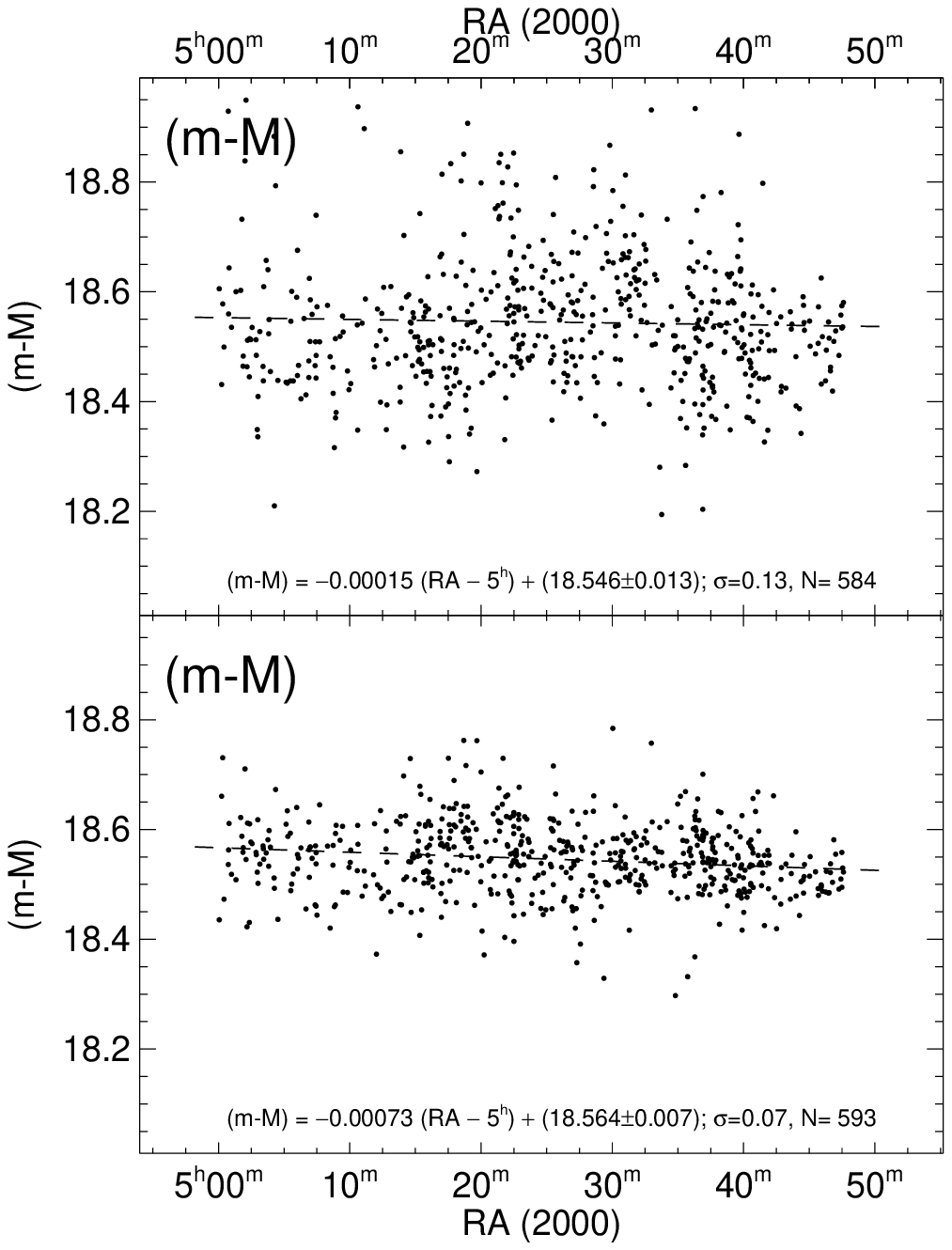}}
\caption{The distance moduli $(m_{V}-M_{V})$ of the individual
  Cepheids along the LMC bar are plotted in function of Right
  Ascension. Upper panel: $M_{V}$ calculated from
  Eqs.~(\ref{eq:PLC:BVlt1}) and (\ref{eq:PLC:BVge1}),
  respectively. Lower panel: $M_{V}$ calculated from
  Eqs~(\ref{eq:PLC:VIlt1}) and (\ref{eq:PLC:VIge1}). The data
  involving $(V\!-\!I)$ show considerably less scatter than those 
  involving $(B\!-\!V)$.}
\label{fig:LMCorientation}
\end{figure}

The present LMC Cepheids being distributed along the bar of the galaxy
offer the possibility to determine the space orientation of the
bar. The P-L-C relation with its small scatter is ideally suited for
this purpose. The moduli $(m_{V}-M_{V})$ are calculated twice from the
only two {\em independent\/} P-L-C relations (see
Table~\ref{tab:PLC:coeff}) 
\begin{eqnarray}
 \label{eq:PLC:BVlt1}
  \log P < 1.0\!: && \!M_{V}=-3.417\log P+1.686(B\!-\!V)-1.922 \\
 \label{eq:PLC:BVge1}
  \log P > 1.0\!: && \!M_{V}=-3.407\log P+1.948(B\!-\!V)-2.034 
\end{eqnarray}
and
\begin{eqnarray}
 \label{eq:PLC:VIlt1}
  \log P < 1.0\!: && \!M_{V}=-3.328\log P+2.434(V\!-\!I)-2.565 \\
 \label{eq:PLC:VIge1}
  \log P > 1.0\!: && \!M_{V}=-3.300\log P+2.517(V\!-\!I)-2.660 
\end{eqnarray}
(Note that the P-L-C relations involving $M_{B},(B\!-\!V)$ and
$M_{I},(V\!-\!I)$ can be derived from the above equations and do not
contain any additional information). The resulting distances are
plotted against Right Ascension in Fig.~\ref{fig:LMCorientation}.
As can be seen, the bar is
amazingly well aligned with the plane of the sky; the eastern end of
the bar is only $0\fm03$ (1.5\%) closer than its western
counterpart \citep[cf.][]{Welch:etal:87,Caldwell:Laney:91}.

   A word of warning is in place: the P-L-C relation can only be used
if it can be proven that the Cepheids under consideration follow the
same P-L {\em and\/} P-C relations. Otherwise any {\em intrinsic\/}
color difference is multiplied by the constant-period slope $\beta$
and erroneously forced upon the magnitudes with detrimental effects on
any derived distances.
It would be particularly objectionable to apply for instance the LMC
P-L-C relation to Galactic Cepheids which have different P-L$_{BVI}$
relations and (necessarily) also different P-C$_{B\!-\!V,B\!-\!I}$
relations.

\section{The instability strip in the luminosity 
         \boldmath{$-$ $T_{\rm e}$} plane}
\label{sec:InstabilityStrip:MTeff}
\begin{figure}[t]
\centering
\resizebox{1.0\hsize}{!}{\includegraphics{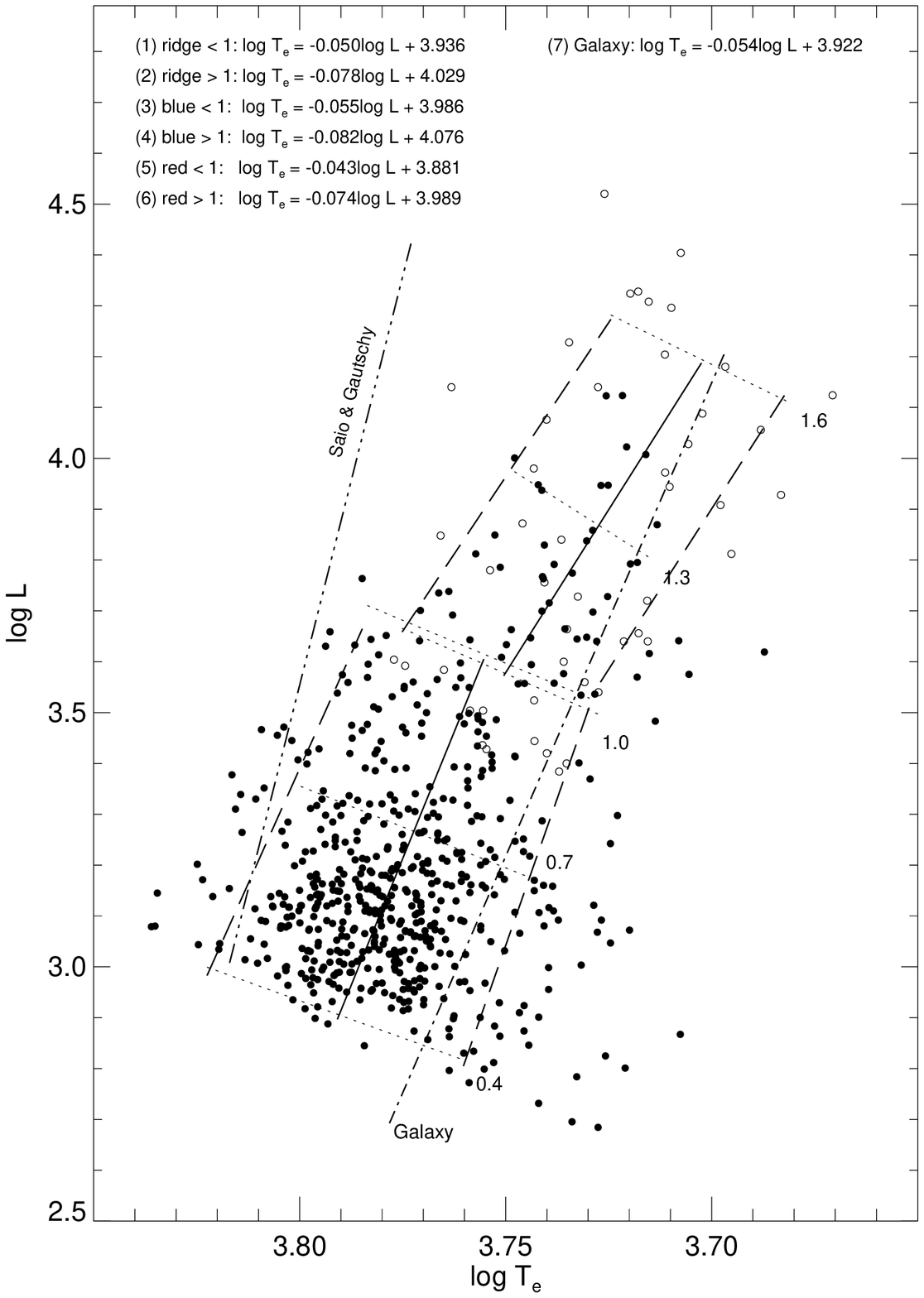}}
\caption{The instability strip in the $\log L - \log T_{\rm e}$
  plane. The individual Cepheids are shown. Symbols as in
  Fig.~\ref{fig:PC:BV}a. The ridge line (full line) and boundaries
  (dashed lines) are transformations of the corresponding lines in
  Fig.~\ref{fig:InstabilityStrip}. The dashed-dotted line is the mean
  relation of the Galaxy. The theoretical line of
  \citet{Saio:Gautschy:98} is dashed-double-dotted.}
\label{fig:LMC:Teff}
\end{figure}

The instability strip in the plane of the CMD in
Fig.~\ref{fig:InstabilityStrip} is converted into the $\log L-\log
T_{\rm e}$ plane by means of the atmospheric models of Cepheids by
\citeauthor*{Sandage:etal:99}. The colors $(B\!-\!V)^0$ and
$(V\!-\!I)^0$ of the ridge line and the boundary lines of the strip as
well as of the individual Cepheids yield by interpolation in their
Table~6 for $\mbox{[Fe/H]}=-0.3$ adopted for LMC, a turbulence of
$1.7\kms$, and variable $\log g$ two temperatures which are averaged.
The dependence of $\log g$ on $\log P$ was determined from their
Table~3 to be
\begin{equation}\label{eq:logg:logP}
     \log g = -1.09 \log P + 2.64,
\end{equation}
which holds well for the ridge line as well as the blue and red
boundary, and which is insensitive to metallicity (cf. their
Table~2). 
For the luminosity $L$ (in solar units) it is assumed that $M_{\rm
  bol}\approx M_{V}$ and $M_{{\rm bol}\sun}=4\fm75$. The resulting
diagram is shown in Fig.~\ref{fig:LMC:Teff}.
The mean strip position as defined by Galactic Cepheids in
\citeauthor*{Tammann:etal:03} is also shown.

     It is obvious in Fig.~\ref{fig:LMC:Teff} that the blueness of LMC
Cepheids is due, in addition to the technical effect of blanketing, to
a real temperature difference. LMC Cepheids with a luminosity
of $\log L = 3.5$ are hotter by about $350\,$K
than their Galactic counterparts, decreasing to $\sim\!80\,$K at $\log
L = 4.0$. 

    Although the temperature shift in the HR diagram of
Fig.~\ref{fig:LMC:Teff} is small at $\Delta \log T_{\rm e}=0.02$
compared with the total width of the strip of 
$\Delta \log T_{\rm e}=0.08$, the reason for the temperature
difference remains obscure. While it is true that the 
temperature of the blue edge of the strip depends on the metal
abundance, $Z$, there is also a strong dependence on the helium
abundance, $Y$.

     The dependence of the blue edge temperature for fundamental
mode Cepheids on $Y$ and $Z$ has been known since the earliest
theoretical work on the low mass RR Lyrae variables by
\citet{Iben:Huchra:71}, \citet{Christy:66},
\citet{Tuggle:Iben:72}, \citet{vanAlbada:Baker:71}, and others into
modern times. The earliest calculations of the $Z,Y$ dependence of the
blue edge of the $L/T_{\rm e}$ strip for stars in the mass and
luminosity range of the classical Cepheids was that of
\citet{Iben:Tuggle:75}. Their results remain nearly definitive today
as shown by the following comparison. 

     \citeauthor*{Sandage:etal:99} compared the position of
the Iben and Tuggle blue edge $L(T_{\rm e},Z,Y)$ locus with the modern 
equation~(6) of \citet{Chiosi:etal:92} for the $Z,Y$ blue edge
dependencies and found identical temperatures to within about
$\Delta \log T_{\rm e} = 0.002$ when reduced to the same values of $Z$
and $Y$. Similar comparisons were made by
\citeauthor*{Sandage:etal:99} for the blue edge positions of
Iben/Tuggle and \citet{Saio:Gautschy:98}, and again were found 
to be nearly identical when reduced to same $Z$ and $Y$
compositions. These comparisons of \citeauthor{Saio:Gautschy:98} with
\citet{Iben:Tuggle:75} and \citet{Chiosi:etal:92} were particularly
important because \citeauthor{Saio:Gautschy:98} used the new OPAL
opacities by \citet{Rogers:Iglesias:92} as extended to lower
temperatures by \citet{Alexander:Ferguson:94}
whereas \citeauthor{Iben:Tuggle:75} and \citeauthor{Chiosi:etal:92}
used the earlier Los Almos Cox-Stewart opacities. Hence, 
differences in the adopted opacity tables have a negligible
effect on the position of the blue edges of the instability
strip.

     Using Eq.~(6) of \citeauthor{Chiosi:etal:92} for the blue edge as 
function of $Z$ and $Y$, and reading the mean mass/luminosity ratio
at intermediate $Z$ and $Y$ values (in any case, the $T_{\rm e}/L$
blue edge position has only a small mass dependence), the temperature
dependence on $Z$ and $Y$ is found to be

\begin{equation}\label{eq:logTe:YZ}
         \log T_{\rm e} = 3.771 + 0.12Y - 0.66Z,      
\end{equation}
at $\log L = 3.5 \; (M_{\rm bol} = -4.0)$. Hence if $Y$ increases as
$Z$ increases, due to the manufacture of helium with higher
metallicity production, as all models of stellar nucleosynthesis
require, the position of the blue edge would remain fixed
(independent of $Y$ and $Z$) if $0.12\Delta Y$ were to equal 
$0.66\Delta Z$ (i.e.\ if $\Delta Y/\Delta Z = 5.5$). There would then
be no temperature shift in Fig.~\ref{fig:LMC:Teff} due to different
$Z$ metallicities. 

     However, the canonical change of $Y$ with $Z$ has the ratio of
$\Delta Y/\Delta Z = 2.2$ (VandenBerg in \citealt{Sandage:etal:03},
hereafter \citeauthor*{Sandage:etal:03}). The Padua astronomers, led by 
Chiosi, (see \citealt{Fagotto:etal:94} for the last of four papers on
the evolutionary tracks for different compositions) use values
close to $2.5$. The Geneva astronomers, led by Maeder, (the last
paper of their four paper series is by \citet{Schaerer:etal:93}) used
$\Delta Y/\Delta Z = 3$. Hence, although there is compensation in the
$T_{e}$ value as $Y$ increases with $Z$, it is not complete if
Eq.~(\ref{eq:logTe:YZ}) is correct. If we use $\Delta Y/\Delta Z = 2.2$,
then for a change by a factor of $1$0 in $Z$ from $0.0188$ (which is
solar) to say $0.0019$, with $Y$ changing  from $0.276$ (solar) to
$0.239$  according to the prescription that $Y = 0.235 + 2.2Z$ given
by \citeauthor*{Sandage:etal:03}, then $\Delta \log T_{\rm e} = 0.007$,
with the low $Z$ case being hotter. This is much smaller than the
requirement from Fig.~\ref{fig:LMC:Teff} for $\Delta T_{\rm e} = 
0.03$ (although in the same direction), but even here the LMC
metallicity is only smaller than the Sun by a factor of 4 not 10. 
Hence, the compensation of $Y$ and $Z$ in Eq.~(\ref{eq:logTe:YZ}) is
the reason that the temperature of the blue edges for the variety of
compositions used by the Padua astronomers and by Iben and Tuggle is
virtually unchanged for variable $Z$. The hotter temperature for LMC
Cepheids relative to those in the Galaxy is therefore not explained by
Eq.~(\ref{eq:logTe:YZ}). 

     Even if $Z$ changes by a factor of 10 in the above example
{\em with $Y$ kept constant for some reason}, the change in $\log
T_{\rm e}$ would only be $\Delta T_{\rm e} = 0.01$, again smaller than
a factor of two or three from what is required by
Fig.~\ref{fig:LMC:Teff}. 

     Of course, a reverse sign variation of a decrease in $Y$ with
increasing $Z$ would solve the dilemma, producing any hotter value
for $T_{\rm e}$ for small $Z$ that the observations require
(Fig.~\ref{fig:LMC:Teff}), but this is highly counter intuitive and
entirely ad hoc. We are left with the dilemma that we do not have a
reasonable explanation of the shift in $T_{\rm e}$ with lower
metallicity in Fig.~\ref{fig:LMC:Teff} based on any 
extant theoretical model. We cannot conclude that the variation
is, in fact due, to metallicity, or at least not to metallicity
alone.

\begin{table}
\begin{center}
\caption{Predicted and observed slopes of the $P$-$L_{V}$ relation
  (see text).}
\label{tab:Te:coeff}
\small
\begin{tabular}{lccc}
\hline
\hline
\noalign{\smallskip}
   {Cepheids in} &
   $c$ &
   $d M_{\rm bol}/d log P$ &
   $d M_{V}/d log P$ \\
   &
   &
   predicted &
   observed \\
\noalign{\smallskip}
\hline
\noalign{\smallskip}
Galaxy          & $-0.054$ & $-3.02$ & $-3.09\pm0.09^{1)}$ \\
LMC, $\log P<1$ & $-0.050$ & $-3.09$ & $-2.96\pm0.06^{2)}$ \\
LMC, $\log P>1$ & $-0.078$ & $-2.76$ & $-2.57\pm0.10^{2)}$ \\
\noalign{\smallskip}
\hline
\multicolumn{3}{l}{$^{1)}$\,from Eq.~(\ref{eq:gal:PL:V}). 
        $^{2)}$\,from Eqs.~(\ref{eq:PL:V:lt1}\,\&\,\ref{eq:PL:V:ge1}).}
\end{tabular}
\end{center}
\end{table}
     It remains to be shown that any change of the {\em slope\/} of
the ridge line of the $\log T_{\rm e}-\log L$ relation translates into
a change of slope of the P-L relation. In
\citeauthor*{Tammann:etal:03} we have shown that the slope of the P-L
relation is given by 
\begin{equation}
\label{eq:dMbol}
d M_{V}/d \log P \approx d M_{\rm bol}/d \log P= -2.5(0.84 -0.68a -3.48c)^{-1},
\end{equation}
where $a$ is the coefficient of the log Mass$-\log L$ dependence, and
$c$ is the slope of the $\log T_{\rm e}-\log L$ relation.
This equation is based on a simple pulsation equation
by \citet{vanAlbada:Baker:73} applicable to RR\,Lyrae stars and, as it
turns out, is also an excellent equation for Cepheids in its
coefficients from the $\log L$, $\log {\frak M}$, and $\log T_{\rm e}$
dependences (see \citeauthor*{Sandage:etal:99} for an extended
discussion). We have used $a=0.30$,  
typical for Cepheids  (cf.\ \citeauthor*{Sandage:etal:99},
Tables~1-5). 
The coefficient $a$ enters only weakly into Eq.~(\ref{eq:dMbol}).
The deduced values of $c$ are shown in Fig.~\ref{fig:LMC:Teff} for the
ridge lines of the short- and long-period LMC Cepheids and of Galactic
Cepheids.  The ridge line values are tabulated in
Table~\ref{tab:Te:coeff} together with the predicted ridge line slopes
from Eq.~(\ref{eq:dMbol}) and the corresponding observed values.
The agreement between predicted and observed slopes is almost within
the errors of the observed slopes. Allowing for some additional errors
in the predicted slopes makes the agreement good. The most significant
result is that the steep slope of the long-period LMC Cepheids in the
$\log T_{\rm e}- \log L$ plane requires a flat slope of their P-L
relation. This is observed, and is the main point of this
Paper. Hence, the different {\em slope\/} of the ridge line 
of the $\log L-\log T_{\rm e}$ relation between LMC and the Galaxy
explains the different slopes to the P-L relation in
Fig.~\ref{fig:PL:final} (lower right panel).

\section{Conclusions and Summary}
\label{sec:conclusion}
The consequences of the differences in the slopes of the P-L
relations for the Galaxy, LMC, and SMC weakens the hope of using
Cepheids to obtain precision galaxy distances. 
Until we understand the reasons for the differences in
the P-L relations and the shifts in the period-color relations,
(after applying blanketing corrections for metallicity
differences), we are presently at a loss to choose which of the
several P-L relations to use (Galaxy, LMC, and SMC) for other galaxies.

     Although we can still hope that the differences may yet be
caused only by variations in metallicity, which can be measured,
this can only be decided by future research such as survey
programs to determine the properties of Cepheids in galaxies such
as M\,33 and M\,101 where metallicity gradients exist across
the image. But until we can prove or disprove that metallicity
difference is the key parameter, we must provisionally {\em assume\/}
that this is the case, and use either the Galaxy P-L relations in
Sect.~\ref{sec:PL:galaxy:revised}
(Eqs.~\ref{eq:gal:PL:B}$-$\ref{eq:gal:PL:I}), or the LMC P-L relations
in Sect.~\ref{sec:PL:LMC} (Eqs.~\ref{eq:PL:B}$-$\ref{eq:PL:I:ge1}), or
those in the SMC from the forthcoming Paper~III, in deriving galaxy
distances from Cepheids. This has now in fact been done elsewhere for
M\,83 \citep{Thim:etal:03}. 

     There are eleven principal research points in this paper.

\noindent
 (1) The period-color (P-C) relations in $(B\!-\!V)^{0}$ and
     $(V\!-\!I)^{0}$ have significantly different slopes for periods
     smaller and larger than 10 days (Figs.~\ref{fig:PC:BV}a and
     \ref{fig:PC:VI}b). The slope of the longer period Cepheids is
     steeper than those with $P < 10$ days at the significance level
     of $3.4\sigma$ in $(B\!-\!V)^{0}$ and $4.1\sigma$ in
     $(V\!-\!I)^{0}$.  

\noindent
 (2) The LMC Cepheids are bluer than Galactic Cepheids by $0\fm07$
     in $(B\!-\!V)^{0}$ at $\log P=0.4$, increasing to $0\fm12$ at
     $\log P=1$. The corresponding differences in $(V\!-\!I)^{0}$ 
     are $0\fm03$ and $0\fm10$
     (Eqs.~\ref{eq:PC:BV:lt1}$-$\ref{eq:PC:VI:ge1} for LMC compared
     with Eqs.~3 and 5 of \citeauthor*{Tammann:etal:03}). Blanketing
     differences due to the lower metallicity of LMC accounts for only
     half of the observed color differences. A temperature difference
     (Figs.~\ref{fig:logPTe} and \ref{fig:LMC:Teff}) accounts for the
     remainder, with the LMC Cepheids being hotter by between $350\;$K  
     and $80\;$K, depending on the period. Theoretical models of the
     position of the fundamental blue edge as functions of metallicity
     and helium abundance cannot account for the temperature
     difference (Eq.~\ref{eq:logTe:YZ}) if $Z$ and $Y$ increase in
     lock step. An explanation is possible if $Y$ and $Z$ are
     anticorrelated (Sect.~\ref{sec:InstabilityStrip:MTeff}), 
     but this is highly counterintuitive based on the models of
     nucleosynthesis.

\noindent
 (3) The P-L relations for LMC in $B$, $V$, and $I$ are all 
     non-linear, each with a break at $\log P = 1$ 
     (Eqs.~\ref{eq:PL:B:lt1}$-$\ref{eq:PL:I:ge1}). The slope
     differences are significant at the at least $3\sigma$
     level. The reality of the break in the P-L relations at 10 days 
     is supported by the breaks in the P-C relations (item 1).
     Necessarily, breaks in the P-C relations require breaks in the
     P-L relations.

\noindent
 (4) The slopes of the LMC P-L relations in the $B$, $V$, and
     $I$ pass bands are different  from the slopes
     of the corresponding relations in the Galaxy as updated in
     \ref{sec:PL:galaxy:revised}. 
     For Cepheids with $\log P<1$ the P-L slope differences between
     LMC and the Galaxy vanish in $B$, but amount to $2\sigma$ and
     $3\sigma$ in $V$ and $I$. The significance of the slope
     differences is even larger for long-period Cepheids with $\log
     P>1$, i.e.\ it increases from $3\sigma$ in $B$ to $5\sigma$ in
     $V$ and $I$. Since most available galaxy distances from Cepheids
     are based on long-period Cepheids ($\log P_{\rm median}\!\sim\!1.4$)
     the choice of which P-L relation is used has a non-negligible
     effect on the derived absorption-corrected moduli by about
     $0\fm2$ (cf. Eq.~40 in \citeauthor*{Tammann:etal:03}) or
     $\sim\!10\%$ in distance. 

\noindent
 (5) The absolute magnitudes of LMC Cepheids are {\em brighter\/} than
     those in the Galaxy by $0\fm42$ to $0\fm32$ in $B$, $V$, and $I$
     at $\log P=0.4$, becoming {\em fainter\/} by $0\fm06$ in $V$ and
     by $0\fm14$ in $I$ at $\log P=1.5$
     (Eqs.~\ref{eq:PL:B:lt1}$-$\ref{eq:PL:I:ge1} compared with
     \ref{eq:gal:PL:B}$-$\ref{eq:gal:PL:I}). 
     All attempts to determine a Cepheid distance of LMC by means of a
     Galactic P-L relation
     \citep[e.g.][]{Fouque:etal:03,Groenewegen:Salaris:03,Storm:etal:04}
     are therefore frustrated. The resulting values of $(m-M)^{0}_{\rm
     LMC}$ would vary between $\sim\!18.16$ and $\sim\!18.63$ (still
     somewhat dependent on pass band) depending on whether Cepheids
     with $\log P=0.4$ or $\log P=1.5$ are compared. (The {\em
     adopted\/} LMC distance of $18.54$ in the present paper rests on
     a compilation of distance determinations which are {\em
     independent\/} of the P-L relation of Cepheids; see
     \citeauthor*{Tammann:etal:03}, Table~6).
     -- There are now, however, good prospects to re-establish
     Cepheids as important distance indicators of LMC by means of
     their individual BBW distances, which depend little on
     metallicity. Work towards this aim is in progress (W.~Gieren,
     private communication).

\noindent
 (6) The reality of the break in Cepheid properties at 10 days
     is shown in various correlations. 
     (a) The Fourier components $R_{21}$ and $\Phi_{21}$, introduced
     by \citet{Simon:Lee:81}, show discontinuities at 10 days
     (Fig.~\ref{fig:break:Rphi}).  
     (b) A similar discontinuity exists in the period-amplitude
     relation, both for LMC and the Galaxy
     (Fig.~\ref{fig:break:Vamp}). 
     (c) The slope of the ridge-line of the color-magnitude diagram
     for the instability strip (Fig.~\ref{fig:InstabilityStrip})
     changes at 10 days, as in the theoretical HR ($L/T_{\rm e}$)
     diagram (Fig.~\ref{fig:LMC:Teff}). 
     (d) Changes at 10 days also occur in other correlations such as
     the slopes of the lines of constant period (Fig.~\ref{fig:alpha}
     and Sect.~\ref{sec:InstabilityStrip:CPL}), the character of the
     color-amplitude correlation in the period range of 10 to 20 days
     (Figs.~\ref{fig:Bamp} and \ref{fig:BampBV}), and in the $R_{21}$
     vs. $\log P$ relation with the data binned by absolute magnitude
     (Fig.~\ref{fig:logPRphi21}). 

\noindent
 (7) The loci of constant period in the CMD
     (Fig.~\ref{fig:InstabilityStrip}) slope toward fainter magnitudes
     as the color changes across the strip from blue to red. Hence, at
     a given period there is a variation in absolute magnitude causing
     the systematic scatter in the P-L relations in $B$, $V$, and
     $I$. The mapping of this effect is particularly strong here using
     the LMC data because the data base over the entire period range
     from 2 to 40 days is superbly large from the OGLE project. The
     slopes of the lines of constant period in the $M_{V}$ P-L relation
     average $1.8$ in $(B\!-\!V)^{0}$ (but vary with period) and
     $2.4$ in $(V\!-\!I)^{0}$ (Fig.~\ref{fig:CPL} and 
     Eqs.~\ref{eq:CPL:BV:lt1}$-$\ref{eq:CPL:VI}). The slope of the
     constant-period lines is of course steeper than this in the
     $M_{B}-(B\!-\!V)$ plane and more shallow in the $M_{I}-(V\!-\!I)$
     plane (Eqs.~\ref{eq:alpha:B:BV} and \ref{eq:alpha:I:VI}), showing
     why the scatter in the P-L relations becomes progressive smaller
     from $B$, through $V$, to $I$, as was originally shown
     \citep{Sandage:58,Sandage:72} as due to the finite color width of
     the instability strip.  

\noindent
 (8) The large data base permits a high-weight mapping of
     several Cepheid characteristics with position in the strip. The 
     data binned in narrow period intervals from 3 to 40 days, (i.e.\
     along lines of constant period in the CMD), show
     (Fig.~\ref{fig:Bamp}) that, except near 10 days, the amplitude is
     largest at the blue edge of the strip, just as for the RR Lyrae
     stars in the low mass instability strip peculiar to them. The
     analogy with the RR Lyrae stars is stronger when the data are
     binned in narrow intervals of absolute magnitude
     (Fig.~\ref{fig:BampBV}) similar to the case of RR Lyrae 
     variables in globular clusters. Fig.~\ref{fig:BampBV} shows that
     the correlation of light-curve amplitude with color (at the
     relevant absolute magnitude from $-2.75$ to brighter than
     $-4.75$) is in the sense that largest amplitude occurs at the
     bluest color for Cepheids fainter than $M_{V}= -4.25$ (periods
     smaller than 7 days), the trend reverses for periods of 10 to 15
     days ($-4.25 < M_{V} < -4.75$) and reverts to the original sense
     for $P > 20$ days as was earlier found by
     \citet{Sandage:Tammann:71} from a much smaller data sample.

\noindent
 (9) The Fourier coefficient $R_{12}$ varies systematically
     with amplitude and in color across the strip as shown in 
     Fig.~\ref{fig:R21BampBV}, binned by period and in
     Fig.~\ref{fig:BVRphi21}, in analogy with the RR Lyrae 
     correlations, binned by absolute magnitude. Large
     $R_{21}$ means that the second term of the Fourier series,
     $\sin 2x$, is large, showing that the light curve is highly
     peaked (approaching a saw tooth for $R_{21} = 0.5$;
     Sect.~\ref{sec:InstabilityStrip:Rphi21}). Small $R_{21}$
     stands for nearly symmetrical light curves, which, by 
     Fig.~\ref{fig:R21BampBV}, occur at small amplitude in all period 
     ranges. 

          Hence, Figs.~\ref{fig:R21BampBV} and \ref{fig:BVRphi21} show
     that the light curve shapes are highly peaked toward the blue
     edge of the instability strip, becoming nearly sinusoidal (and of
     small amplitude) toward the red. {\em $R_{12}$ varies
     systematically across the strip}. This is an important result
     concerning the physics of pulsation. Large amplitude with its
     near saw-tooth shape (small interval in phase between minimum and
     maximum light) means that the pulsation is more strongly driven
     into the non-linear regime at the blue edge than toward the red
     edge. Hence, the prediction is that the PV work diagram in the
     pulsation thermodynamics must have a larger non-dissipative area
     for Cepheids near the blue edge than near the red edge
     \citep[cf.][]{Christy:66}.  

\noindent
 (10) By necessity, from these correlations of $R_{21}$ with color
      (Figs.~\ref{fig:R21BampBV} and \ref{fig:BVRphi21}), and from the
      correlations of color and period, (Fig.~\ref{fig:logPBVI}) there
      must be a correlation of $R_{21}$ with period at constant
      absolute magnitude. This is confirmed in
      Fig.~\ref{fig:logPRphi21}. The reason is obvious (by inspecting
      the CMD of Fig.~\ref{fig:InstabilityStrip}), once it is known
      that $R_{21}$ varies systematically with color across the
      strip. 

\noindent
 (11) Application of the PLC relation to the entirety of the OGLE LMC
      data gives the {\em relative\/} distance to each individual LMC
      Cepheid. This permits determination of the orientation of the
      bar in the line of sight. The eastern edge of the bar is only
      $0\fm03$ closer than the western edge
      (Fig.~\ref{fig:LMCorientation}), confirming earlier conclusions
      by \citet{Welch:etal:87} and \citet{Caldwell:Laney:91}.

     Finally we reiterate that the unexpected slope difference between
the P-L relations particularly of long-period Cepheids in the Galaxy
and LMC can be understood as the consequence of the different
gradients in the luminosity-temperature diagram, yet the reason {\em
why\/} these gradients are different remains unknown.

\begin{acknowledgements}
G.\,A.\,T. and B.\,R. thank the Swiss National Science Foundation for
valuable support. The authors thank Dres. D.~Bersier, R.~Buser,
A.~Gautschy, S.~Kanbur, C.~Ngeow, M.~Samland, N.~Simon, A.~Udalski,
and N.~Walborn for helpful discussions and informations. Much of this
paper was written while G.A.T. was a visitor at the Space Telescope
Science Institute in Baltimore. He thanks the Director and the Staff
for their hospitality and for a most stimulating period of time.
\end{acknowledgements}



\clearpage
\onecolumn
\setcounter{table}{0}
\scriptsize
\begin{longtable}{lcccccccccccc}
\caption{LMC Cepheids with photoelectric photometry\label{tab:LMCother}.}\\
\hline
\hline
\noalign{\smallskip}
 \multicolumn{1}{c}{Cepheid} &
 \multicolumn{1}{c}{$\log P$} &
 \multicolumn{1}{c}{$V$} &
 \multicolumn{1}{c}{$(B\!-\!V)$}&
 \multicolumn{1}{c}{$(V\!-\!I)$}&
 \multicolumn{1}{c}{$E(B\!-\!V)$} &
 \multicolumn{1}{c}{Source} &
 \multicolumn{1}{c}{$R_{V}$} &
 \multicolumn{1}{c}{$V^{0}$} &
 \multicolumn{1}{c}{$(B\!-\!V)^{0}$}&
 \multicolumn{1}{c}{$(V\!-\!I)^{0}$} &
 \multicolumn{1}{c}{${\cal A}_{V}$} &
 \multicolumn{1}{c}{${\cal A}_{B}$} \\
 \multicolumn{1}{c}{(1)}  & \multicolumn{1}{c}{(2)}  &
 \multicolumn{1}{c}{(3)}  & \multicolumn{1}{c}{(4)}  &
 \multicolumn{1}{c}{(5)}  & \multicolumn{1}{c}{(6)}  &
 \multicolumn{1}{c}{(7)}  & \multicolumn{1}{c}{(8)}  &
 \multicolumn{1}{c}{(9)}  & \multicolumn{1}{c}{(10)} &
 \multicolumn{1}{c}{(11)} & \multicolumn{1}{c}{(12)} &
  \multicolumn{1}{c}{(13)} \\
\noalign{\smallskip}
\hline
\noalign{\smallskip}
\endfirsthead
\caption{(Continued)}\\
\hline
\hline
\noalign{\smallskip}
 \multicolumn{1}{c}{Cepheid} &
 \multicolumn{1}{c}{$\log P$} &
 \multicolumn{1}{c}{$V$} &
 \multicolumn{1}{c}{$(B\!-\!V)$}&
 \multicolumn{1}{c}{$(V\!-\!I)$}&
 \multicolumn{1}{c}{$E(B\!-\!V)$} &
 \multicolumn{1}{c}{Source} &
 \multicolumn{1}{c}{$R_{V}$} &
 \multicolumn{1}{c}{$V^{0}$} &
 \multicolumn{1}{c}{$(B\!-\!V)^{0}$}&
 \multicolumn{1}{c}{$(V\!-\!I)^{0}$} &
 \multicolumn{1}{c}{${\cal A}_{V}$} &
 \multicolumn{1}{c}{${\cal A}_{B}$} \\
 \multicolumn{1}{c}{(1)}  & \multicolumn{1}{c}{(2)}  &
 \multicolumn{1}{c}{(3)}  & \multicolumn{1}{c}{(4)}  &
 \multicolumn{1}{c}{(5)}  & \multicolumn{1}{c}{(6)}  &
 \multicolumn{1}{c}{(7)}  & \multicolumn{1}{c}{(8)}  &
 \multicolumn{1}{c}{(9)}  & \multicolumn{1}{c}{(10)} &
 \multicolumn{1}{c}{(11)} & \multicolumn{1}{c}{(12)} &
 \multicolumn{1}{c}{(13)} \\
\noalign{\smallskip}
\hline
\noalign{\smallskip}
\endhead
\noalign{\smallskip}
\hline
\endfoot
\noalign{\smallskip}
\hline
\multicolumn{13}{c}{Sources:
         (1) \citealt{Martin:etal:79},
         (2) \citealt{Madore:85},
         (3) \citealt{Laney:Stobie:94},
         (4) \citealt{Caldwell:Coulson:85},}\\
\multicolumn{13}{c}{(5) \citealt{Martin:80},
                    (6) \citealt{Caldwell:etal:86},
                    (7) \citealt{Sebo:etal:02}.}
\endlastfoot
ROB\,44    & 0.408 & 16.36 & 0.58    & \nodata & 0.10 &  1         & 3.13 & 16.05 & 0.48    & \nodata & 0.80    & 1.45    \\
HV\,5541   & 0.415 & 16.06 & 0.56    & \nodata & 0.06 &  3         & 3.12 & 15.87 & 0.50    & \nodata & 0.91    & \nodata \\
ROB\,24    & 0.429 & 16.22 & 0.51    & \nodata & 0.10 &  1         & 3.10 & 15.91 & 0.41    & \nodata & 1.20    & 1.57    \\
HV\,12225  & 0.478 & 16.22 & 0.66    & \nodata & 0.06 &  1,3       & 3.17 & 16.03 & 0.60    & \nodata & 0.50    & 0.63    \\
HV\,2353   & 0.492 & 15.37 & 0.53    & \nodata & 0.10 &  1         & 3.11 & 15.06 & 0.43    & \nodata & \nodata & 0.69    \\
ROB\,25    & 0.529 & 16.03 & 0.60    & \nodata & 0.10 &  1         & 3.14 & 15.72 & 0.50    & \nodata & 0.52    & 0.70    \\
HV\,12765  & 0.535 & 15.29 & 0.57    & \nodata & 0.10 &  1         & 3.13 & 14.98 & 0.47    & \nodata & \nodata & 0.57    \\
HV\,12747  & 0.556 & 15.76 & 0.56    & \nodata & 0.06 &  1,3       & 3.12 & 15.57 & 0.50    & \nodata & 0.37    & 0.70    \\
ROB\,29    & 0.569 & 15.80 & 0.58    & \nodata & 0.10 &  1         & 3.13 & 15.49 & 0.48    & \nodata & 0.54    & 0.57    \\
HV\,12720  & 0.635 & 15.74 & 0.63    & \nodata & 0.10 &  1         & 3.16 & 15.42 & 0.53    & \nodata & \nodata & 1.50    \\
HV\,12869  & 0.653 & 15.06 & 0.62    & \nodata & 0.10 &  1         & 3.15 & 14.74 & 0.52    & \nodata & \nodata & 0.80    \\
ROB\,22    & 0.669 & 15.55 & 0.70    & \nodata & 0.10 &  1         & 3.19 & 15.23 & 0.60    & \nodata & 0.82    & 1.42    \\
HV\,12231  & 0.677 & 15.43 & 0.64    & \nodata & 0.10 &  1         & 3.16 & 15.11 & 0.54    & \nodata & \nodata & 1.27    \\
HV\,6093   & 0.680 & 15.35 & 0.64    & \nodata & 0.06 &  1,3       & 3.16 & 15.16 & 0.58    & \nodata & \nodata & 1.16    \\
HV\,12077  & 0.703 & 15.30 & 0.61    & \nodata & 0.10 &  1         & 3.15 & 14.99 & 0.51    & \nodata & \nodata & 1.00    \\
HV\,12079  & 0.841 & 14.94 & 0.66    & \nodata & 0.10 &  1         & 3.17 & 14.62 & 0.56    & \nodata & \nodata & 1.14    \\
HV\,935    & 0.849 & 15.02 & 0.71    & 0.80    & 0.10 &  5         & 3.19 & 14.70 & 0.61    & 0.67    & 0.68    & \nodata \\
HV\,1000   & 0.859 & 15.04 & 0.76    & 0.76    & 0.10 &  5         & 3.21 & 14.72 & 0.66    & 0.63    & 0.78    & \nodata \\
HV\,12976  & 0.895 & 14.99 & 0.74    & \nodata & 0.10 &  1         & 3.20 & 14.67 & 0.64    & \nodata & \nodata & 1.76    \\
HV\,5730   & 0.897 & 15.12 & 0.85    & 0.85    & 0.10 &  5         & 3.25 & 14.79 & 0.75    & 0.72    & 0.65    & \nodata \\
HV\,6104   & 0.902 & 14.85 & 0.70    & 0.81    & 0.10 &  5         & 3.19 & 14.53 & 0.60    & 0.68    & 0.80    & \nodata \\
HV\,12581  & 0.904 & 15.15 & 0.83    & 0.85    & 0.10 &  5         & 3.24 & 14.83 & 0.73    & 0.72    & 0.60    & \nodata \\
HV\,12700  & 0.911 & 14.84 & 0.74    & 0.75    & 0.03 &  1,3,4     & 3.20 & 14.74 & 0.71    & 0.71    & \nodata & 0.90    \\
HV\,12823  & 0.919 & 14.57 & 0.55    & \nodata & 0.10 &  1         & 3.12 & 14.26 & 0.45    & \nodata & \nodata & 1.13    \\
HV\,2738   & 0.920 & 14.66 & 0.86    & 0.61    & 0.10 &  5         & 3.26 & 14.33 & 0.76    & 0.48    & 0.85    & \nodata \\
HV\,2854   & 0.936 & 14.64 & 0.72    & 0.78    & 0.05 &  1,3,4     & 3.19 & 14.48 & 0.67    & 0.72    & \nodata & 0.85    \\
HV\,2733   & 0.941 & 14.69 & 0.62    & 0.76    & 0.12 &  1,3,4     & 3.15 & 14.31 & 0.50    & 0.61    & \nodata & 0.66    \\
HV\,12452  & 0.941 & 14.79 & 0.71    & \nodata & 0.10 &  1         & 3.19 & 14.47 & 0.61    & \nodata & \nodata & 1.33    \\
HV\,12717  & 0.947 & 14.69 & 0.70    & \nodata & 0.10 &  1         & 3.19 & 14.37 & 0.60    & \nodata & \nodata & 1.25    \\
HV\,12816  & 0.960 & 14.53 & 0.56    & 0.72    & 0.08 &  1,3,4,6   & 3.12 & 14.28 & 0.48    & 0.62    & 0.69    & 1.08    \\
HV\,971    & 0.968 & 14.43 & 0.68    & \nodata & 0.10 &  1         & 3.18 & 14.11 & 0.58    & \nodata & \nodata & 1.02    \\
HV\,2510   & 0.973 & 14.85 & 0.71    & 0.77    & 0.10 &  5         & 3.19 & 14.53 & 0.61    & 0.64    & 0.85    & \nodata \\
HV\,2301   & 0.978 & 13.94 & 0.84    & \nodata & 0.10 &  1         & 3.25 & 13.62 & 0.74    & \nodata & \nodata & 0.80    \\
HV\,12474  & 0.993 & 14.65 & 0.69    & \nodata & 0.10 &  1         & 3.18 & 14.33 & 0.59    & \nodata & \nodata & 0.87    \\
HV\,6105   & 1.019 & 14.91 & 0.79    & \nodata & 0.10 &  1         & 3.23 & 14.59 & 0.69    & \nodata & \nodata & 1.05    \\
HV\,5551   & 1.021 & 15.00 & 0.76    & 0.86    & 0.10 &  5         & 3.21 & 14.68 & 0.66    & 0.73    & 0.65    & \nodata \\
HV\,2432   & 1.038 & 14.23 & 0.52    & \nodata & 0.10 &  1         & 3.11 & 13.92 & 0.42    & \nodata & 0.74    & 0.73    \\
HV\,12248  & 1.038 & 14.47 & 0.75    & \nodata & 0.10 &  1         & 3.21 & 14.15 & 0.65    & \nodata & \nodata & 1.80    \\
HV\,2864   & 1.041 & 14.67 & 0.80    & 0.89    & 0.07 &  1,3,4,6   & 3.23 & 14.44 & 0.73    & 0.80    & 0.81    & 1.24    \\
HV\,12716  & 1.051 & 14.70 & 0.76    & \nodata & 0.10 &  1         & 3.21 & 14.38 & 0.66    & 0.81    & \nodata & 1.24    \\
HV\,2662   & 1.082 & 14.46 & 0.88    & 0.83    & 0.10 &  5         & 3.27 & 14.13 & 0.78    & 0.70    & 0.50    & \nodata \\
HV\,2580   & 1.089 & 14.01 & 0.81    & 0.87    & 0.10 &  7         & 3.24 & 13.69 & 0.71    & 0.74    & \nodata & \nodata \\
HV\,12253  & 1.099 & 14.41 & 0.76    & \nodata & 0.10 &  1         & 3.21 & 14.09 & 0.66    & 0.65    & \nodata & 1.46    \\
HV\,874    & 1.103 & 14.45 & \nodata & 0.71    & 0.10 &  5         & 7.27 & 13.72 & \nodata & 0.58    & 1.10    & 1.70    \\
HV\,2527   & 1.112 & 14.61 & 0.83    & 0.86    & 0.10 &  1,7       & 3.24 & 14.29 & 0.73    & 0.73    & \nodata & 1.52    \\
HV\,2260   & 1.114 & 14.85 & 0.86    & 0.96    & 0.14 &  1,3,4,7   & 3.26 & 14.39 & 0.72    & 0.78    & \nodata & 1.42    \\
HV\,997    & 1.119 & 14.55 & 0.86    & 0.98    & 0.11 &  1,3,4,7   & 3.26 & 14.19 & 0.75    & 0.84    & \nodata & 1.52    \\
HV\,2579   & 1.128 & 13.99 & 0.66    & 0.75    & 0.10 &  1,7       & 3.17 & 13.67 & 0.56    & 0.62    & \nodata & 1.58    \\
HV\,2352   & 1.134 & 14.19 & 0.70    & 0.84    & 0.11 &  1,3,4,7   & 3.19 & 13.84 & 0.59    & 0.70    & \nodata & 1.17    \\
HV\,955    & 1.138 & 14.07 & 0.73    & \nodata & 0.10 &  1         & 3.20 & 13.75 & 0.63    & \nodata & \nodata & 1.27    \\
HV\,2463   & 1.145 & 14.22 & 0.77    & 0.87    & 0.10 &  7         & 3.22 & 13.90 & 0.67    & 0.74    & \nodata & \nodata \\
HV\,5655   & 1.153 & 14.52 & 0.94    & 0.95    & 0.10 &  7         & 3.29 & 14.19 & 0.84    & 0.82    & \nodata & \nodata \\
HV\,2324   & 1.160 & 14.35 & 0.85    & 0.92    & 0.06 &  1,4       & 3.25 & 14.15 & 0.79    & 0.84    & \nodata & 1.44    \\
HV\,12471  & 1.200 & 14.68 & 0.99    & \nodata & 0.06 &  1,3       & 3.31 & 14.48 & 0.93    & 1.05    & 0.88    & 1.39    \\
HV\,2262   & 1.200 & 14.29 & 0.75    & 0.96    & 0.10 &  7         & 3.21 & 13.97 & 0.65    & 0.83    & \nodata & \nodata \\
HV\,2549   & 1.209 & 13.70 & 0.72    & \nodata & 0.06 &  1,3       & 3.19 & 13.51 & 0.66    & 0.71    & \nodata & 1.63    \\
HV\,2580   & 1.229 & 13.96 & 0.75    & 0.86    & 0.11 &  1,3,4     & 3.21 & 13.61 & 0.64    & 0.72    & \nodata & 1.33    \\
HV\,2261   & 1.237 & 13.26 & 0.69    & 0.74    & 0.10 &  7         & 3.18 & 12.94 & 0.59    & 0.61    & \nodata & \nodata \\
HV\,2836   & 1.246 & 14.62 & 0.98    & 1.10    & 0.19 &  1,3,4,7   & 3.31 & 13.99 & 0.79    & 0.86    & \nodata & 1.49    \\
HV\,1005   & 1.272 & 13.96 & 0.82    & \nodata & 0.10 &  1,4       & 3.24 & 13.64 & 0.72    & 0.84    & \nodata & 1.65    \\
HV\,2793   & 1.283 & 14.09 & 1.01    & 1.08    & 0.10 &  1,3,4     & 3.32 & 13.76 & 0.91    & 0.95    & \nodata & 1.59    \\
U\,11      & 1.303 & 13.99 & 0.86    & \nodata & 0.10 &  1         & 3.26 & 13.66 & 0.76    & \nodata & 0.74    & 1.47    \\
U\,1       & 1.353 & 14.10 & 0.98    & \nodata & 0.10 &  1         & 3.31 & 13.77 & 0.88    & \nodata & \nodata & 1.71    \\
HV\,2749   & 1.364 & 14.73 & 1.24    & \nodata & 0.10 &  1         & 3.42 & 14.39 & 1.14    & \nodata & 1.05    & 1.32    \\
HV\,878    & 1.367 & 13.53 & 0.67    & 0.85    & 0.06 &  1,2,3,7   & 3.17 & 13.34 & 0.61    & 0.77    & 1.45    & 1.92    \\
HV\,886    & 1.380 & 13.33 & 0.83    & \nodata & 0.06 &  1,2,3     & 3.24 & 13.14 & 0.77    & 0.62    & 1.28    & 2.08    \\
HV\,1013   & 1.382 & 13.83 & 1.04    & 0.96    & 0.12 &  1,3,4     & 3.34 & 13.43 & 0.92    & 0.81    & \nodata & 1.49    \\
HV\,1003   & 1.388 & 13.25 & 0.71    & \nodata & 0.06 &  1,3       & 3.19 & 13.06 & 0.65    & 0.57    & 1.05    & 1.84    \\
HV\,889    & 1.412 & 13.71 & 0.98    & \nodata & 0.06 &  1,2,3     & 3.31 & 13.51 & 0.92    & \nodata & 0.89    & 1.66    \\
HV\,12815  & 1.417 & 13.48 & 0.95    & 0.99    & 0.08 &  1,3,4,6,7 & 3.30 & 13.22 & 0.87    & 0.89    & 1.08    & 1.84    \\
HV\,902    & 1.421 & 13.22 & 0.70    & \nodata & 0.10 &  1         & 3.19 & 12.90 & 0.60    & \nodata & 1.27    & 2.10    \\
HV\,1023   & 1.425 & 13.75 & 0.97    & 1.03    & 0.07 &  1,3,4,7   & 3.30 & 13.52 & 0.90    & 0.94    & \nodata & 1.47    \\
HV\,2251   & 1.447 & 13.10 & 0.75    & \nodata & 0.10 &  1,3       & 3.21 & 12.78 & 0.65    & 0.44    & 1.15    & 1.70    \\
HV\,2540   & 1.449 & 13.81 & 1.20    & 1.04    & 0.10 &  7         & 3.41 & 13.47 & 1.10    & 0.91    & \nodata & \nodata \\
HV\,8036   & 1.453 & 13.60 & 0.90    & \nodata & 0.10 &  1,3       & 3.27 & 13.27 & 0.80    & 0.68    & \nodata & 2.00    \\
HV\,872    & 1.475 & 13.69 & 0.96    & 0.98    & 0.10 &  7         & 3.30 & 13.36 & 0.86    & 0.85    & \nodata & \nodata \\
HV\,875    & 1.482 & 13.04 & 0.83    & 0.87    & 0.10 &  1,2,7     & 3.24 & 12.72 & 0.73    & 0.74    & 0.45    & 1.00    \\
HV\,1002   & 1.484 & 12.94 & 0.76    & 0.78    & 0.00 &  1,2,3,4   & 3.21 & 12.94 & 0.76    & 0.78    & 1.35    & 2.10    \\
HV\,899    & 1.492 & 13.43 & 0.94    & 1.10    & 0.11 &  1,3,4     & 3.29 & 13.07 & 0.83    & 0.96    & 1.24    & 1.84    \\
HV\,882    & 1.503 & 13.42 & 0.76    & 0.98    & 0.10 &  1,2,7     & 3.21 & 13.10 & 0.66    & 0.85    & 1.07    & 1.68    \\
HV\,873    & 1.536 & 13.52 & 1.09    & 1.36    & 0.10 &  1,7       & 3.36 & 13.18 & 0.99    & 1.00    & 0.71    & 1.08    \\
HV\,881    & 1.553 & 13.11 & 1.04    & 0.92    & 0.10 &  1,2,7     & 3.34 & 12.78 & 0.94    & 0.79    & 1.26    & 1.95    \\
HV\,2294   & 1.563 & 12.66 & 0.82    & 0.95    & 0.06 &  1,2,3,4   & 3.24 & 12.47 & 0.76    & 0.87    & 1.25    & 2.10    \\
HV\,879    & 1.566 & 13.35 & 1.03    & 1.06    & 0.06 &  1,3,4     & 3.33 & 13.15 & 0.97    & 0.98    & \nodata & 1.98    \\
HV\,909    & 1.575 & 12.75 & 0.80    & 1.01    & 0.07 &  1,2,3     & 3.23 & 12.52 & 0.73    & 0.92    & 0.92    & 1.45    \\
HV\,2257   & 1.592 & 13.04 & 0.96    & 1.04    & 0.06 &  1,2,3,4   & 3.30 & 12.84 & 0.90    & 0.96    & 1.30    & 2.06    \\
HV\,2338   & 1.625 & 12.78 & 0.94    & 0.95    & 0.07 &  1,3,4     & 3.29 & 12.55 & 0.87    & 0.86    & 1.14    & 1.94    \\
HV\,877    & 1.655 & 13.35 & 1.20    & 1.19    & 0.11 &  1,2,3,4   & 3.41 & 12.98 & 1.09    & 1.05    & 0.69    & 1.01    \\
HV\,900    & 1.677 & 12.78 & 0.92    & 0.92    & 0.09 &  1,3       & 3.28 & 12.48 & 0.83    & 0.80    & 1.08    & 1.74    \\
HV\,953    & 1.680 & 12.28 & 0.87    & 0.90    & 0.09 &  1,2,3,4   & 3.26 & 11.99 & 0.78    & 0.78    & 0.95    & 1.60    \\
HV\,2369   & 1.684 & 12.61 & 0.96    & 1.02    & 0.10 &  1,3,4     & 3.30 & 12.28 & 0.86    & 0.89    & 1.13    & 1.75    \\
HV\,270100 & 1.872 & 11.79 & 0.96    & \nodata & 0.06 &  3         & 3.30 & 11.59 & 0.90    & \nodata & \nodata & \nodata \\
HV\,2827   & 1.897 & 12.30 & 1.24    & 1.08    & 0.08 &  1,3,4     & 3.42 & 12.03 & 1.16    & 0.98    & \nodata & 0.94    \\
HV\,5497   & 1.995 & 11.93 & 1.20    & 1.13    & 0.09 &  1,2,3,4   & 3.41 & 11.62 & 1.11    & 1.01    & 0.56    & 0.95    \\
HV\,2883   & 2.033 & 12.41 & 1.23    & 1.08    & 0.01 &  1,2,4     & 3.42 & 12.38 & 1.22    & 1.07    & 1.17    & 1.92    \\
HV\,2447   & 2.077 & 12.00 & 1.25    & 1.11    & 0.04 &  1,2,4     & 3.43 & 11.86 & 1.21    & 1.06    & 0.57    & 0.89    \\
HV\,883    & 2.127 & 12.14 & 1.19    & 1.10    & 0.09 &  1,2,4     & 3.40 & 11.83 & 1.10    & 0.98    & 1.23    & 1.80    \\
\end{longtable}

\end{document}